\documentclass[AMA,STIX1COL]{WileyNJD-v2}

\usepackage{yaacro}
\usepackage{graphicx, subfigure}
\usepackage{tikz}
\usetikzlibrary{positioning, fit, shadows, calc}
\usepackage[autolanguage]{numprint}
\let\n=\numprint
\npfourdigitnosep
\hypersetup{draft}
\usepackage[section]{placeins}

\def\revcolor{black}
\usepackage{caption}
\newcommand{\rev}[1]{{\color{\revcolor}#1}}
\newcommand{\revcaption}{\captionsetup{labelfont={color=\revcolor},textfont={color=\revcolor}}}

\articletype{Article Type}%

\received{dd mm 20yy}
\revised{dd mm 20yy}
\accepted{dd mm 20yy}

\begin{acgroupdef}[list=acronimos]
    \acdef{CACC}{Cooperative Adaptive Cruise Control}
    \acdef{RL}{Reinforcement Learning}
    \acdef{DRL}{Deep Reinforcement Learning}
    \acdef{RRL}{Robust Reinforcement Learning}
    \acdef{SRL}[CRL]{Conventional Reinforcement Learning}
    \acdef{CNN}[\rev{ANN}]{\rev{Artificial Neural Network}}
    \acdef{DDPG}{Deep-Deterministic Policy Gradient}
    \acdef{DQN}{Deep $Q$ Network}
    \acdef{DPPO}{Distributed Proximal Policy Optimization}
    \acdef{SAC}{Soft Actor-Critic}
    \acdef{ReLU}{Rectified Linear Unit}
    \acdef{DAG}{Directed Acyclic Graph}
    \acdef[variantof=DAG]{DAGs}{Directed Acyclic Graphs}
    \acdef{PF}{predecessor following}
    \acdef{PFL}{predecessor following leader}
    \acdef{TPF}{two predecessors following}
    \acdef{TPFL}{two predecessors following leader}
    \acdef{CNPq}{Conselho Nacional de Desenvolvimento Científico e Tecnológico}
    \acdef{CAPES}{Coordenação de Aperfeiçoamento de Pessoal de Nível Superior}
    \acdef{FAPEMIG}{Fundação de Amparo à Pesquisa do Estado de Minas Gerais}
\end{acgroupdef}

\begin{document}

\title{Robust Longitudinal Control for Vehicular Autonomous Platoons Using Deep Reinforcement Learning
%\protect\thanks{This is an example for title footnote.}
}

\author[1]{Armando Alves Neto*}
\author[1]{Leonardo Amaral Mozelli}
\authormark{Alves Neto A. and Mozelli L. A.}

\address[1]{\orgdiv{Dep. of Electronics Engineering}, \orgname{Universidade Federal de Minas Gerais}, \orgaddress{\state{Minas Gerais}, \country{Brazil}}}

\corres{*Armando Alves Neto, Dep. of Electronics Engineering, UFMG. \email{aaneto@cpdee.ufmg.br}}

%\presentaddress{This is sample for present address text this is sample for present address text}

%%%%%%%%%%%%%%%%%%%%%%%%%%%%%%%%%%%%%%%%%%%%%%%%%%%%%%%%%%%%%%
\abstract[Summary]{In the last few years, researchers have applied machine learning strategies in the context of vehicular platoons to increase the safety and efficiency of cooperative transportation. \acd{RL} methods have been employed in the longitudinal spacing control of \acd{CACC} systems, but to date, none of those studies have addressed problems of disturbance rejection in such scenarios. Characteristics such as uncertain parameters in the model and external interferences may prevent agents from reaching null-spacing errors when traveling at cruising speed. On the other hand, complex communication topologies lead to specific training processes that can not be generalized to other contexts, demanding re-training every time the configuration changes.
Therefore, in this paper, we propose an approach to generalize the training process of a vehicular platoon, such that the acceleration command of each agent becomes independent of the network topology. Also, we have modeled the acceleration input as a term with integral action, such that the \acd{CNN} is capable of learning corrective actions when the states are disturbed by unknown effects. We illustrate the effectiveness of our proposal with experiments using different network topologies, uncertain parameters, and external forces. Comparative analyses, in terms of the steady-state error and overshoot response, were conducted against the state-of-the-art literature. The findings offer new insights concerning generalization and robustness of using \acd{RL} in the control of autonomous platoons.}


% \jnlcitation{\cname{%
% \author{Williams K.}, 
% \author{B. Hoskins}, 
% \author{R. Lee}, 
% \author{G. Masato}, and 
% \author{T. Woollings}} (\cyear{2016}), 
% \ctitle{A regime analysis of Atlantic winter jet variability applied to evaluate HadGEM3-GC2}, \cjournal{Q.J.R. Meteorol. Soc.}, \cvol{2017;00:1--6}.}

\maketitle

%\footnotetext{\textbf{Abbreviations:} ANA, anti-nuclear antibodies; APC, antigen-presenting cells; IRF, interferon regulatory factor}

%%%%%%%%%%%%%%%%%%%%%%%%%%%%%%%%%%%%%%%%%%%%%%%%%%%%%%%%%%%%%%
% operacoes
\newcommand{\escalar}[1]{\ensuremath{\mathit{#1}}}
\newcommand{\vetor}[1]{\ensuremath{\boldsymbol{#1}}}
\newcommand{\matriz}[1]{\ensuremath{\mathbf{\uppercase{#1}}}}
\newcommand{\conjunto}[1]{\ensuremath{\mathcal{\uppercase{#1}}}}
\newcommand{\distribuicao}[1]{\ensuremath{\mathcal{\uppercase{#1}}}}
\newcommand{\transpose}{\ensuremath{{}^T}}%{\ensuremath{{}^\intercal}}
\newcommand{\espaco}[1]{\ensuremath{\mathbb{\MakeUppercase#1}}}

%%%%%%%%%%%%%%%%%%%%%%%%%
% constantes
\newcommand{\nrobots}{\escalar{n}}

\newcommand{\mass}{\escalar{m}}
\newcommand{\wheel}{\escalar{r}}
\newcommand{\grav}{\escalar{g}}
\newcommand{\dragcoef}{\escalar{C}}
\newcommand{\friction}{\escalar{\zeta}}
\newcommand{\efficiency}{\escalar{\eta}}
\newcommand{\density}{\escalar{\rho}}
\newcommand{\inertialdelay}{\escalar{\varsigma}}
\newcommand{\slope}{\escalar{\phi}}

\newcommand{\rewardgain}{\escalar{\kappa}}
\newcommand{\penaltygain}{\escalar{\beta}}
\newcommand{\discount}{\escalar{\gamma}}
\newcommand{\updateTarget}{\escalar{\tau}}

\newcommand{\dist}[1]{\vetor{\delta}_{#1}}
\newcommand{\xspacing}[1]{\escalar{d}_{#1}}

\newcommand{\Ts}{T_{s}}%{\dot{\uin}}
%%%%%%%%%%%%%%%%%%%%%%%%%
% variaveis
\newcommand{\pos}{\escalar{p}}
\newcommand{\ipos}[1]{\pos_{#1}}

\newcommand{\vel}{\escalar{v}}
\newcommand{\ivel}[1]{\vel_{#1}}

\newcommand{\accel}{\escalar{a}}
\newcommand{\iaccel}[1]{\accel_{#1}}

\newcommand{\uin}{\escalar{u}}
\newcommand{\iuin}[1]{\uin_{#1}}
\newcommand{\duin}{\Delta\uin}%{\dot{\uin}}
\newcommand{\iduin}[1]{\duin_{#1}}

\newcommand{\win}{\escalar{\epsilon}}
\newcommand{\iwin}[1]{\win_{#1}}

\newcommand{\thr}{\escalar{T}}
\newcommand{\ithr}[1]{\thr_{#1}}
\newcommand{\thrRef}{\widetilde{\thr}}

\newcommand{\xv}{\vetor{x}}
\newcommand{\ixv}[1]{\xv_{#1}}
\newcommand{\dixv}[1]{\dot{\xv}_{#1}}

\newcommand{\statev}{\vetor{s}}
\newcommand{\action}{\vetor{a}}
\newcommand{\layers}{\escalar{l}}
\newcommand{\neuron}{\escalar{z}}
\newcommand{\bias}{\vetor{b}}

\newcommand{\error}{\escalar{e}}
\newcommand{\emean}{\bar{\error}}
\newcommand{\ev}{\vetor{e}}

\newcommand{\yout}{\escalar{y}}
\newcommand{\iyout}[1]{\yout_{#1}}

%%%%%%%%%%%%%%%%%%%%%%%%%
% matrices
\newcommand{\Amat}{\matriz{A}}
\newcommand{\Bmat}{\matriz{B}}
\newcommand{\Cmat}{\matriz{C}}
\newcommand{\Kmat}{\matriz{K}}
\newcommand{\Pmat}{\matriz{P}}
\newcommand{\Ident}[1]{\matriz{I}_{#1}}
\newcommand{\Wmat}{\matriz{W}}
\newcommand{\Gammamat}{\matriz{\Gamma}}

%%%%%%%%%%%%%%%%%%%%%%%%%
% conjunto
\newcommand{\neighSet}{\conjunto{N}}
\newcommand{\stateSet}{\conjunto{S}}
\newcommand{\actionSet}{\conjunto{A}}
\newcommand{\Reais}{\ensuremath{\espaco{R}}}

%%%%%%%%%%%%%%%%%%%%%%%%%
% function
\newcommand{\reward}[1][\cdot]{r\left(#1\right)}
\newcommand{\policysym}{\mu}
\newcommand{\policy}[1][\cdot]{\policysym\left(#1\right)}
\newcommand{\policyexp}[1][\cdot]{\policysym'\left(#1\right)} % exploration policy
\newcommand{\policybest}[1][\cdot]{\policy[#1]}%{\policysym^{\star}\left(#1\right)}
\newcommand{\activation}[1][\cdot]{\sigma\left(#1\right)}
\newcommand{\transition}[1][\cdot]{p\left(#1\right)}
\newcommand{\residue}[1][\cdot]{\varrho\left(#1\right)}
\newcommand{\upbound}{\escalar{\bar{k}}}
\newcommand{\lowbound}{\underbar{\escalar{k}}}

\newcommand{\rand}[1][\cdot]{\textrm{rand}\left(#1\right)}

%%%%%%%%%%%%%%%%%%%%%%%%%
% distribuições

\newcommand{\noise}{\distribuicao{N}}

%%%%%%%%%%%%%%%%%%%%%%%%%%%%%%%%%%%%%%%%%%%%%%%%%%%%%%%%%%%%%%
%%%%%%%%%%%%%%%%%%%%%%%%%%%%%%%%%%%%%%%%%%%%%%%%%%%%%%%%%%%%%%
\section{Introduction}
\label{sec:introduction}

In the last two decades, the interest in the study of autonomous platoons for transportation missions has grown significantly, in such a way that many researchers have focused on developing control and decision-making strategies for these scenarios. 
\rev{This interest is motivated by the fact that, from an economic point of view, land transport of goods is a critical issue for many countries, and \emph{vehicular platoons} present many advantages, such as greater throughput and safe navigation, higher passenger comfort, lower energy consumption, and emission of pollutants, among others.}
More recently, traditional planning and control methods have been replaced by new machine learning algorithms, opening up a new range of possibilities to improve safety and efficiency in transportation systems. But despite the rapid progress, most papers have focused more on the internal aspects of neural networks, and less on the practical aspects of the real world \rev{such as uncertainties in vehicle parameters (mass, friction, and inertial delay), team heterogeneity (different types of vehicles or communication flows), and the existence of external perturbations}. So, this brings us to the need of thinking about more robust and general strategies in the context of multi-agent systems.

\begin{figure}[t]
    \centering
    \includegraphics[width=.6\linewidth]{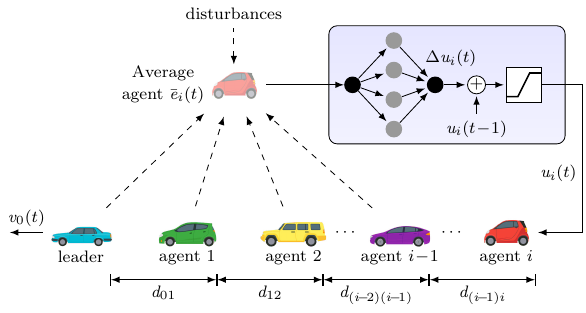}
    \caption{A robust \protect\ac{DRL} proposed approach for linear spacing control in heterogeneous vehicular platoons.}
    \label{fig:problem}
\end{figure}

In this paper, we deal specifically with the problem of null spacing error in the longitudinal spacing control of \ace{CACC} systems. The main idea is to provide a controller that \emph{learns} how to regulate the space between two subsequent agents in a vehicular platoon, such that: $i$) the training stage of the \ac{CNN} is independent of the communication topology among the vehicles; $ii$) the output rejects bounded external disturbances, and $iii$) it operates under bounded parametric uncertainties for the model dynamics.
Our proposed approach is illustrated in Fig.~\ref{fig:problem}, where vehicle $i$ runs a \ac{DRL} controller that receives as input $\emean_{i}(t)$ (the sum of average spacing, speed, and acceleration differences of vehicle $i$ to all its neighbors), and outputs an incremental acceleration command $\iduin{i}(t)$, subsequently integrated and limited by a saturation.

Here, we aim for the following contributions:
\begin{itemize}
    \item since the controller receives as input an error based on the average difference between the states of the vehicle $i$ and its neighbors, we can model nearby teammates as a single-vehicle \rev{subjected to model uncertainties}. It allows us to perform the training process in a simplified scenario with only two vehicles, later generalizing its application to other topologies without the need for a new training process.
    
    \item to handle fluctuations in the average error calculated across all neighbors \rev{(caused by the different dynamics of the heterogeneous neighbors)}, we had to exchange the output of the network from $\iuin{i}(t)$ (commonly used in the literature of \ac{RL} strategies for platoons) to $\iduin{i}(t)$. This modification endowed the \ac{CNN} with the capability to adjust the acceleration command every time the system is disturbed by unpredictable forces. This output must be integrated to obtain $\iuin{i}(t)$ and the resulting command is employed to avoid undesirable effects.
    
    \item concerning the previous topics, we offer a theoretical formalization to justify the improved performance of the proposed modification, in terms of ensuring null spacing error in steady-state conditions, even when subjected to different communication topologies, external forces (such as elevation in the road), \rev{or uncertainty in the model parameters (like mass, tire friction, drag coefficient or inertial delay, among others)}.
\end{itemize}

\rev{When compared to the state-of-the-art literature, our method is completely \emph{model free} in the sense that it doesn't need to know the communication topology of the team. As will be discussed later, other works assume fixed and known communication architectures, which makes the training stage more difficult and less generalizable. On the other hand, similarly to other papers, we still need to have an estimate of the vehicle parameters in order to realize an approximated dynamic cancellation, although, in our case, it doesn't need to be perfect. If the parameters deviate from the nominal values, the proposed approach can mantain a similar performance.} Our strategy is also capable of approximating null steady-state error under the existence of external disturbances and parametric uncertainties, while other papers assume perfect communication and precisely known vehicle dynamics.

The remainder of this paper is structured as follows. In Sec.~\ref{sec:related_work}, we present a review of the current literature. In Sec.~\ref{sec:methodology}, we formalize the problem and discuss the environment setting, while in Sec.~\ref{sec:proposed_method}, we present the proposed methodology. Simulated experiments in Sec.~\ref{sec:experiments} illustrate the effectiveness of our approach when subjected to generalized environments. Finally, in Sec.~\ref{sec:conclusion}, we draw some conclusions and discuss avenues for future investigations.

%%%%%%%%%%%%%%%%%%%%%%%%%%%%%%%%%%%%%%%%%%%%%%%%%%%%%%%%%%%%%%
%%%%%%%%%%%%%%%%%%%%%%%%%%%%%%%%%%%%%%%%%%%%%%%%%%%%%%%%%%%%%%
\section{Related Work}
\label{sec:related_work}

In the field of Transportation Systems, cooperative platoons present many advantages such as greater transport throughput and safe navigation, lower energy consumption and emission of pollutants, and higher passenger comfort, among others. In this context, the recent work of Dey et al. \cite{Dey2016Review} summarizes the literature, where the authors categorize the area in three main issues: $i$) \emph{communication framework}, $ii$) \emph{driver interference}, and $iii$) \emph{control strategies}.

Regarding the control of \ace{CACC} systems, many subtopics have been addressed, ranging from kinematic\cite{Korouglu2017IJRNC}/dynamic\cite{Wang2021IJRNC} modeling to control policy designing, whose analysis incorporates imperfect network\cite{Shen2022TITS}, human-driver interaction\cite{Liu2022TITS}, external disturbances\cite{Gao2016Robust}, and fault tolerance\cite{Godinho:22}, factors that can impair the team stability. Therefore, the development of robust techniques is crucial for the operation in real-world scenarios.
%
%But, although many papers in the literature have concentrated on uncertainty systems, 

Most of the \emph{classical} control laws depend on identifying a mathematical model, many times assuming linear dynamics or using linearization approaches to handle nonlinearities.
A good example is the problem of longitudinal spacing control addressed in Zheng et al. \cite{Zheng2015Stability}, where the authors proposed a distributed protocol to stabilize a platoon with different communication topologies. This work has been extended in Zheng et al. \cite{Zheng2019Cooperative} and Neto et al. \cite{Neto2019Control} to deal with heterogeneous teams, and in Souza et al. \cite{Souza2019Stability} to deal with time-delayed communication. Alternatively, some researchers concentrate on developing nonlinear and/or adaptive control strategies, handling nonlinearities directly instead of resorting to some kind of feedback linearization. For example, a combination of sliding mode control and adaptive terms is proposed in Guo et al.\cite{Guo2017IJRNC}. Results are extended in Guo et al.\cite{Guo2019IJRNC}, where the approximation capabilities of Neural Networks are used to estimate online the uncertain parameters. In both approaches, the spacing error converges to a small neighborhood around the origin, but it cannot be zeroed. Also, there is the possibility of a shattering effect, and the gain design relies on some heuristics.

An important topic that influences the choice of the control strategy, with impacts on the system stability, is the communication topology. Several papers in the literature have used \ac{DAG} models with homogeneous\cite{Bian2019Behavioral} and heterogeneous\cite{Zheng2019Cooperative} platoons to improve the team robustness. \ac{DAGs} are directed graphs with no directed cycles that, in the case of \ac{CACC} systems, provide good performance when using consensus algorithms\cite{Zhang2017Effect}.
Directed graphs have been considered in Santini et al.\cite{Santini2019Platooning}, where the authors formally prove the stability of a control law under topology changes to accommodate new vehicles or to disengage them.
In this paper, we also deal with \ac{DAGs} to model the communication topology of the platoon, as will be further discussed.

In contrast with analytic methods, recent papers are using alternative machine learning approaches - such as \ace{RL} - in the operation of \ac{CACC}s seeking to reduce the dependence on the vehicular models.
Although the literature on \ac{RL} applied to cooperative teams of vehicles has grown significantly in the last five years, the first works in the area are almost two decades old. Concerning spacing control strategies, there are two papers, Luke et al. \cite{Luke2006Decentralized, Luke2008Reinforcement}, using state-action-value functions $Q(s, a)$ to model and solve the problem of longitudinal control of consecutive vehicles in a linear formation.
More recently, with the advent of deep neural networks, techniques based on \ace{DRL} start to be applied to several problems in Robotics \cite{Mnih2015Humanlevel}. In \ac{CACC} systems, Wei et al. \cite{Wei2018Design} proposed a Supervised \ac{RL} approach for the longitudinal control of one leader-follower pair. Their main idea was to learn input commands by interacting with a supervised network trained to emulate the behavior of a human driver. Although interesting, the work considers only one type of network topology, the \ace{PF}, and neglects the existence of external disturbances.

In the sequence, Chu and Kalabić \cite{Chu2019Model} proposes the use of the \ac{DDPG}, a model-free off-policy \ac{RL} method, to compute continuous actions to stabilize a platoon composed of both, human-driven and autonomous vehicles. But although the existence of a human in the control loop may eventually represent a problem, its behavior was modeled according to a deterministic \emph{optimal velocity} function, which makes the problem relatively easier and limited.
In Liu et al. \cite{Liu2020Platoon}, the authors use deep $Q$-networks and a consensus algorithm to learn the longitudinal control of a platoon in a distributed manner. The main idea is that each vehicle trains its $Q$-network by interacting with its front and back neighbors (called bi-directional topology) before sending the network parameters to other agents. With the same information provided by its teammates, it uses a consensus protocol that will make the entire team agree about the optimal $Q$-network.

Haotian et al. \cite{Haotian2021Connected} also address the problem of control in mixed teams, composed of autonomous and human-driven vehicles. Here it uses the \ac{DPPO} algorithm with three main objectives: improve car following efficiency, enhance string stability and decrease fuel consumption. This is a good example of how powerful \ac{DRL} approaches can be since, by decomposing the platoon into subsystems with less than 5 followers, all these objectives can be minimized according to a single (and simple) reward function.
%A diminuição do string stability precisa conhecer a aceleração do lider.
%
A similar scenario is studied in Li et al. \cite{Li2021Reinforcement}, where the main goal is to reduce energy consumption and collision by regulating the acceleration of the vehicle.
Qu et al. \cite{Qu2020Jointly} also use the \ac{DDPG} algorithm to find a policy that simultaneously reduces traffic oscillations and improves electric energy consumption in the spacing control of successive and connected vehicles.

\rev{As far as we know, Zheng’s work \cite{Zheng2015Stability} and many other traditional control methods dealing with longitudinal platoon control also use prior knowledge of the vehicle model to cancel the non-linearity of disturbing effects, such as drag and friction. The main idea is to linearize such models to apply linearized consensus protocols. In this sense, the current \ac{DRL} approaches employ these models to learn linearized vehicles. Although at this point our approach matches others in the literature, our robust policy allows dealing with uncertainties in the model's parameters, which causes an imperfect linearization.
Another important topic is that all papers in the state-of-the-art literature for \ac{DRL} control of platoons heavily depend on the network topology. Therefore, they must know not only the vehicles' parameters but also the communication graph among the agents. If these premises do not hold, the other approaches may underperform or even not work at all. Since our protocol is robust to different topologies, it is model-free in terms of the network structure, while others are not.}

In summary, all the previously mentioned works based on \ac{RL} present some common characteristics. First, they emulate the dynamic behavior of the vehicles neglecting the existence of external disturbances, such as wind and road slope, among others. Second, they assume only homogeneous teams of robots, with known parameters free of uncertainties. And finally, and most important, they use predefined communication topologies in the training stage of the algorithm, which limits its application to platoons with the same configuration. For simplicity reasons, most of the papers use the \ac{PF} topology or others in which the vehicle only communicates with its front and back neighbors.
These three characteristics prevent the learning method from being generalized to situations in which the team is heterogeneous, subject to disturbances, or with topologies different from the one used in the training stage. In other words, they are prone to the phenomenon of overfitting.

Currently, the generalization in \ac{DRL} methods is a topic of increasing interest for the learning community, once it aims to produce policies that can be efficiently extended to unseen or untrained situations. Among many categories of methods for tackling generalization in \ac{RL}, are those based on the idea of changing the optimization objective to improve robustness \cite{Kirk2021Survey}. Robust \acd{RL} approaches generally use the \emph{worst-case} scenario to maximize the obtained police under the existence of model misspecifications\cite{Perrusquia2020IJRNC}, i.e. external disturbances \cite{Mankowitz2020Robust}. The problem with such strategies is that they perform well when perturbation occurs, but they are not optimized for regular conditions. Furthermore, the designer must know all possible worst-case situations to learn robust behaviors, which can be difficult in practice.

Then, concerning the current literature, we propose a \ac{DRL} capable of reacting to disturbances in the input states by computing input increments to the current acceleration command of a vehicle. As will be illustrated with experiments in Sec.~\ref{sec:experiments}, this approach is more robust and generalizable than classical consensus protocol and conventional \ac{DRL} approaches that act directly on the acceleration command.
\rev{Our proposal is similar to the one presented in \cite{Haotian2021MFMARL}, where the authors provide a framework for training a multi-robot systems control policy based on the \emph{mean} information of all neighboring agents. The MF-MARL method uses this \emph{mean} information as part of the state vector to feedback to the agent with the reward in a decentralized manner. But, as a consequence, the agent needs to know the network topology during training, making the solution less generalizable and breaking the model-free requirement. Also, the authors assume a team composed only of homogeneous agents, ruled by a discrete police $\pi(\cdot)$ (with a finite set of states and actions). They are not interested in external disturbances, uncertainty in the agents' parameters, or changing topologies.
On the other hand, our main goal is to provide a policy robust to such perturbations. Then, we use only one neighbor in the training stage and the average of all neighbors in the execution stage, employing an integration term to face the discrepancies between both stages. In addition, we use a continuous police $\mu(\cdot)$, and our team can be a heterogeneous one.
}

%%%%%%%%%%%%%%%%%%%%%%%%%%%%%%%%%%%%%%%%%%%%%%%%%%%%%%%%%%%%%%
%%%%%%%%%%%%%%%%%%%%%%%%%%%%%%%%%%%%%%%%%%%%%%%%%%%%%%%%%%%%%%
\section{Methodology}
\label{sec:methodology}

It is hard to model complex or high nonlinear systems accurately. In the case of vehicular platoons, many factors interfere with the design of a precise model: exit/entrance of new vehicles; coexistence of human and robotic drivers; varying road conditions; communication breakdowns. Therefore, instead of painstakingly characterizing a complete platoon to account for so many factors and then proposing a distributed control law, we can use statistical and machine learning methods to directly design an effective set of controllers, which handles robustly the missing dynamics, uncertainties, and disturbances.

\ace{RL} algorithms are a good choice in this scenario, once they can be regarded as methods where an agent interacts with the environment through a control deterministic policy $\policy$ while it only receives partial information concerning the effectiveness of this interaction (reward/punishment). Over time, the agent refines its interplay with the environment, improving the capacity of making good decisions - maximization of the expected return from the reward function $\reward$.
When concerning \ac{CACC} systems, \ac{RL} methods are generally used in the following problem:

\begin{problem}[\acd{RL} in platooning control]
    Consider a vehicular platoon composed of $\nrobots$ agents following one leader, whose communication topology can be modeled as a \acd{DAG}. To control the inter-spacing distance among the teammates, we have to find a decentralized policy $\policybest$ for each agent $i$ that maps the difference between its states and the states of its neighbors $\neighSet_{i}$ into acceleration/braking commands.
    \label{prob:platoon_RL}
\end{problem}

As previously discussed, most of the current literature on platoon control considers the environment as a fixed structure in the sense that the same topology and number of neighbors are used in both, the training and the execution stages. 
In this context, our proposed approach consists of some key improvements concerning the control scheme and the training method, whose objective is to simultaneously promote generalization and robustness. 
Therefore, we propose a more general setting, as described below:

\begin{problem}[Topology generalization]
    By using a \acd{DRL} approach, the main challenge is to solve Prob.\ref{prob:platoon_RL} with a $\policybest$ that is also able to operate in environments distinct from those used during the training stage. We henceforth refer to this as the \emph{generalization problem}. 
    \label{prob:generalization}
\end{problem}

In this paper, we are interested in the generalization problem when the platoon operates under topologies distinct from the trained topology. Also, we aim to include disturbances in the system, such as road conditions (slopes, wind) and parametric uncertainties (mass, power train). To do so, let us first present the formalization.

%%%%%%%%%%%%%%%%%%%%%%%%%%%%%%%%%%%%%%%%%%%%%%%%%%%%%%%%%%%%%%
\subsection{Environment setting}
\label{subsec:environment}

We begin by considering the longitudinal dynamics of vehicles described in Gao et al. \cite{Gao2016Robust}, which is a variation of Zheng et al. \cite{Zheng2015Stability} concerning parametric uncertainties and external disturbances. Since we consider rigid and symmetrical platforms, free of yaw moments and sliding of tires, a vehicle $i$ can be given by:
\begin{align}
  \ivel{i}(t) &= \dot{\ipos{i}}(t), \nonumber\\
  \mass_{i} \iaccel{i}(t) &= \frac{\efficiency_{i}}{\wheel_{i}}\ithr{i}(t) - \frac{1}{2} \density \dragcoef_{i} \big( \ivel{i}(t) + \vel_{w}(t)\big)^2 - \mass_{i} \grav \friction \cos \slope(t) - \mass_{i} \grav \sin \slope(t), \label{eq:nonlinear_system} \\
  \ithr{i}(t) &= \thrRef_{i}(t) - \inertialdelay_{i} \dot{\thr}_{i}(t), \nonumber
\end{align}
\noindent where $\ipos{i}$, $\ivel{i}$ and $\iaccel{i}$ are position, speed and acceleration, respectively, along the $X$-axis. Also, there is the mass $\mass_{i}$, motor efficiency $\efficiency_{i}$, tire radius $\wheel_{i}$, gravity acceleration $\grav$, friction constant $\friction$, air density $\density$, and drag coefficient $\dragcoef_{i}$. Variables $\vel_{w}$ and $\slope$ are external parameters corresponding to wind speed and road slope, while the last first-order dynamic equation relates the torque $\ithr{i}$ of the vehicle propulsion and braking system with the desired torque $\thrRef_{i}$, associated to the power-train time constant $\inertialdelay_{i}$.

Supposing we know all constant parameter values, and assuming zero $\vel_{w}$ and $\slope$, we can use the feedback linearization in Zheng et al.\cite{Zheng2015Stability}, 
\begin{align}
    \label{eq:Tdes}
    \thrRef_{i}(t) = \frac{\wheel_{i}}{\efficiency_{i}} \bigg[ \frac{1}{2} \density \dragcoef_{i} \ivel{i}(t) \big(2 \inertialdelay_{i} \iaccel{i}(t)\!+\!\ivel{i}(t) \big) %\nonumber\\
    \!+\! \mass_{i} \grav \friction_{i} \!+\! \mass_{i} \iuin{i}(t) \bigg],
\end{align}
\noindent to transform the nonlinear longitudinal dynamics of $i$ into the third-order linear time-invariant model
\begin{align} 
    \label{eq:EEi}
    \dixv{i}(t) &= \Amat_{i} \ixv{i}(t) + \Bmat_{i} \iuin{i}(t) = 
    \def\arraycolsep{5pt}
    \begin{bmatrix}
        0 & 1 &  0 \\[.1cm] 
        0 & 0 & 1 \\[.1cm] 
        0 & 0 & -\dfrac{1}{\inertialdelay_{i}}
    \end{bmatrix} %\ixv{i}(t) + 
    \begin{bmatrix}
      \ipos{i} \\[.1cm] 
      \ivel{i} \\[.1cm]
      \iaccel{i}
    \end{bmatrix} +
    \begin{bmatrix}
      0 \\[.1cm]
      0 \\[.1cm]
      \dfrac{1}{\inertialdelay_{i}} 
    \end{bmatrix} \iuin{i}(t),
\end{align}
\noindent with $\ixv{i} = \begin{bmatrix} \ipos{i} & \ivel{i} & \iaccel{i} \end{bmatrix}\transpose$ and $\iuin{i}$ representing a new (acceleration/braking) input command.
Here it is important to highlight that most of the current literature on classic control and \ac{RL} approaches are based on this linear representation. However, once we are interested in addressing the problem of robust control, we will consider most of the vehicle parameters as being uncertain, assuming the existence of limited wind speed and road slope.
In this sense, the new system can be given by
\begin{align}
    \label{eq:EEi_incerto}
    \dixv{i}(t) &= %\Amat_{i} \ixv{i}(t) + \Bmat_{i} \iuin{i}(t) = 
    \def\arraycolsep{5pt}
    \begin{bmatrix}
        0 & 1 &  0 \\[.1cm] 
        0 & 0 & 1 \\[.1cm] 
        0 & 0 & -\dfrac{1}{\inertialdelay_{i}}
    \end{bmatrix} \ixv{i}(t) + 
    \begin{bmatrix}
      0\\[.1cm]
      0 \\[.1cm]
      \dfrac{1}{\inertialdelay_{i}} 
    \end{bmatrix} \iuin{i}(t) +
    \begin{bmatrix}
      0 \\[.1cm]
      0 \\[.1cm]
      1
    \end{bmatrix} \iwin{i}(t),\\[.2cm]
    \label{eq:out_EE_incerto}
    \iyout{i}(t) &= \Cmat \ixv{i}(t), %= 
    % \begin{bmatrix}
    %     1 & 1 & 1
    % \end{bmatrix} \ixv{i}(t), \nonumber
\end{align}
\noindent where $\iyout{i}$ is the observed output and $\iwin{i}$ is a term that incorporates all uncertainties related to unknown values such as mass, drag coefficient, wind speed, road inclination, and power-train inertia, among others \cite{Gao2016Robust}.
%
% To simulate this environment in a digital computer for purposes of training in a \ac{DRL} algorithm, we need to discretize the vehicle model:
% %
% \begin{align}
%     \ixv{i}(k+1) &= \Amat_{i}^d\ixv{i}(k)+\Bmat_{i}^d\iuin{i}(k) +\Bmat_{w,i}^d\iwin{i}(k)\\
%     &=
%     \def\arraycolsep{5pt}
%     \begin{bmatrix}
%         1 & \Ts &  0 \\[.1cm] 
%         0 & 1 & \Ts \\[.1cm]
%         0 & 0 & 1-\dfrac{\Ts}{\inertialdelay_{i}} 
%     \end{bmatrix}\ixv{i}(k) +
%     \begin{bmatrix}
%       0 \\[.1cm]
%       0 \\[.1cm]
%       \dfrac{\Ts}{\inertialdelay_{i}} 
%     \end{bmatrix} \iuin{i}(k) +
%     \begin{bmatrix}
%       0 \\[.1cm]
%       0 \\[.1cm]
%       \Ts
%     \end{bmatrix} \iwin{i}(k), \nonumber\\
%     %
%     \iyout{i}(k) &= \Cmat_{i}^d \ixv{i}(k) = \begin{bmatrix}
%     1 & 1 & 1     
%     \end{bmatrix} \ixv{i}(k),
%     \label{eq:Cd_x}
% \end{align}
% %
% %\aaneto{Que tal colocar $\iuin{i}(k) = \iuin{i}(k-1) + \iduin{i}(k) \Ts$, e deixar claro que nossa entrada é $\iduin{i}(k)$?}
% %
% %\moz{Acho que isso dilui a importância da mudança de controlador da contribuição se já vier neste ponto. Revisor pode pensar que é o normal.}
% %
% \noindent where the Euler method has been employed considering a time-step of $\Ts$.
% %\moz{Euler tem problema de estabilidade, usar ele mesmo?}

\begin{figure}[t]
    \centering
    \includegraphics[width=.6\linewidth]{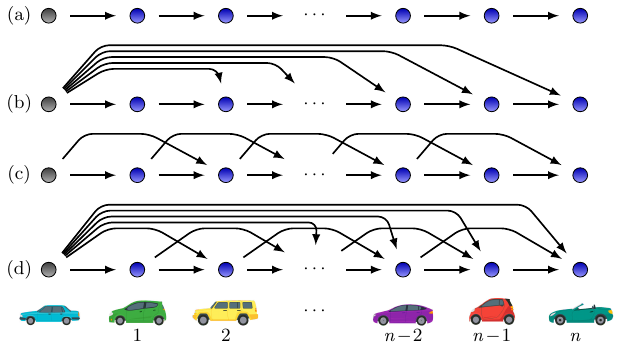}
    \caption{Most common look-ahead topologies for vehicular platoons: (a) \protect\ace{PF}, (b) \protect\ace{PFL}, (c) \protect\ace{TPF}, (d) \protect\ace{TPFL} -- adapted from Zheng et al.\cite{Zheng2019Cooperative}.}
    \label{fig:topologies}
\end{figure}

%%%%%%%%%%%%%%%%%%%%%%%%%%%%%%%%%%%%%%%%%%%%%%%%%%%%%%%%%%%%
Based on the nonlinear (or linearized) dynamics, the next step is to establish a communication topology among the platooning agents and use all information provided by nearby teammates to decide about the action $\iuin{i}(t)$, which is generally performed by each agent in a decentralized manner.
Similar to most of the papers in the literature, we have modeled the platoon connectivity as a time-invariant \ace{DAG}, a finite directed graph with no closed cycles that presents a topological ordering among the vertices. This graph structure allows describing the so-called \emph{look-ahead topologies}, where some of the most common are illustrated in Fig.~\ref{fig:topologies}. An important characteristic of this type of network, in terms of robustness, is that they are free from the influence of successors' oscillations since the vehicle only receives information from its predecessors\cite{Zheng2019Cooperative}.
%
%Here, we assume a time-invariant \ac{DAG} topology with a fixed number $\nrobots$ of vehicles.

Most of the current literature on \ac{RL} applications in platoons are also concentrated on the \ace{PF} topology, Fig.\ref{fig:topologies}(a), the simplest one \cite{Luke2006Decentralized, Luke2008Reinforcement,Wei2018Design,Chu2019Model}. This can be explained by the fact that training with only one neighbor is more easy and fast since it decreases the input dimension of the \ac{CNN}. In other words, the state space used to describe the platoon is smaller. However, an important disadvantage is that the vehicle becomes less robust to uncertainties and more susceptible to external attacks or failures of its single neighbor\cite{Dey2016Review}.
More rarely, some papers have used the \ace{PFL} structure \cite{Li2021Reinforcement}, incorporating the leader information in the control protocol of all agents, which is not very realistic for long platoons. But the fact is that when training \ac{DRL} approaches, one must define the network structure a priori, which breaks the idea of having a complete \emph{model free} policy. Also, some network structures can be complex and too hard to learn. %These concerns lead us to the following problem:
Therefore, our first goal is to propose a modification to generalize the \ac{DRL} strategy, such that it can be applied to all \ac{DAG} topologies, including those in Fig.~\ref{fig:topologies}.

%%%%%%%%%%%%%%%%%%%%%%%%%%%%%%%%%%%%%%%%%%%%%%%%%%%%%%%%%%%%%%
\subsubsection{Typical state/action representation}

In the current literature, consensus protocols applied to the problem of longitudinal spacing control\cite{Zheng2015Stability, Zheng2019Cooperative,Neto2019Control,Souza2019Stability} are generally represented by the form: 
\begin{equation}
    \label{eq:lei_consensus}
    \iuin{i}(t) = -\sum_{j \in \neighSet_{i}} \Kmat \ev_{ij}(t) = -\sum_{j \in \neighSet_{i}} \Kmat
    \begin{bmatrix}
        \ipos{i}(t) - \ipos{j}(t) - \xspacing{ij}\\
        \ivel{i}(t) - \ivel{j}(t)\\
        \iaccel{i}(t) - \iaccel{j}(t)
    \end{bmatrix},
\end{equation}
\noindent with $\xspacing{ij}$ being the desired inter-space among the vehicles, $\neighSet_{i}$ the set with neighbors of $i$, and $\Kmat$ a constant gain vector. These strategies use as the state, a vector containing position, speed, and acceleration error.

Following this principle, recent \ac{DRL} methods addressing longitudinal control of platoons also use as the most common variables the position and speed errors from the vehicle itself and its neighbors, which leads to kinematic models with an acceleration input command \cite{Haotian2021Connected, Li2021Reinforcement, Liu2020Platoon}. Other papers incorporate the dynamics of the vehicles by adding the acceleration error in the state vector, using as input a reference signal for the power-train system\cite{Wei2018Design, Chu2019Model} as presented in Eq.~\eqref{eq:nonlinear_system}.
In the last case, the increase in the state dimension makes the \ac{DRL} algorithm more difficult to train, but the output policy is more ``realistic'' since it can learn the motor time constant while computing $\iuin{i}(t)$.

\rev{In any case, when using \ac{DRL} techniques, the linear (consensus) protocol is replaced by the state/action representation:
\begin{align}
    \statev_i(t) &= \big\{ \ev_{ij}(t) : \ev_{ij}(t) ~~\forall j \in \neighSet_{i} \big\},\\
    \action_i(t) &= \policybest[\statev_i(t)],
    \label{eq:commoon_policy}
\end{align}
\noindent where the state $\statev_i(t)$ is a composition of the errors to other vehicles, and the action $\action_i(t)$ is given by a policy that approximates the optimum behavior, according to some cost (reward) function, only for the specific $\neighSet_{i}$ used in the training stage. In practice, this policy is implemented as a deep neural network, as described in the following. The problem with the such formulation is that $\statev_i(t)$ depends on the number of neighbors (network topology), which limits its application to the trained platoon. This problem will be treated further with the proposal of a novel state/action representation.}

%%%%%%%%%%%%%%%%%%%%%%%%%%%%%%%%%%%%%%%%%%%%%%%%%%%%%%%%%%%%%%
\subsubsection{Control scheme}

The basic architecture of a \ace{CNN} consists of: $i$) an input layer, which receives the state $\statev$, $ii$) $\layers_{\max}$ hidden layers, and $iii$) an output layer, which provides the action $\action$. Individual neurons are indexed according to the number of layer $\layers$ and their position within the layer $j$. So, each neuron produces as output a scalar $\neuron^\layers_j$, which are grouped into the vector $\neuron^\layers = [\neuron_1^\layers, ~ \cdots, ~\neuron_{N_\layers}^\layers]\transpose$. 
In the input layer, $\neuron^0 = \statev$, and in the hidden layers
\begin{align}
    \neuron^{\layers} = \activation[\Wmat^\layers \neuron^{\layers-1} + \bias^\layers], \text{~with~~} \layers = 1, \ldots, \layers_{\max},
    \label{eq:hidden_layers}
\end{align}
\noindent where $\activation$ is the set of activation functions, $\Wmat^\layers \in \Reais^{N_\layers \times N_\layers}$ is a weighting matrix, and $\bias^\layers \in \Reais^{N_\layers}$ is a bias vector, all of which compose the network parameters. Consequently, the output is given as a linear combination of the output of the last hidden layer:
\begin{align}
    \action = \sum^{N_\layers}_{j=1} \alpha_j \neuron^\layers_j + c = \sum^{N_\layers}_{j=1} \alpha_j \activation[\Wmat^{\layers_{\max}} \neuron^{\layers_{\max-1}} + \bias^{\layers_{\max}}] + c.
    \label{eq:out_layers}    
\end{align}

In a nutshell, the basic process of \ac{DRL} can be regarded as a Markov decision process represent by a 4-tuple $\{\stateSet, \actionSet, \transition, \reward\}$, where $\stateSet \in \Reais^{N}$ is the state space, $\actionSet \in \Reais^{M}$ is the action space, $\transition[\statev, \statev'] : \stateSet \times \actionSet \times \stateSet \rightarrow \Reais$ is the transition dynamics from state $\statev$ to the state $\statev'$ in the next step given action $\action$, and $\reward : \stateSet \times \actionSet \rightarrow \Reais$ is the immediate reward function, received by the agent by performing the transition. The accumulated reward at the end of this process is given by
\begin{align}
    J_N = \sum^K_{k=0} \discount^k \reward[\statev_k, \statev_{k+1}],
    \label{eq:accumulated_reward}
\end{align}
\noindent where $\discount \in [0 \ldots 1]$  is a discount factor. 

The general objective of a \ac{DRL} algorithm is to train a \ac{CNN} using it to obtain a policy $\policy[\statev]$ that maximizes the cumulative reward $J_N$ while attaining state $\statev_N$. 
%
%In this paper, the policy will serve as a controller for the input $\iduin{i}(k)$ in the vehicles traveling in the platoon, as illustrated in Fig.~\ref{fig:problem}. The details of the policy are presented in the following sections, associating them with the attributes of the platoon control.

%%%%%%%%%%%%%%%%%%%%%%%%%%%%%%%%%%%%%%%%%%%%%%%%%%%%%%%%%%%%%%
\subsubsection{Reward function}

The reward function $\reward$ is a very important portion of \ac{DRL} approaches since it is used to adjust the policy. In other words, given a set of control objectives, the algorithm will search for a $\policybest$ that maximizes $\reward$. Once the longitudinal spacing control in \ac{CACC} systems is a problem of regulation with continuous inputs, most of the current literature employs reward functions that are a weighted sum of the errors in position, speed, and sometimes acceleration \cite{Wei2018Design, Chu2019Model, Haotian2021Connected, Li2021Reinforcement}. While reducing the error distance between consecutive vehicles in steady-state is generally the main goal, other papers use the reward function to reduce fuel consumption or to increase the string stability of the team.
But typically, in all these works, the main control objective is ensuring null steady-state spacing error among the vehicles, and the minimum representation for $\reward$ can be given by:
\begin{equation}
    \reward[t] = \exp \left\{ 
        - \begin{bmatrix} \ev_{ij}(t)\transpose & \iuin{i}(t) \end{bmatrix} 
        \begin{bmatrix} & \Gammamat & & 0 \\ 0 & 0 & 0 & \penaltygain \end{bmatrix} 
        \begin{bmatrix} \ev_{ij}(t) \\ \iuin{i}(t) \end{bmatrix}
    \right\} ~~\forall j \in \neighSet_{i},
    \label{eq:reward}
\end{equation}
\noindent where $\Gammamat$ is a positive diagonal matrix and $\penaltygain > 0$ is an action penalty coefficient. Once again, we can see that the reward, like the adjusted control policy, depends on the network topology, breaking the idea of model-free representation. But even in this context, most of the current literature on \ac{DRL} platoons use off-policy approaches to find $\policybest$.

%%%%%%%%%%%%%%%%%%%%%%%%%%%%%%%%%%%%%%%%%%%%%%%%%%%%%%%%%%%%%%
\subsection{Deep-Deterministic Policy Gradient}

To define the best policy $\policybest$, we have used the \ace{DDPG}\cite{Lillicrap2016ContinuousCW}, an off-policy actor-critic-based approach capable of learning continuous space policies (actor) and the corresponding $Q$-value function (critic) by using approximation functions. \rev{The \ac{DDPG} has been proposed as an extension of the \ac{DQN} \cite{Mnih2015Humanlevel}, becoming one of the first \ac{DRL} approaches to deal with continuous actions. It has already been employed in the longitudinal spacing control of \ac{CACC} systems with good results \cite{Chu2019Model, Qu2020Jointly}, although other methods such as \ac{DQN}, \ac{DPPO} and \ac{SAC} have also been used.}

The algorithm uses an experience replay buffer to minimize correlations between samples during the training stage, and separate target networks to improve the stability of the learning process.
It also uses a parameter $\updateTarget$ to \emph{slowly} update the weights $\theta'$ of the target networks based on the learned networks, such that:
\begin{equation}
    \theta' = \updateTarget \theta + (1 - \updateTarget) \theta',
\end{equation}
\noindent with $\updateTarget \ll 1$. This means that the lower the $\updateTarget$ value, the more slowly and stably $\theta'$ tracks $\theta$.

The \ac{DDPG} strategy explores the environment by adding noise to the exit of the actor policy
\begin{equation}
    \policyexp[\statev] = \policy[\statev|\theta] + \noise,
\end{equation}
\noindent where $\noise$ is often modeled as an Ornstein-Uhlenbeck noise process due to its efficiency in the exploration and control of dynamic systems. Concerning the accumulated reward of Eq.~\eqref{eq:accumulated_reward}, the actor network is then updated by following the policy gradient:
\begin{equation}
    \nabla_{\theta\policy} J \approx \frac{1}{N} \sum_{i} \nabla_{a} Q(\statev, \action | \theta^{Q}) \big| _{\statev = \statev_{i}, \action=\policy[\statev_{i}]} \nabla_{\theta^{\policy}} \policy[\statev | \theta^{\policy}] \big| _{\statev_{i}}.
\end{equation}

%%%%%%%%%%%%%%%%%%%%%%%%%%%%%%%%%%%%%%%%%%%%%%%%%%%%%%%%%%%%%%
%%%%%%%%%%%%%%%%%%%%%%%%%%%%%%%%%%%%%%%%%%%%%%%%%%%%%%%%%%%%%%
\section{Proposed method}
\label{sec:proposed_method}

Here we describe our main contributions to address Prob.~\ref{prob:generalization}. The methodology described in the previous section is modified to improve the generalization of the training stage and enhance the ability of the policy to compensate for parametric deviations and external factors in platoon control.  

%%%%%%%%%%%%%%%%%%%%%%%%%%%%%%%%%%%%%%%%%%%%%%%%%%%%%%%%%%%%%%
\subsection{Alternative training environment}

Since we are considering \ac{DAG} topologies, actions and states of successor agents have no impact on the vehicles ahead of them. Therefore, regarding the point of view of a single agent $i$, the effective environment consists of interactions with its predecessors in $\neighSet_{i}$. In the literature on linear control for platoons modeled as \ac{DAG}, a well-known fact is that better connectivity with the leader or with more neighbors can improve individual performance. In this sense, as will be illustrated in the experiments, the \ac{TPF} topology is better than the \ac{PF}, and the \ac{TPFL} surpasses the \ac{TPF}. At the same time, these improvements in terms of performance can degrade the generalization properties. 

For instance, the reader is referred to the work of Godinho et al.\cite{Godinho:22}, examining the problem of re-configurable platoons with arbitrary entrances and exits of vehicles. The paper demonstrates that given a stabilizing matrix $\Kmat$ in Eq.~\eqref{eq:lei_consensus} for the \ac{PF} topology, then this same controller can stabilize platoons with distinct topologies, possibly with more than one neighbor. However, the opposite is not always true.
Transposing this knowledge into the general context of \acd{DRL}, the use of specific environments can lead to overfitting, resulting in great errors in new scenarios. In the traditional scope, a given policy $\policybest$ capable of stabilizing a platoon in the \ac{PFL} topology may not be useful to the \ac{PF} topology, for instance. Even worse, it can lead to an unstable platoon, with unbounded spacing errors.  
Therefore, we propose a new training environment consisting of a couple of agents, one leader and one follower. We conjecture that a well-adjusted \ac{DRL} control for this parsimonious environment can be properly generalized when more information about extra predecessors is included, therefore avoiding overfitting. %Considering how humans drive, this proposition is intuitively adequate. It is harder to follow a big truck, that blocks some line of sight, relying only on local information than when the driver follows another car and has more information about the road ahead. 

To present our new state/action representation, let us go back to the linear case and reformulate Eq.~\eqref{eq:lei_consensus}, such that:
\begin{align}
    \iuin{i}(t) &= - |\neighSet_{i}| ~ \Kmat 
    \begin{bmatrix}
        \ipos{i}(t)\\
        \ivel{i}(t)\\
        \iaccel{i}(t)
    \end{bmatrix}
    + \sum_{j \in \neighSet_{i}} \Kmat
    \begin{bmatrix}
        \ipos{j}(t) + \xspacing{ij}\\
        \ivel{j}(t)\\
        \iaccel{j}(t)
    \end{bmatrix}, \nonumber\\[.2cm]
    & \rev{= - \Kmat 
    \begin{bmatrix}
        \ipos{i}(t)\\
        \ivel{i}(t)\\
        \iaccel{i}(t)
    \end{bmatrix}
    + \frac{1}{|\neighSet_{i}|} \sum_{j \in \neighSet_{i}} \Kmat
    \begin{bmatrix}
        \ipos{j}(t) + \xspacing{ij}\\
        \ivel{j}(t)\\
        \iaccel{j}(t)
    \end{bmatrix}
    + \residue[\neighSet_{i}].}
    \label{eq:lei_consensus_reescrita}
\end{align}
Here, one can see that the first two terms of Eq.~\eqref{eq:lei_consensus_reescrita} represent the error between the states of agent $i$ and the average states of all its $|\neighSet_{i}|$ neighbors, while $\residue[\neighSet_{i}]$ is a residual unknown value. In other words, it is possible to feed the system with a protocol that considers all nearby vehicles as a single agent which approximates the average value of the states, except for a compensation term $\residue[\neighSet_{i}]$.

Bearing this in mind, it is possible to think of a \ac{DRL} strategy in which a vehicle learns to follow one single \emph{average virtual} neighbor, such that its training stage will be the simplest possible. 
When applied to vehicles in the \ac{PF} topology, this virtual agent will be played by the only neighbor of $i$ and $\residue[\neighSet_{i}] = 0$. However, for different topologies, it will represent the average value of all nearby vehicles, and $\residue[\neighSet_{i}]$ will be a disturbance in our controller. On one hand, the average term in Eq. \eqref{eq:lei_consensus_reescrita} will provide more information about the platoon conditions. With more neighbors, however, possible noise or missing information will be filtered out. On the other hand, there is the approximation residue that must be compensated. Therefore, if we can provide a policy robust to limited disturbances, a \ac{CNN} trained just with two vehicles might be generalized to other topologies. 
This topic will be discussed in the next sections. For now, we propose the output in Eq.~\eqref{eq:out_EE_incerto} as the combination of position, velocity, and acceleration errors to compose our state $\statev(t)$ as the average error:
\begin{equation}
    \emean_i(t) = \iyout{i}(t) - \frac{1}{|\neighSet_{i}|} \sum_{j \in \neighSet_{i}} \Big(y_j(t) - \xspacing{ij} \Big).
    \label{eq:mean_error}
\end{equation}

Another clear advantage of the proposed environment is the reduction in terms of the state-space dimension. Since this dimension affects the convergence and training time, the reduction can promote faster tuning. So, to compensate for the environment simplicity, the reduced time spent allows training with more complex episodes and scenarios. Since we chose to use a more compact training environment (model) we intend to compensate for the uncertainties and disturbances with some improvements in the control policy, as described in the following sections.

\subsection{Alternative state/action and reward representation}

In the literature on vehicular platoons, as far goes the authors' knowledge, the actions selected by the \ac{DRL} agent are the direct values of the input command in \eqref{eq:EEi}, given by the policy \eqref{eq:commoon_policy}.
As shown in the literature, this policy can be regarded as nonlinear static controller, which can provide a good transitory regime in the longitudinal control of vehicles. In the general case, since the platoon topology can be complex and there many states interconnected, the theory of nonlinear systems can provide answers regarding the stability and steady-state response, see for instance\cite{Kim2018NN,Yin2022TAC} and references therein.  

In this paper, besides the modifications explained in the environment, we choose to simplify the state/action and the structure of the policy. 
\rev{By choosing $\Cmat = [1~1~1]$ in Eq.~\eqref{eq:out_EE_incerto}, we use Eq.~\eqref{eq:mean_error} to define our novel state and action representation as:
\begin{align}
    \statev_i(t) &= \emean_i(t) = \ipos{i} + \ivel{i} + \iaccel{i} - \frac{1}{|\neighSet_{i}|} \sum_{j \in \neighSet_{i}} \Big( \ipos{j} + \ivel{j} + \iaccel{j} - \xspacing{ij} \Big),\\
    \action_i(t) &= \policy[\emean_i(t)].
    \label{eq:control_policy_new}
\end{align}
}

Activation functions for \ac{CNN}s (ReLU, sigmoid, tanh, etc) are generally nonlinear, and the compound of successive activation functions $\activation$ within the layers results in policies with static (memory-less) non-linearities, satisfying the following assumptions:
\begin{assumption}
    Let the nonlinear mapping $\policy: \Reais \rightarrow \Reais$, with $\policy[0] = 0$. Considering $\lowbound < \upbound <\infty$, we assume that:
    \[
        (\policy[\statev]-\lowbound\statev)(\policy[\statev]-\upbound\statev)\leq 0,~\forall s \in \Reais,
    \]
    \noindent i.e., the policy is \emph{sector bounded} within $[\lowbound,\upbound]$.
    \label{assu:bounded}
\end{assumption}

\begin{assumption}
    The nonlinear mapping $\policy: \Reais \rightarrow \Reais$ locally satisfies Assumption~\ref{assu:bounded} with $0 < \lowbound < \upbound$.
    \label{assu:local}
\end{assumption}

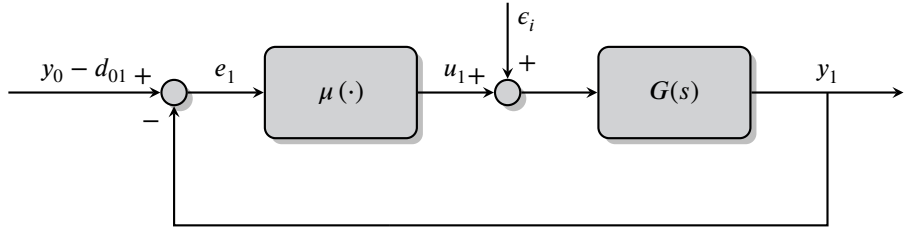
\begin{figure}
    \centering
    \tikzstyle{box} = [draw, minimum width=2cm, minimum height=1.2cm, align=center, rounded corners, fill=lightgray, drop shadow]
    \tikzstyle{sum} = [draw, circle, minimum size=0.3cm, fill=gray!50, fill=lightgray, drop shadow]
    
    \begin{tikzpicture}[thick]
 
    \node [] (input) at (0,0){};
     
    \node[sum, right=2cm of input] (sum) {};
    
    \node [box, right=1cm of sum]  (controller) {$\policy$};
    
    \node[sum, right=1cm of controller] (sum2) {};
    
    \node [box, right=1cm of sum2] (system) {$G(s)$};
    
    \node [below right= 1cm and -0.5cm of controller]  (feedback) {};
    
    \node [above = 1cm of sum2]  (disturbance) {};
    
    \draw[-stealth] (input.east) -- (sum.west) node[midway,above]{$y_0-\xspacing{01}$};
    
    \draw[-stealth] (controller.east) -- (sum2.west) 
        node[midway,above]{$u_1$};
        
    \draw[-stealth] (disturbance.south) -- (sum2.north) 
        node[near start,right]{$\iwin{i}$};
        
    \draw[-stealth] (sum2.east) -- (system.west) ;
        
    \draw[-stealth] (sum.east) -- (controller.west) 
        node[midway,above]{$e_1$};
        
    \draw[-stealth] (system.east) -- ++ (2,0) 
        node[midway](output){}
        node[midway,above]{$y_1$};
        
    \draw[-] (output.center) |- (feedback.center) ;
    
    \draw[-stealth] (feedback.center) -| (sum.south) ;
     
    \node[above left=-1pt of sum.west]{$+$}; 
    
    \node[below left=2pt of sum.south] {$-$}; 
    
    \node[above left=-1pt of sum2.west]{$+$};
    
    \node[above right=-1pt of sum2.80]{$+$};
 
\end{tikzpicture}

    \caption{Nonlinear control structure for the case of the training environment.}
    \label{fig:nonlinear_loop}
\end{figure}

These interesting characteristics of \ac{CNN}s motivated the development of methodologies to estimate tight bounds that encapsulate the nonlinear mapping, to certify classifiers\cite{Zhang2018Neur} and safety\cite{Fazlyab2022TAC}. Considering the training environment, with one virtual neighbor (reference) and one follower, the control loop of Fig.~\ref{fig:nonlinear_loop} can be proposed, where:
\begin{align}
    G(s) = \Cmat_{i}(s\Ident{3} -\Amat_{i})^{-1}\Bmat_i = \dfrac{s^2+s+1}{\inertialdelay_{i} s^2\left(s+\dfrac{1}{\inertialdelay_{i}}\right)}.
    \label{eq:Gs}
\end{align}

%%%%%%%%%%%%%%%%%%%%%%%%%%%%%%%%%%%%%%%%%%%%%%%%%%%%
%%%%%%%%%%%%%%%%%%%%%%%%%%%%%%%%%%%%%%%%%%%%%%%%%%%%

This control loop is closely associated with the general problem of feedback interconnection of a linear system and a static cone bounded nonlinearity, the so-called \emph{Lur'e problem}, and its stability properties studied according to the absolute stability theory.
The following Lemma allows us to investigate the stability using frequency domain information of the environment $G(s)$: 

\begin{lemma}[Adapted from Sandberg and Johnson\cite{Sandberg92}]
    \label{lem:circ_nyq}
    Consider a single-input–single-output system of the form \eqref{eq:EEi_incerto}, where $\{\Amat_i, \Bmat_i, \Cmat_i\}$ is a minimal realization of $G(s)$, and Assumption~\ref{assu:bounded} (or \ref{assu:local}) holding true. Then, the system is absolutely stable (with a finite domain) if the following conditions are satisfied:
    \begin{enumerate}
        \item the Nyquist plot of $G(s)$ does not enter the circle $O_r$, with radius $r=1/2[\lowbound^{-1}-\upbound^{-1}]$, centered at $(-1/2[\lowbound^{-1}+\upbound^{-1}],0)$.
        \item the Nyquist plot of $G(s)$ encircles the circle $O_r$ $m$ times in the counterclockwise direction, where $m$ is the number of unstable poles of $G(s)$ .
    \end{enumerate}
\end{lemma}

In the framework proposed, the reference signal in Fig.~\ref{fig:nonlinear_loop} is replaced by the average term in \eqref{eq:mean_error}, so the analysis remains the same. The idea is to use only the sector information about $\policy$, the bounds $\lowbound$ and $\upbound$ instead of the complete description, to investigate the stability, which is a strong result since it applies to a large class of static nonlinearities. The drawback is that the result can be conservative, since no explicit information about $\policy$ is used. 

\rev{Another relevant concept present in the study of autonomous platoons is string stability. There are many definitions in the area, as cataloged by the survey\cite{Feng2019ARC}. Some definitions can be immediately applied to platoon systems with general topologies, whereas other definitions are restricted to the predecessor-following scheme. Lets focus on the Time-domain String Stability (TSS), whose advantages are no linearity assumption and being suitable to many topologies.

Under assumption \ref{assu:bounded} (or \ref{assu:local}), the dynamics of the states for an interconnection of vehicles with the $\policy$ in a look-ahead platoon, can be rewritten as:

\begin{equation}
    \dot{\statev_{i}}(t) = f_i(\statev_i(t),\statev_{ij}(t)), ~\forall i = \{1,2,\cdots,\nrobots\},~ \forall j\in \neighSet_i \label{eq:interconnection_state},
\end{equation}

\noindent leading to the following result:

\begin{lemma}[Adapted from Feng et al.\cite{Feng2019ARC}]
The origin $\statev_{i} =0$, $i \in \nrobots$, has TSS if given any $\varphi>0$ there exists $\psi>0$  such that $\|\statev_{i}(0)\|_\infty < \varphi \Rightarrow \sup_i \|\statev_{i}(\cdot)\|<\psi$.
\end{lemma}

This result is connected to the Weak Coupling Theorem for string stability\cite{Feng2019ARC} and the demonstration follows similar lines as in \cite{Swaroop1996}. Lemma~\ref{lem:circ_nyq} imposes that $\dot{\statev_{i}}(t) = f_i(\statev_i(t),0)$ is stable and assumption~\ref{assu:bounded} (or \ref{assu:local}) indicates that $f(\cdot)$ is Lipschitz in its arguments.}

%If Lemma~\ref{lem:circ_nyq} holds, the then TSS of the platoon is guaranteed, according to the following result: 

% An alternative to the graphical approach of Lemma~\ref{lem:circ_nyq} is to resort to the Lyapunov stability and the LMI formulation, which can be numerically tracked as the following feasibility problem:
% %
% \begin{lemma}[Adapted from...]
%     \label{lem:circ_lmi}
%     Consider a single-input–single-output system of the form \eqref{eq:EEi_incerto}, where $\{\Amat_i,\Bmat_i, \Cmat_i\}$ is a minimal realization of $G(s)$ and Assumptions~\ref{assu:bounded} and \ref{assu:local} hold true. Then, the system is absolutely stable if there exist matrix $P\geq0$ and scalar $\gamma>0$ such that the following LMI is satisfied:
%     %
%     \begin{align}
%         \begin{bmatrix}
%             \Amat_i^T \Pmat + \Pmat \Amat_i -2\gamma \psi \Cmat_i^T \Cmat_i & \Pmat \Bmat_i +\gamma\nu \Cmat_i^T\\
%             \Bmat_i^T \Pmat +\gamma\nu \Cmat_i & -2\gamma \\
%         \end{bmatrix} \leq 0,
%     \end{align}
%     %
%     \noindent where $\psi = \lowbound+\upbound$ and $\nu = \lowbound\upbound$.
% \end{lemma}

Although these stability certificates are important, the problem of reference tracking, disturbances rejection, and parametric uncertainties in the context of \ac{DRL} platoon control and training generalization remains to be addressed.
\begin{problem}[Robustness Problem]
    \label{prob:robustness}  
    By using a Deep Reinforcement Learning approach, the main challenge is to solve Problems \ref{prob:platoon_RL} and \ref{prob:generalization} with a policy $\policy$ that is also capable of compensating factors neglected during the training stage (such as disturbances, parametric uncertainties, distinct communication topologies), ensuring null steady-state error in terms of inter-vehicle spacing.
\end{problem}

% Consider once again the platoon of Prob.~\ref{prob:generalization}, where each agent is ruled by the policy $\policybest$ trained by using the environment composed by agent $i$ and its virtual average neighbor described by Eq.~\eqref{eq:mean_error}. According to Eq.~\eqref{eq:lei_consensus_reescrita}, we can assume that topologies different from the \ace{PF} will cause a disturbance in the control protocol. In addition, Eq.~\eqref{eq:nonlinear_system} can also incorporate other sources of model misspecifications, such as external disturbances and parametric uncertainties. Then, concerning all these interference, neglected in the training stage, our second goal is to propose a control scheme capable of rejecting such disturbances and ensuring null steady-state error.
  
Since the nonlinear controller $\policy$ is static, the properties for reference tracking or disturbance rejection mainly stem from the environment characteristics, specifically from the system type\footnote{Number of poles at the origin (integrators).}, as discussed in\cite{Sandberg92,Sandberg92TAC}.    
  
\begin{lemma}[Adapted from Sandberg and Johnson\cite{Sandberg92TAC}]
    Consider that the configuration shown in Fig.~\ref{fig:nonlinear_loop} satisfies the conditions of Lemma~\ref{lem:circ_nyq}. If $r=y_0-\xspacing{01}$ approaches a limit $r^{\infty}$ as $t\rightarrow \infty$, then $\emean_{i}^{\infty} = \lim_{t\rightarrow \infty} \emean_{i}(t)$ exists. Moreover $\emean_{i}^{\infty}\neq0$ if and only if $r^\infty\neq0$ and $m=0$, where $m$ is the number of integrator poles of $G(s)$. 
\end{lemma}

%\rev{Furthermore, if the error convergence is guaranteed, a stronger concept can be drawn, the Asymptotic Time-Domain String Stability (ATSS):

Since the proposed model possesses two integrator poles, according to Eq.\eqref{eq:Gs}, any stabilizing policy can guarantee null error in steady-state for step and ramp inputs. Nevertheless, since the policy is memory-less it does not carry these properties in the case of disturbances signals $\iwin{i}(t)$. 
For this reason, we proposed an improved modification to the policy. In practical terms, the actions selected by the policy under the \ac{DRL} training are incremental values for the input command in \eqref{eq:EEi}, instead of the direct values:
\begin{align}
    \iuin{i}(t) &= \textrm{sat}\left[ \int_{0}^{t} \policy[\emean_{i}(t)] dt \right], \nonumber\\
                &\approx \textrm{sat}\left[\sum^{t}_{j=0} \policy[\emean_{i}(j)] \Ts\right] = \textrm{sat}\left[\iuin{i}(t-\Ts) + \iduin{i}(t) \Ts\right],
    \label{eq:incremental_new_policy}
\end{align}
\noindent with $\Ts$ being a small time-step discretization, and
\begin{equation}
    \textrm{sat}\left[\uin\right] =  
    \begin{cases}
         \iuin{max} & \text{if~} \uin > \iuin{max} \\
         \iuin{min} & \text{if~} \uin < \iuin{min} \\
         \uin & \text{otherwise}
    \end{cases},
\end{equation}
\noindent being the saturation function, used to limit the acceleration command into the bounds of the real system.

% Another equivalent representation of Eq.~\eqref{eq:incremental_new_policy} is:

% \begin{align}
%     u_i(k)  &=  \sum^{k}_{j=0} \policy[\emean_{i}(j)]
%     \label{eq:integral_new_policy}
% \end{align}

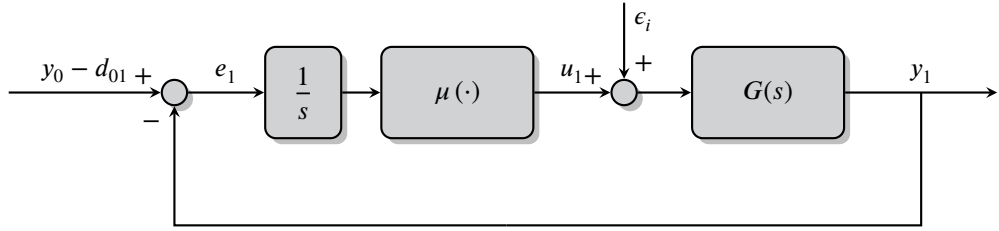
\begin{figure}
    \centering
    \tikzstyle{box} = [draw, minimum width=2cm, minimum height=1.2cm, align=center, rounded corners, fill=lightgray, drop shadow]
    \tikzstyle{sum} = [draw, circle, minimum size=0.3cm, fill=gray!50, fill=lightgray, drop shadow]
    
    \begin{tikzpicture}[thick]
 
    \node [] (input) at (0,0){};
     
    \node[sum, right=2cm of input] (sum) {};
    
    \node [box, minimum width=1cm, right=1cm of sum]  (integral) {$ \dfrac{1}{s}$};
    
    \node [box, right=.5cm of integral]  (controller) {$\policy$};
    
    \node[sum, right=1cm of controller] (sum2) {};
    
    \node [box, right=.7cm of sum2] (system) {$G(s)$};
    
    \node [below right= 1cm and -0.5cm of controller]  (feedback) {};
    
    \node [above = 1cm of sum2]  (disturbance) {};
    
    \draw[-stealth] (input.east) -- (sum.west) node[midway,above]{$y_0-\xspacing{01}$};
    
    \draw[-stealth] (controller.east) -- (sum2.west) 
        node[midway,above]{$u_1$};
        
    \draw[-stealth] (disturbance.south) -- (sum2.north) 
        node[near start,right]{$\iwin{i}$};
        
    \draw[-stealth] (sum2.east) -- (system.west) ;
        
    \draw[-stealth] (sum.east) -- (integral.west) 
        node[midway,above]{$e_1$};
        
    \draw[-stealth] (integral.east) -- (controller.west) 
        node[midway,above]{};
        
    \draw[-stealth] (system.east) -- ++ (2,0) 
        node[midway](output){}
        node[midway,above]{$y_1$};
        
    \draw[-] (output.center) |- (feedback.center) ;
    
    \draw[-stealth] (feedback.center) -| (sum.south) ;
     
    \node[above left=-1pt of sum.west]{$+$}; 
    
    \node[below left=2pt of sum.south] {$-$}; 
    
    \node[above left=-1pt of sum2.west]{$+$};
    
    \node[above right=-1pt of sum2.80]{$+$};
 
\end{tikzpicture}

    \caption{Proposed nonlinear control structure for the case of the training environment.}
    \label{fig:new_nonlinear_loop}
\end{figure}

Some integral representations are not suitable for computational implementation because of memory overhead, however, they enables us to devise the configuration shown in Fig.~\ref{fig:new_nonlinear_loop}. Considering the modified control loop and assuming that the $\policy \in \mathcal{C}^1$, which is a mild assumption considering \ac{CNN} with continuous activation functions, the following result can be established:
\begin{theorem}[Adapted from Sandberg and Johnson\cite{Sandberg92}]
    \label{theo:steady_state_lin_to_nolin}
    Consider that the configuration shown in Fig.~\ref{fig:nonlinear_loop} satisfies the conditions of Lemma~\ref{lem:circ_nyq} when $G(s)$ is replaced by $s^{-1}G(s)$. Let $\emean^0_{i}(t)$ be $\emean_{i}(t)$ when the policy $\policy$ is replaced by a linear constant, $\escalar{k}$. Thus: 
    \begin{equation}
        \emean_{i}^0(t) \rightarrow 0 \Rightarrow \emean_{i}(t) \rightarrow 0.     
    \end{equation}
\end{theorem}
Therefore, to guarantee the reference tracking properties it remains to show that the conditions of Theo.~\ref{theo:steady_state_lin_to_nolin} are satisfied for the training environment. There are 3 integrator poles in $s^{-1}G(s)$, so for a linear policy $\policy[\emean_i(t)] = \escalar{k}\emean_i(t)$ is straightforward to show that $\emean^0_{i}(t)\rightarrow 0$ for a proper gain. 

%%%%%%%%%%%%%%%%%%%%%%%%%%%%%%%%%%%%%%%%%%%%%%%%%%%%
%%%%%%%%%%%%%%%%%%%%%%%%%%%%%%%%%%%%%%%%%%%%%%%%%%%%

Since we have modified the state representation of our \ac{DRL} algorithm, we have also used a different reward function:
\begin{equation}
    \reward[t] = \exp \left\{ 
        - \begin{bmatrix} \emean_i(t) & \iuin{i}(t) \end{bmatrix} 
        \begin{bmatrix} \rewardgain & 0 \\ 0 & \penaltygain \end{bmatrix} 
        \begin{bmatrix} \emean_i(t) \\ \iuin{i}(t) \end{bmatrix}
    \right\},
    \label{eq:alternative_reward}
\end{equation}
\noindent where $\rewardgain$ is a positive gain.

Based on these results, the proposed augmented policy, which provides incremental values for the input command, has the same advantages as the traditional policies in terms of reference tracking, which provides a direct value for the input command. Additionally, the proposed method can improve robustness, since it can benefit from the good features of having an integral term inside the controller to compensate for disturbances and parametric uncertainties in the control loop.

\rev{Under these circumstances, the dynamics of the states for an interconnection of vehicles with the augmented $\policy$ in a look-ahead platoon, can be rewritten as:

\begin{equation}
    \dot{\statev_{i}}(t) = \hat{f}_i(\statev_i(t),\statev_{ij}(t)), ~\forall i = \{1,2,\cdots,\nrobots\},~ \forall j\in \neighSet_i \label{eq:interconnection_state}.
\end{equation}

Given convergence properties, it is also possible to discuss another concept: Asymptotically Time-domain String Stability (ATSS).

\begin{lemma}[Adapted from Feng et al.\cite{Feng2019ARC}]
The origin $\statev_{i} =0$, $i \in \nrobots$, has ATSS if it has TSS and $\sup_i \|\statev_{i}\|_\infty \rightarrow 0$.
\end{lemma}

Theorem~\ref{theo:steady_state_lin_to_nolin} provides the conditions to show that $\dot{\statev_{i}}(t) = \hat{f}_i(\statev_i(t),0)$ is asymptotically stable. According to the Weak Coupling Theorem,  any interconnection of asymptotically stable systems is string stable, if the interconnections are sufficiently weak. Therefore, since each vehicle $i$ is endowed with the same stabilizing policy, following a reference given by the average states of all its $\neighSet_i$ neighbors, the proposed approach can guarantee ATSS.}

By these arguments, we expect that the proposed modification can solve Prob.~\ref{prob:robustness} more adequately than other policies in the literature. To illustrate these advantages, an extensive set of experimental results is presented in the following.

%%%%%%%%%%%%%%%%%%%%%%%%%%%%%%%%%%%%%%%%%%%%%%%%%%%%%%%%%%%%%%
%%%%%%%%%%%%%%%%%%%%%%%%%%%%%%%%%%%%%%%%%%%%%%%%%%%%%%%%%%%%%%
\section{Experiments}
\label{sec:experiments}

The main objective of this section is to illustrate that our method, hereinafter referred to as \rev{\emph{\ac{RRL}}} is less susceptible to the existence of model misspecifications than others. Therefore, we present some simulated experiments comparing the platoon behavior with another two approaches. 

The first one is the distributed consensus protocol in Zheng et al. \cite{Zheng2015Stability} given by Eq.~\eqref{eq:lei_consensus}, where the authors define the acceleration command $\iuin{i}$ by computing the error between the state of the vehicle $\ixv{i}$ and the state of its neighbors $\ixv{j}$. For all trials, we have set $\Kmat = [1, 2, 1]$, values extracted from the above-mentioned paper.

The second one is a \ac{DRL} approach inspired in the current literature \cite{Wei2018Design, Chu2019Model, Liu2020Platoon, Li2021Reinforcement}, where the authors use \acd{RL} algorithms to compute continuous policies that transform the error between the states of the agents, $\ixv{i}$ and $\ixv{j}$, into acceleration commands $\iuin{i}$. Although all these works present different optimization reward functions (learn human-driver behavior, avoid collisions, save fuel, etc) and different network topologies (\ac{PF} and \ac{PFL}), in common they perform a longitudinal spacing control task. Therefore, we synthesized them all in a single benchmark strategy, hereinafter referred to only as \rev{\emph{\ac{SRL}}}, that uses a \acd{DDPG} algorithm with the reward function in Eq.~\eqref{eq:reward}.

\begin{figure}[!ht]
    \centering
    \includegraphics[width=.8\linewidth]{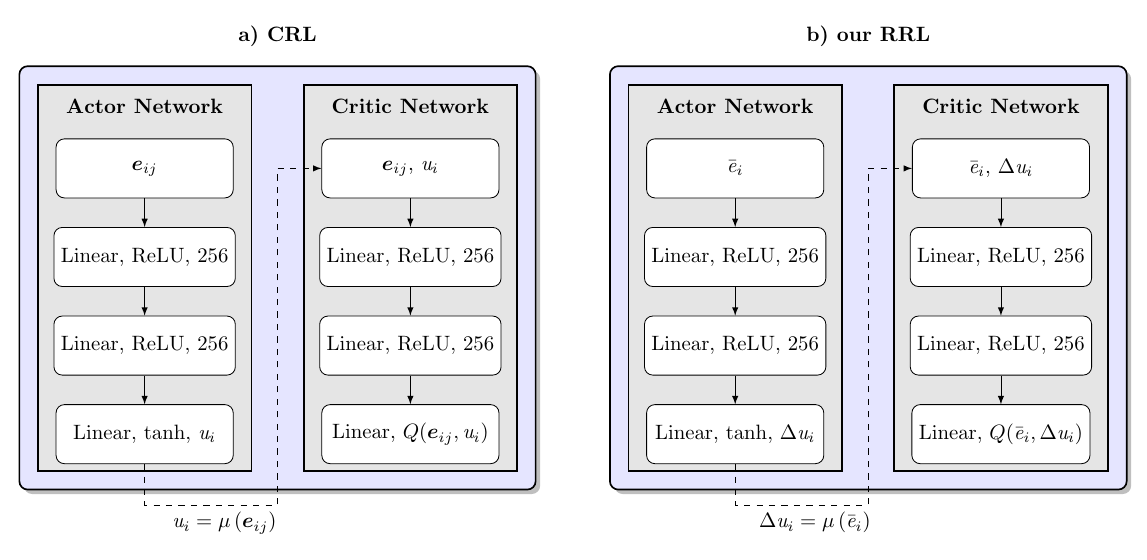}
    \caption{Actor-critic model structures: a) \protect\ace{SRL} network; and b) the proposed \protect\ace{RRL} network. Both modules present two hidden fully-connected layers with 256 neurons each.}
    \label{fig:actor-critic_models}
\end{figure}

The vehicles' nonlinear models and \ac{DDPG} algorithms have been implemented using the \emph{Pytorch} and \emph{Numpy} libraries for \emph{Python 3} language, running at \emph{Ubuntu 20.04}.
Figure \ref{fig:actor-critic_models} presents the actor-critic network structures for both, \rev{\ac{SRL} and \ac{RRL}} strategies, where each module is composed of two hidden fully-connected layers with 256 neurons each, connected through \ac{ReLU} activation. 
Input states $\statev$ are given by $\ev_{ij} \in \Reais^3$ and $\emean_{i} \in \Reais$, respectively, while output actions $\action \in \{-1, 1\}$ are computed via the hyperbolic tangent (\emph{tanh}) activation function and re-scaled for $\{\iuin{min}, \iuin{max}\}$ in the \rev{\ac{SRL}} and for $\{\iduin{min}, \iduin{max}\}$ in the \rev{\ac{RRL}}.

\begin{figure}[!ht]
    \centering
    \includegraphics[width=.5\linewidth]{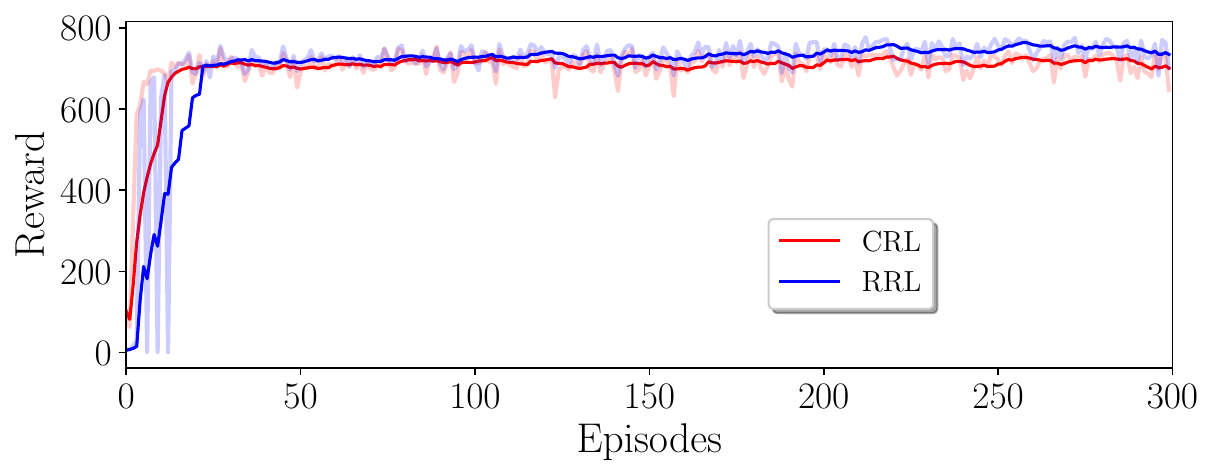}
    \caption{Moving average of the reward for the \protect\ace{SRL}, Eq.~\eqref{eq:reward}, in red, and the \protect\ace{RRL}, Eq.~\eqref{eq:alternative_reward}, in blue.}
    \label{fig:rewards}
\end{figure}

In the \ac{DDPG} implementation, we have used a discounting factor $\discount = \n{.99}$, and $\tau = \n{.005}$ for the soft target updates, while the neural network parameters were adjusted with a learning rate of \n{.0001} for both actor and critic. Meanwhile, the Ornstein-Uhlenbeck distribution $\noise$ was defined with null mean and standard deviation of $\n{0.15}$.
\rev{We have used 300 episodes in the training stage, with 1000 steps in \n[s]{50} of simulation. The vehicle $i$ and its front ``virtual leader'' start in relatively random states, with the last one always at the forefront position and commanded to accelerate and decelerate according to the following function:
\begin{align*}
    \iuin{0}(t) =
    \begin{cases}
        \rand[-0.5,1.5] $~m/s$^2 & \text{if $\rand[0,1] \leq \n{0.01}$,}\\
        \iuin{0}(t-1) & \text{otherwise},\\
    \end{cases}
\end{align*}
\noindent where $\iuin{0}(0) = 0$ and $\rand[a,b]$ is random uniform value selected between $[a,b]$. It means that, for each episode, the leader has a chance of 1 in 100 time-steps to vary its speed with limited acceleration.}
Figure \ref{fig:rewards} presents the moving average reward evolution for both evaluated methods. The \rev{\ac{SRL}}  is based on Eq.~\eqref{eq:reward} with $\Gammamat = [ 1~ 1~ 1]$ and $\penaltygain = \n{0.2}$, while our \rev{proposed \ac{RRL}}  uses Eq.~\eqref{eq:alternative_reward} with $\rewardgain = \n{1.}$ and $\penaltygain = \n{0.2}$.
From the training perspective, it is possible to observe that both approaches rapidly learn a relatively good regulation tendency. The \rev{\ac{SRL}} makes it faster and with less noise, possibly because it is acting directly on $\iuin{i}(t)$. Nevertheless, the proposed \rev{\ac{RRL}} surpasses after approximately 50 episodes and sustains this improvement until the end.

In the following section, we start by testing the influence of the topology structure in the proposed controller. In the sequence, we added external disturbances and uncertainties in the dynamic model of the vehicles to evaluate the system stability. In this execution phase, for all experiments, the leader's acceleration profile was given by:
\begin{align*}
    \iuin{0}(t) =
    \begin{cases}
        \n[m/s^2]{1.} & \text{if~~} \n[s]{5} < t \leq \n[s]{10},\\ %\\
        \n[m/s^2]{0.} & \text{otherwise}.\\
    \end{cases}
\end{align*}
\rev{\noindent Here, the main idea is to evaluate the step response and the steady-state behavior of the \ac{DRL} controller.
Table \ref{tab:model_parameters_nominal} presents the nominal heterogeneous values for the parameters of the nonlinear system \eqref{eq:nonlinear_system}. External parameters were first set to $\vel_{w} = 0$ and $\slope = 0$ (disturbances and uncertainty will be discussed later). For all experiments, the spacing distance was set to $\xspacing{ij} = \n[m]{10}$.
\begin{table}[!ht]
    \centering
    \revcaption
    \caption{Nominal heterogeneous values for the vehicle parameters, extracted from Zheng et al.\cite{Zheng2019Cooperative}.}
    \rev{
    \begin{tabular}{c|c}
        \addlinespace
        \hline
        \multicolumn{2}{c}{\bf Nominal parameter values}\\
        \hline
        $\mass_{i}  =  \n{1500} + \n{100}i$~[kg] & 
        $\wheel_{i} = \n{0.25} + \n{0.005}i$~[m] \\%[.1cm]
        $\efficiency_{i} = \n{80} + i$~[\%] &
        $\inertialdelay_{i} = \n{0.3} + \n{0.02}i$ \\%[.1cm]
        $\dragcoef_{i} = \n{0.4} + \n{0.01}i$  & 
        $\friction_{i} = \n{0.015} + \n{0.001}i$ \\%[.1cm]
        $\density = \n{1.23}$~[kg/m$^3$] & 
        $\grav = \n{9.78}$ [m/s$^2$] \\%[.1cm]
        $\{\iuin{min}, \iuin{max}\} = \{-3, 3\}$ & 
        $\{\iduin{min}, \iduin{max}\} = \{-30, 30\}$ \\
        \hline
    \end{tabular}
    }
    \label{tab:model_parameters_nominal}
\end{table}
}
%%%%%%%%%%%%%%%%%%%%%%%%%%%%%%%%%%%%%%%%%%%%%%%%%%%%%%%%%%%%%%
\subsection{Analysis with different topologies} %-- ablation study?

\begin{figure}[!ht]
    \centering
    \subfigure[\protect\ac{PF} topology, Zheng et al. \cite{Zheng2015Stability}.]{
        \includegraphics[width=.32\linewidth]{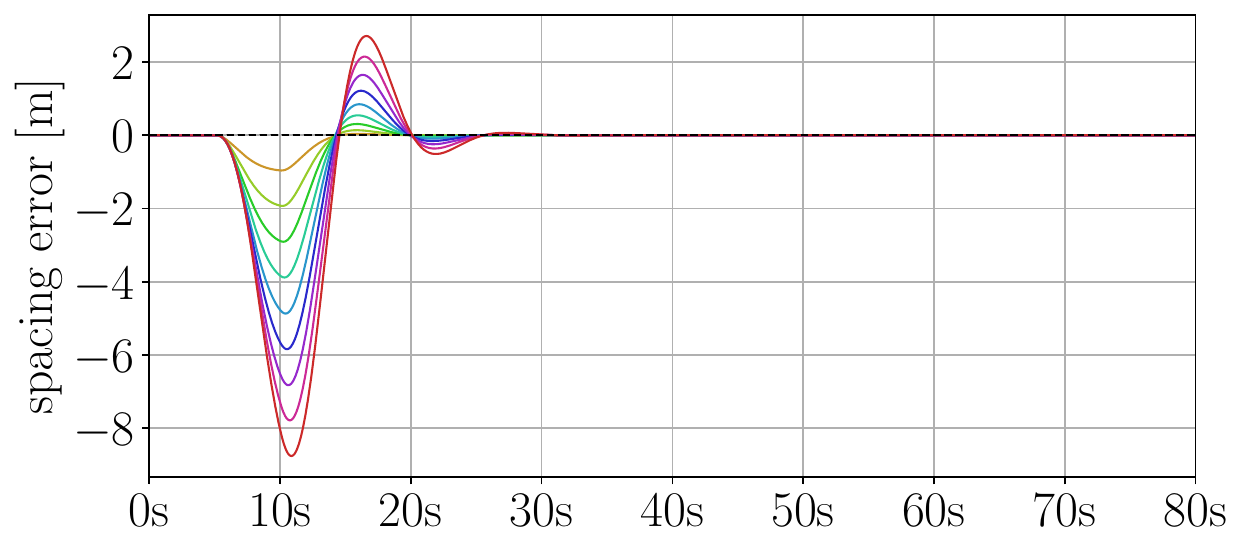}
        \label{subfig:PF_topology_Zheng}
    }
    \subfigure[\protect\ac{PF} topology, \protect\ac{SRL}.]{
        \includegraphics[width=.32\linewidth]{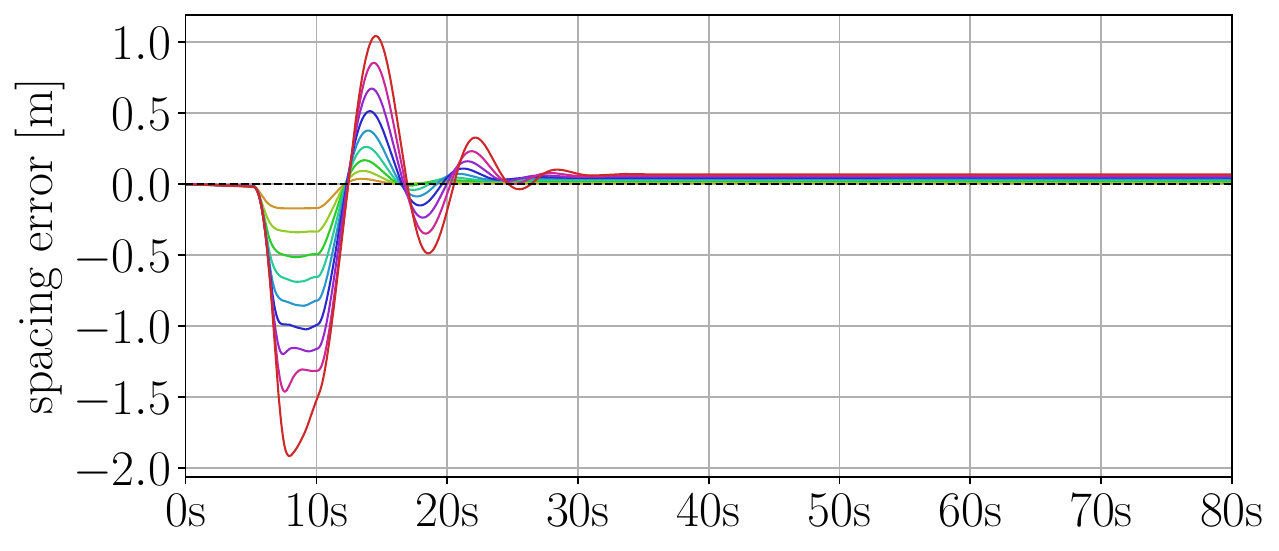}
        \label{subfig:PF_topology_RL}
    }
    \subfigure[\protect\ac{PF} topology, proposed \protect\ac{RRL}.]{
        \includegraphics[width=.32\linewidth]{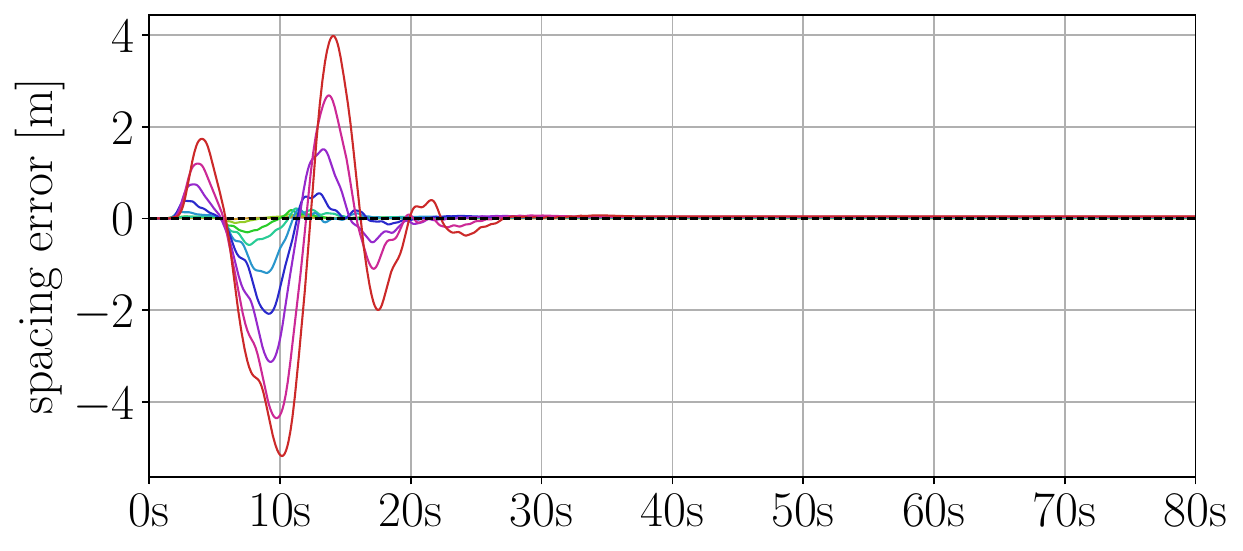}
        \label{subfig:PF_topology_Antiwindup}
    }
    \caption{Impact of the \protect\ac{PF} topology: a) conventional consensus strategy; b) \protect\ac{SRL} method; and c) our proposed \protect\ac{RRL} approach.}
    \label{fig:PF_topologies}
\end{figure}

\begin{figure}[!ht]
    \centering
    \subfigure[\protect\ac{PFL} topology, Zheng et al. \cite{Zheng2015Stability}.]{
        \includegraphics[width=.32\linewidth]{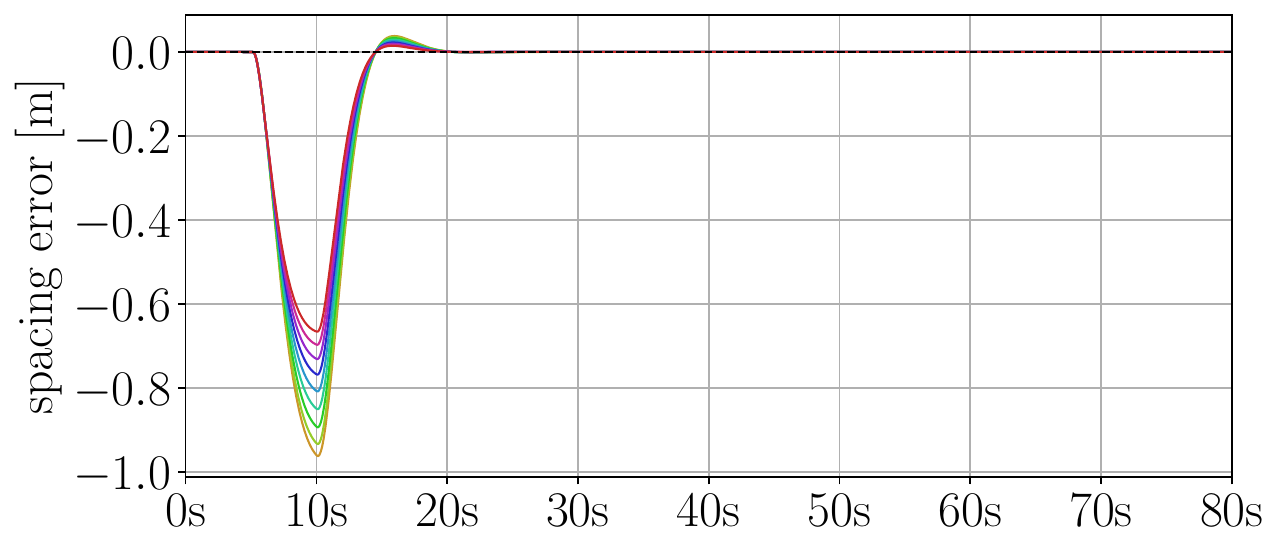}
        \label{subfig:PFL_topology_Zheng}
    }
    \subfigure[\protect\ac{PFL} topology, \protect\ac{SRL}.]{
        \includegraphics[width=.32\linewidth]{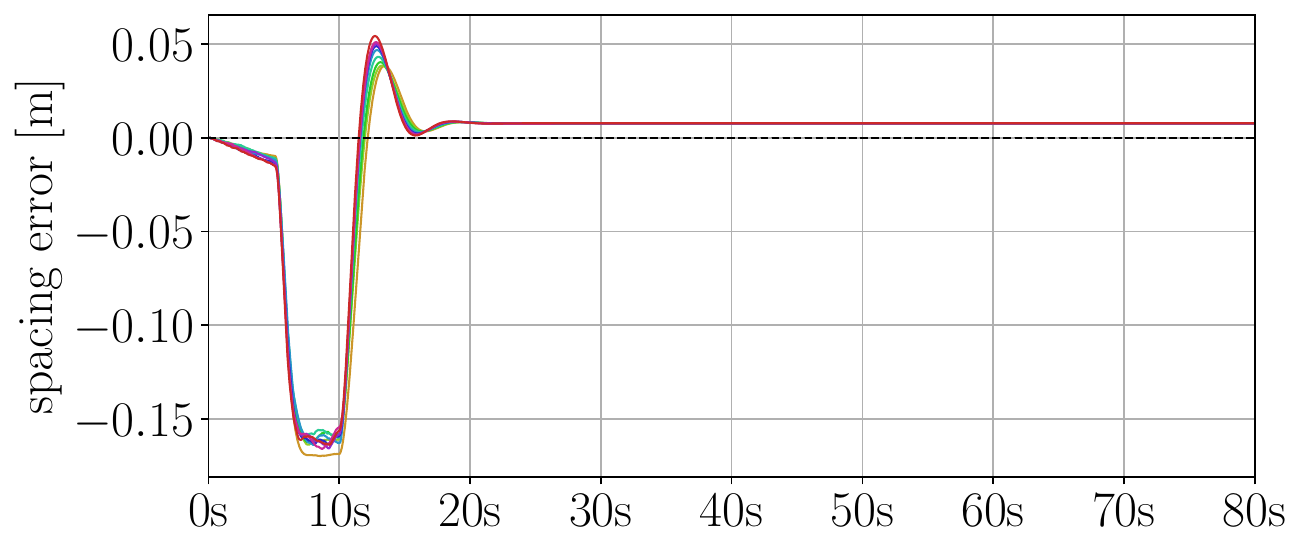}
        \label{subfig:PFL_topology_RL}
    }
    \subfigure[\protect\ac{PFL} topology, proposed \protect\ac{RRL}.]{
        \includegraphics[width=.32\linewidth]{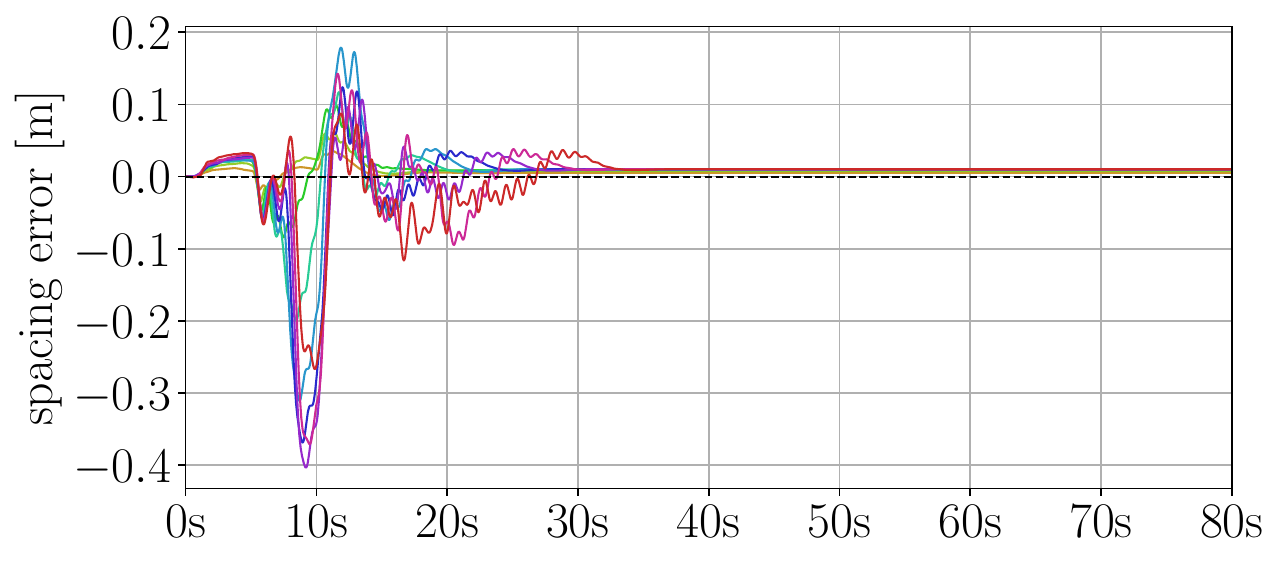}
        \label{subfig:PFL_topology_Antiwindup}
    }
    \caption{Impact of the \protect\ac{PFL} topology: a) conventional consensus strategy; b) \protect\ac{SRL} method; and c) our proposed \protect\ac{RRL} approach.}
    \label{fig:PFL_topologies}
\end{figure}

\begin{figure}[!ht]
    \centering
    \subfigure[\protect\ac{TPF} topology, Zheng et al. \cite{Zheng2015Stability}.]{
        \includegraphics[width=.32\linewidth]{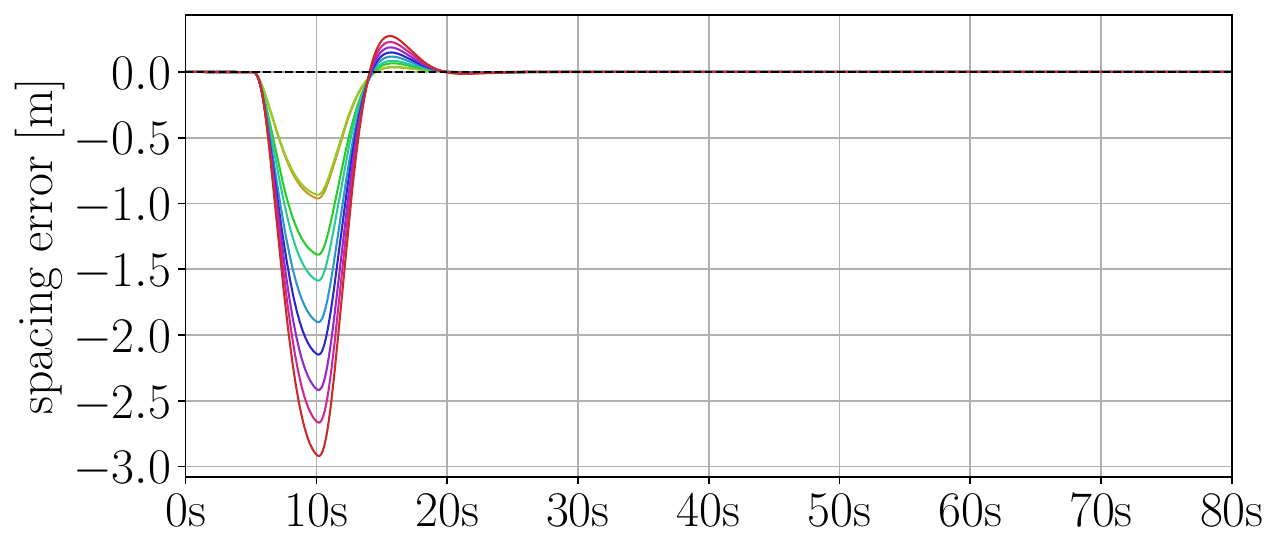}
        \label{subfig:TPF_topology_Zheng}
    }
    \subfigure[\protect\ac{TPF} topology, \protect\ac{SRL}.]{
        \includegraphics[width=.32\linewidth]{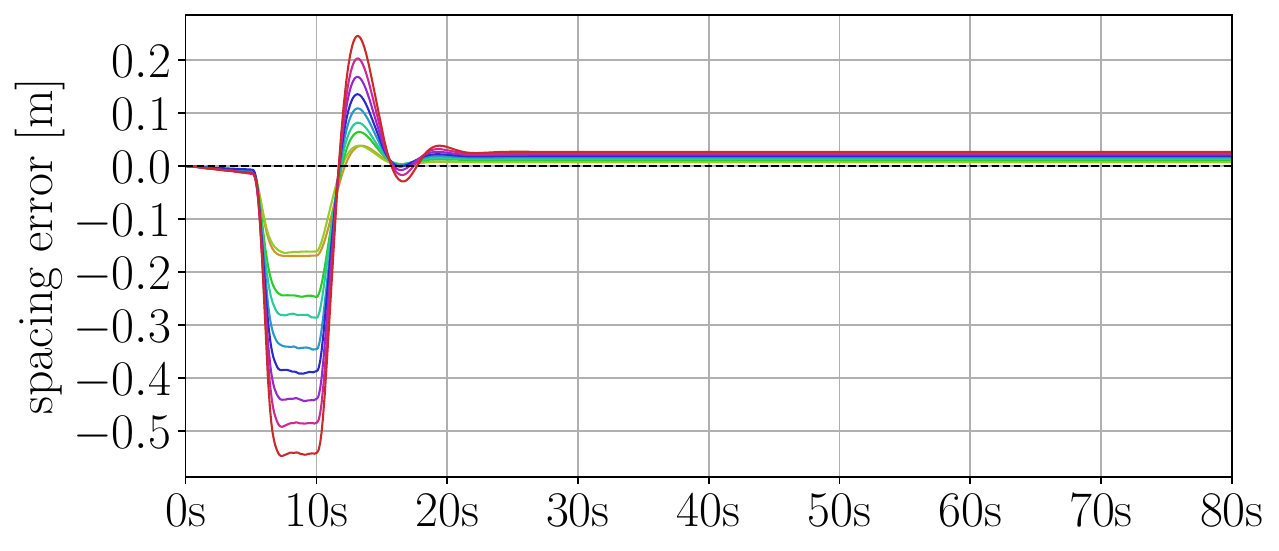}
        \label{subfig:TPF_topology_RL}
    }
    \subfigure[\protect\ac{TPF} topology, proposed \protect\ac{RRL}.]{
        \includegraphics[width=.32\linewidth]{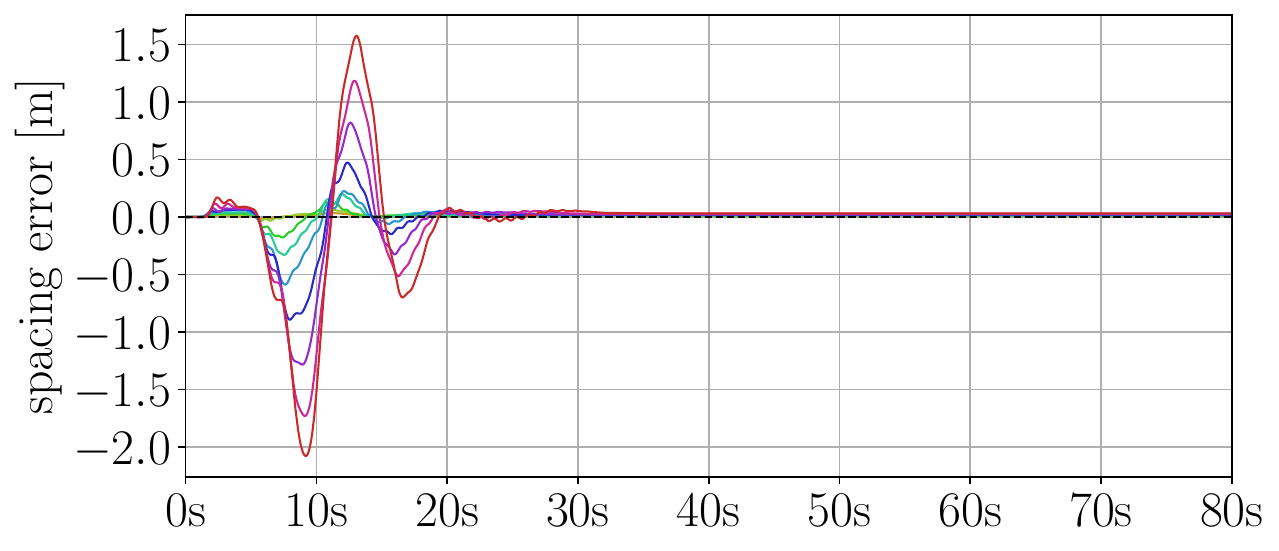}
        \label{subfig:TPF_topology_Antiwindup}
    }
    \caption{Impact of the \protect\ac{TPF} topology: a) conventional consensus strategy; b) \protect\ac{SRL} method; and c) our proposed \protect\ac{RRL} approach.}
    \label{fig:TPF_topologies}
\end{figure}

\begin{figure}[!ht]
    \centering
    \subfigure[\protect\ac{TPFL} topology, Zheng et al. \cite{Zheng2015Stability}.]{
        \includegraphics[width=.32\linewidth]{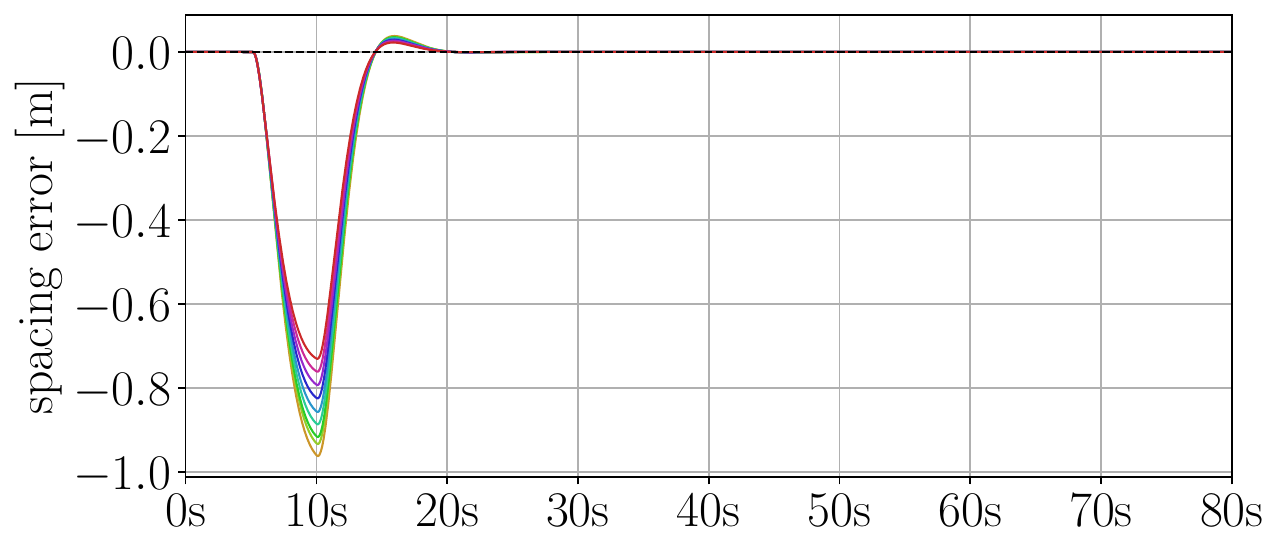}
        \label{subfig:TPFL_topology_Zheng}
    }
    \subfigure[\protect\ac{TPFL} topology, \protect\ac{SRL}.]{
        \includegraphics[width=.32\linewidth]{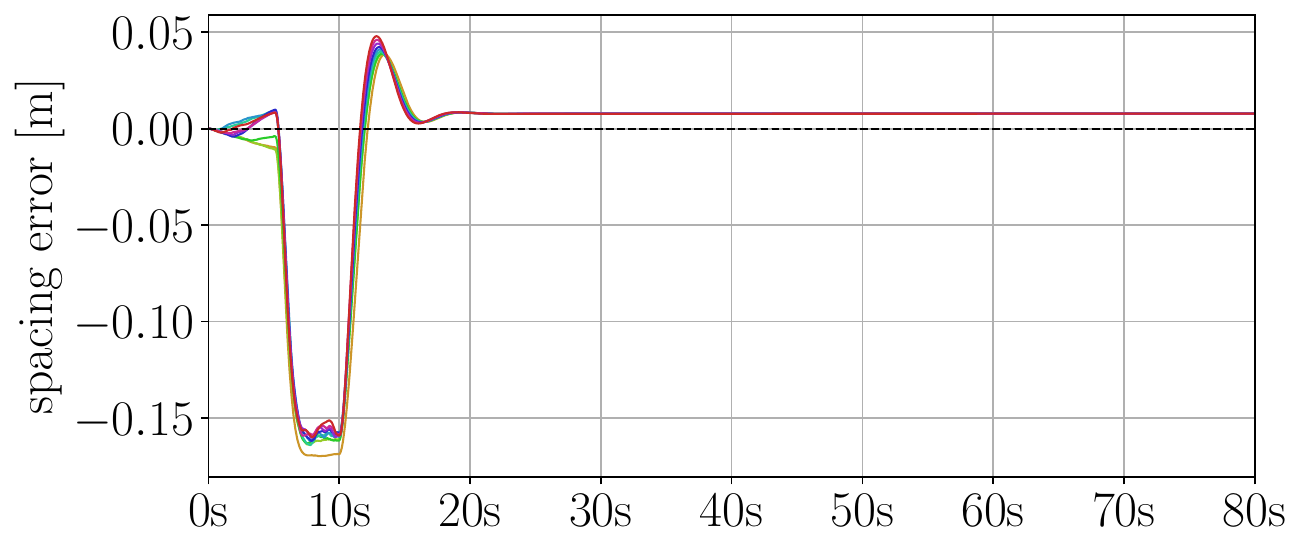}
        \label{subfig:TPFL_topology_RL}
    }
    \subfigure[\protect\ac{TPFL} topology, proposed \protect\ac{RRL}.]{
        \includegraphics[width=.32\linewidth]{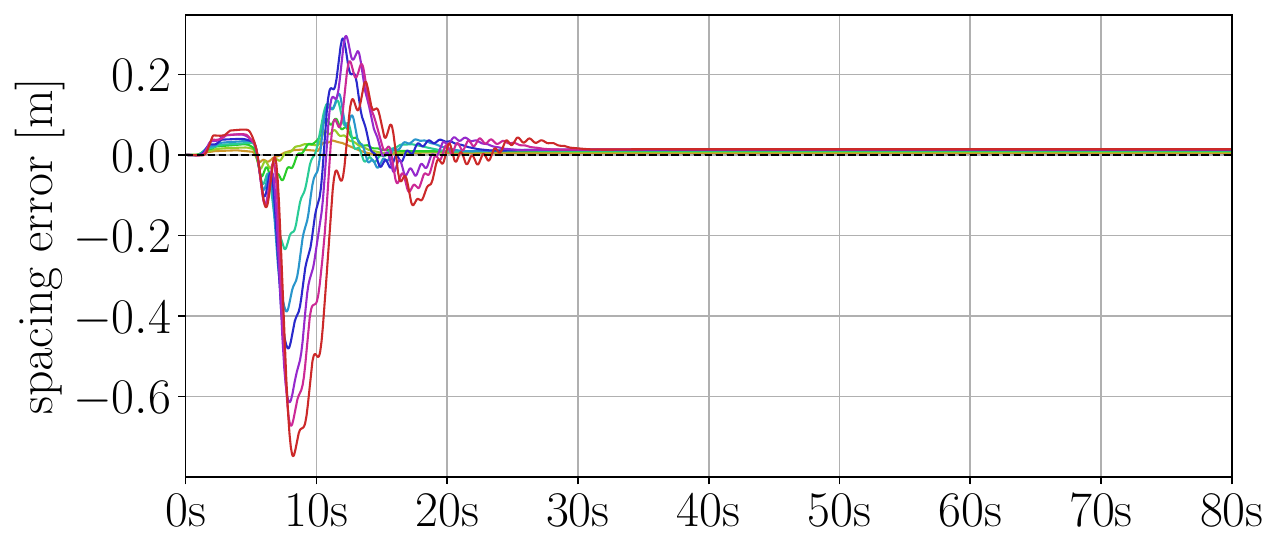}
        \label{subfig:TPFL_topology_Antiwindup}
    }
    \caption{Impact of the \protect\ac{TPFL} topology: a) conventional consensus strategy; b) \protect\ac{SRL} method; and c) our proposed \protect\ac{RRL} approach.}
    \label{fig:TPFL_topologies}
\end{figure}

Let us start by testing the influence of the communication network in the response to our proposed policy. Concerning a platoon composed of ten vehicles (one leader and nine followers), starting with a null spacing error, we ran experiments for all topologies in Fig.~\ref{fig:topologies}. 
Figures \ref{fig:PF_topologies}, \ref{fig:PFL_topologies}, \ref{fig:TPF_topologies} and \ref{fig:TPFL_topologies} present a comparison among the consensus control law in Zheng et al. \cite{Zheng2015Stability}, the conventional approach (\ac{SRL}), and our proposed strategy (\ac{RRL}) for the \ace{PF}, \ace{PFL}, \ace{TPF}, and \ace{TPFL} topologies, respectively.

For the \ac{PF}, one can notice that all three approaches were able to stabilize the system, ensuring null error in steady-state. That was expected, once the method in Zheng et al. \cite{Zheng2015Stability} is stable for all topologies (in the absence of disturbances) and the \ac{DRL} approaches were trained with exactly one predecessor neighbor.
Concerning the dynamic step response, the consensus protocol presented an oscillatory behavior, with a large overshoot, as can be seen in Fig.~\ref{subfig:PF_topology_Zheng}. This may be a consequence of a not careful adjustment of the gains in the control law. On the other hand, as illustrated in Figures \ref{subfig:PF_topology_RL} and \ref{subfig:PF_topology_Antiwindup}, both \ac{DRL} algorithms provide smaller oscillatory overshoots, although this was not taken directly into account in the reward function.   

When the leader is incorporated into the communication structure of all vehicles, \ac{PFL} topology, it is possible to see it helps to improve the dynamic response of all evaluated methods, decreasing oscillations and overshoots in comparison to previous simulations. However, Fig.~\ref{subfig:PFL_topology_RL} also shows that the \ac{SRL} approach starts to present a little steady-state error, while the other two techniques, the consensus protocol in Fig.~\ref{subfig:PFL_topology_Zheng} and the \ac{RRL} with integral action in Fig.~\ref{subfig:PFL_topology_Antiwindup}, remain stable and undisturbed.

The same behavior can be observed in Figures \ref{fig:TPF_topologies} and \ref{fig:TPFL_topologies} for the \ac{TPF} and \ac{TPFL}, respectively.
The obvious conclusion here is that, as the number of neighboring vehicles increases, higher is the difference between the trained and execution environments. These tests illustrate that it is possible to generalize a \ac{DRL} policy by feeding it not with the states of all nearby agents, but with the mean ``virtual'' neighbor, making the controller robust in steady-state conditions to different communication topologies.

On the other hand, as higher the number of neighbors, the more efficient the policy becomes, as claimed in the work of Godinho et al.\cite{Godinho:22} for the consensus protocol. One can observe that, when comparing the profiles in Fig.~\ref{fig:PF_topologies} with those in Fig.~\ref{fig:TPF_topologies}, or those in Fig.~\ref{fig:PFL_topologies} with Fig.~\ref{fig:TPFL_topologies}, the overshoots are clearly smaller for two neighbors.

%%%%%%%%%%%%%%%%%%%%%%%%%%%%%%%%%%%%%%%%%%%%%%%%%%%%%%%%%%%%%%
\subsection{Analysis under uncertainties}

In the following experiments, we demonstrate that our method is also capable of ensuring null spacing error even when the feedback linearization proposed in Zheng et al.\cite{Zheng2015Stability} is imperfect. It happens when the parameters of the vehicle are known with some level of uncertainty, generally because some of them are difficult to measure (like friction or motor efficiency) or because the team presents heterogeneous characteristics.

\begin{figure}[!ht]
    \centering
    \subfigure[\protect\ac{PF} topology, Zheng et al. \cite{Zheng2015Stability}.]{
        \includegraphics[width=.32\linewidth]{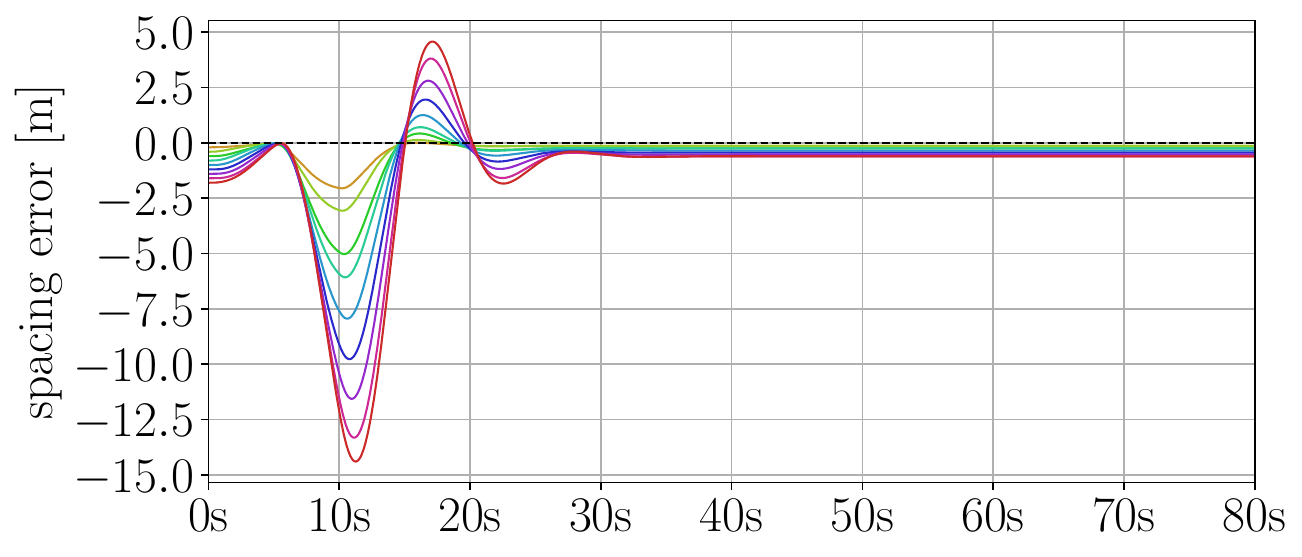}
        \label{subfig:PF_uncertainties_Zheng}
    }
    \subfigure[\protect\ac{PF} topology, \protect\ac{SRL}.]{
        \includegraphics[width=.32\linewidth]{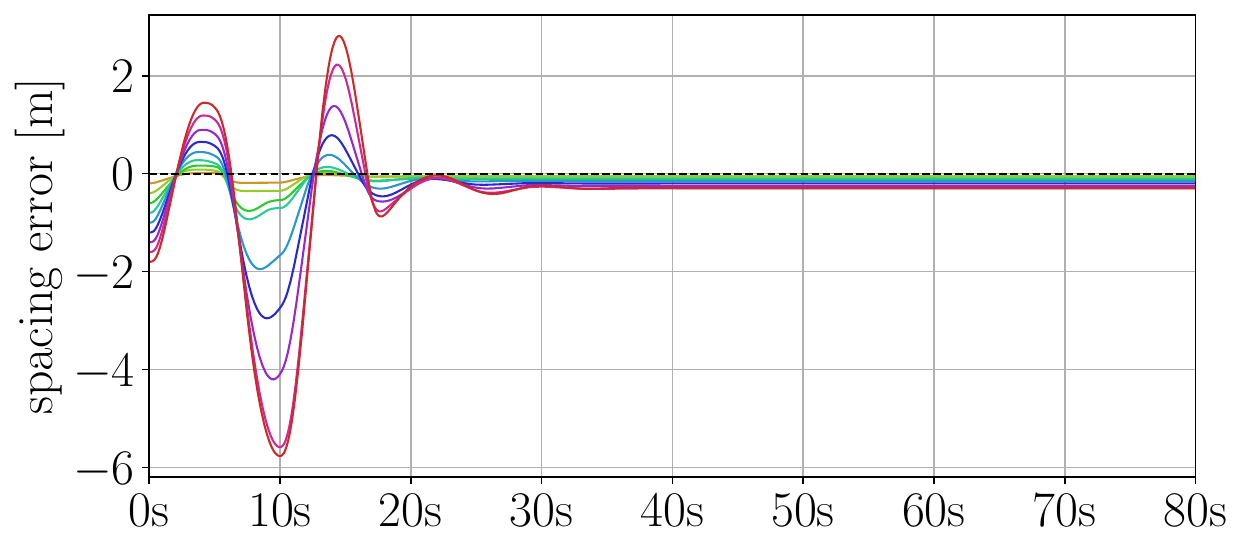}
        \label{subfig:PF_uncertainties_RL}
    }
    \subfigure[\protect\ac{PF} topology, proposed \protect\ac{RRL}.]{
        \includegraphics[width=.32\linewidth]{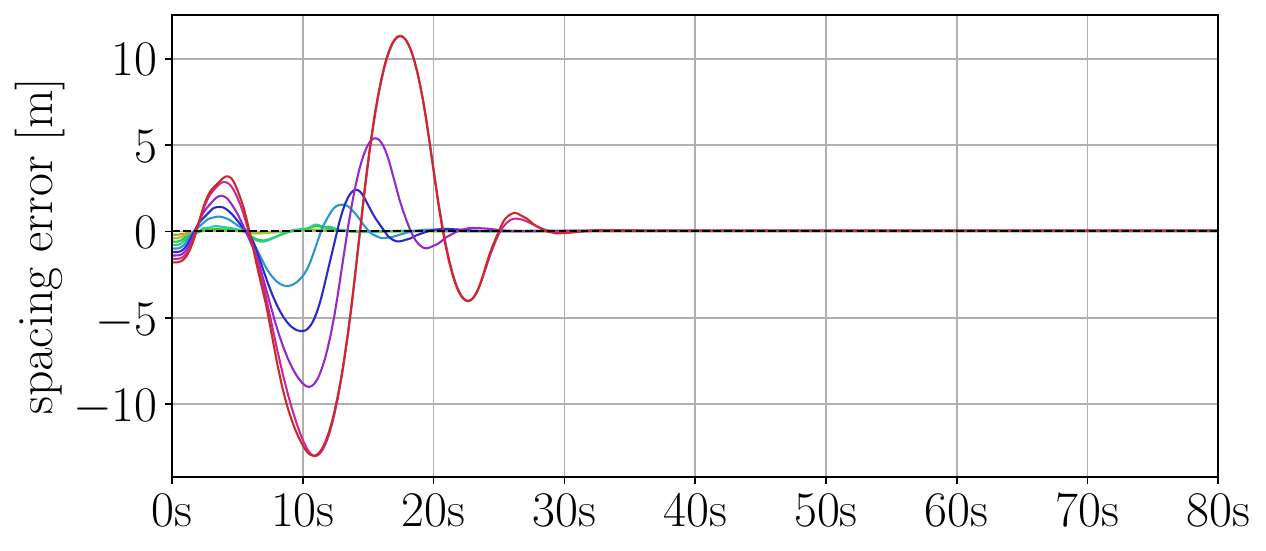}
        \label{subfig:PF_uncertainties_Antiwindup}
    }
    \caption{Impact of uncertainties in the parameters of the vehicles for the \protect\ac{PF} topology: a) conventional consensus strategy; b)  \protect\ac{SRL} method; and c) our proposed \protect\ac{RRL} approach.}
    \label{fig:PF_uncertainties}
\end{figure}

\begin{figure}[!ht]
    \centering
    \subfigure[\protect\ac{PFL} topology, Zheng et al. \cite{Zheng2015Stability}.]{
        \includegraphics[width=.32\linewidth]{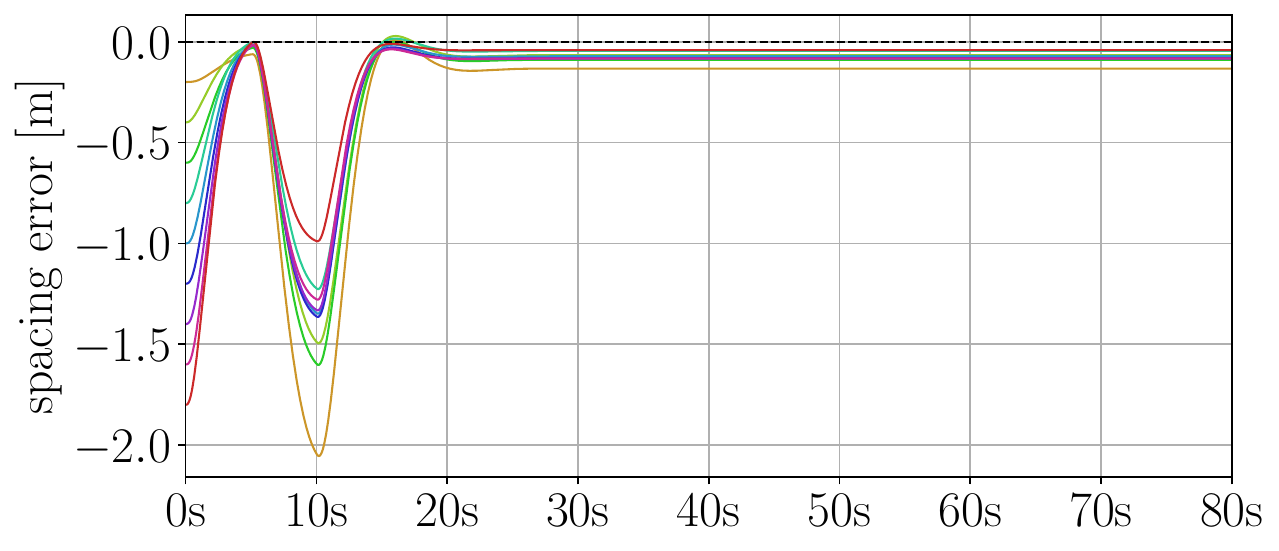}
        \label{subfig:PFL_uncertainties_Zheng}
    }
    \subfigure[\protect\ac{PFL} topology, \protect\ac{SRL}.]{
        \includegraphics[width=.32\linewidth]{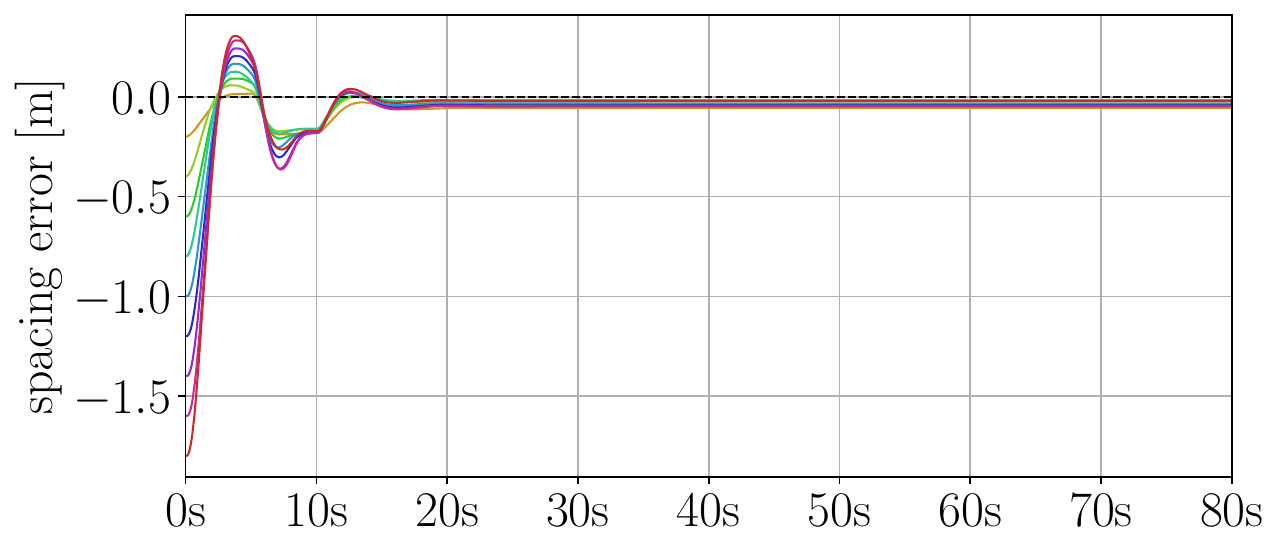}
        \label{subfig:PFL_uncertainties_RL}
    }
    \subfigure[\protect\ac{PFL} topology, proposed \protect\ac{RRL}.]{
        \includegraphics[width=.32\linewidth]{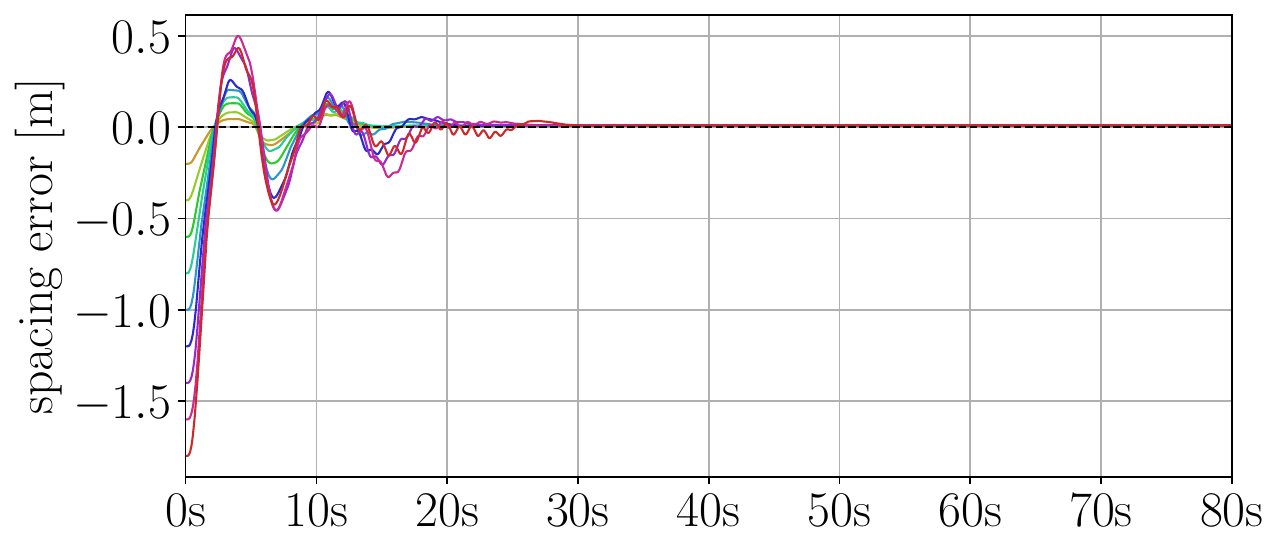}
        \label{subfig:PFL_uncertainties_Antiwindup}
    }
    \caption{Impact of uncertainties in the parameters of the vehicles for the \protect\ac{PFL} topology: a) conventional consensus strategy; b)  \protect\ac{SRL} method; and c) our proposed \protect\ac{RRL} approach.}
    \label{fig:PFL_uncertainties}
\end{figure}

\begin{figure}[!ht]
    \centering
    \subfigure[\protect\ac{TPF} topology, Zheng et al. \cite{Zheng2015Stability}.]{
        \includegraphics[width=.32\linewidth]{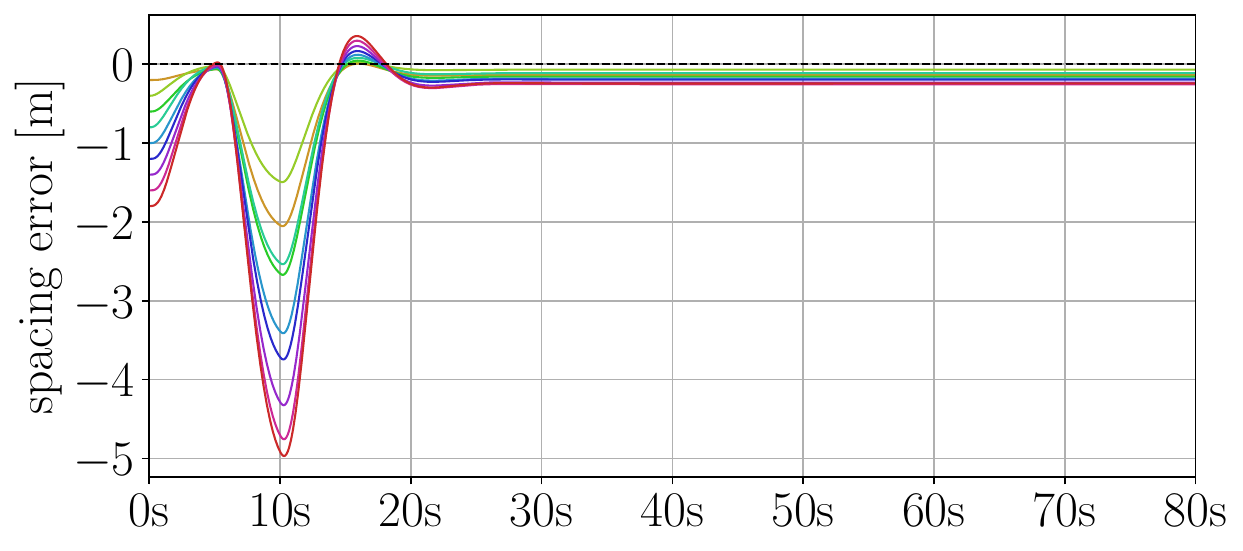}
        \label{subfig:TPF_uncertainties_Zheng}
    }
    \subfigure[\protect\ac{TPF} topology, \protect\ac{SRL}.]{
        \includegraphics[width=.32\linewidth]{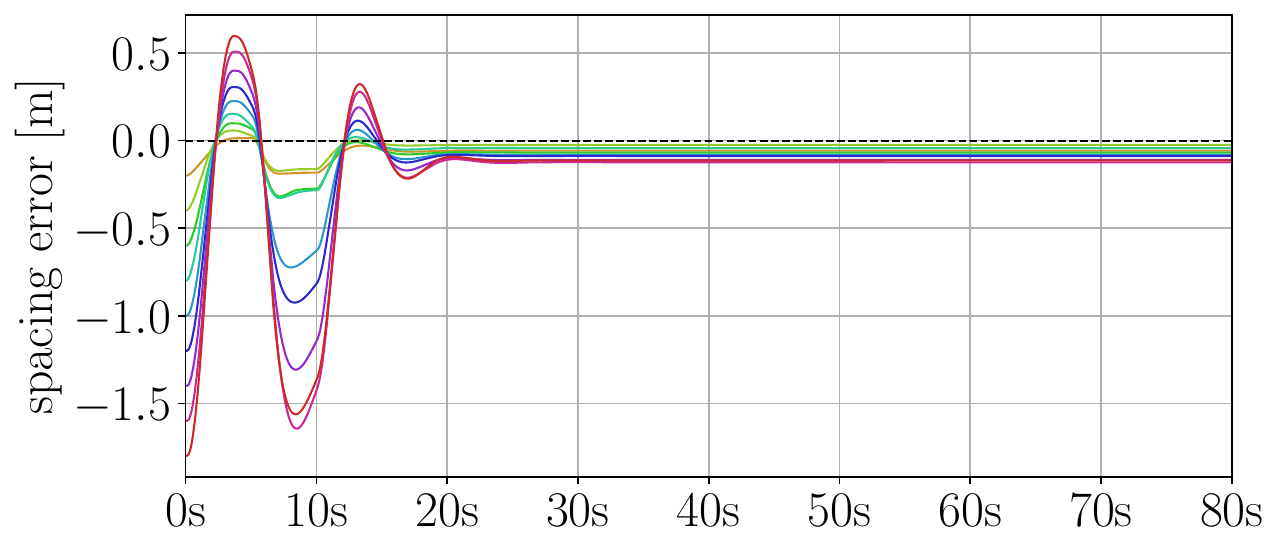}
        \label{subfig:TPF_uncertainties_RL}
    }
    \subfigure[\protect\ac{TPF} topology, proposed \protect\ac{RRL}.]{
        \includegraphics[width=.32\linewidth]{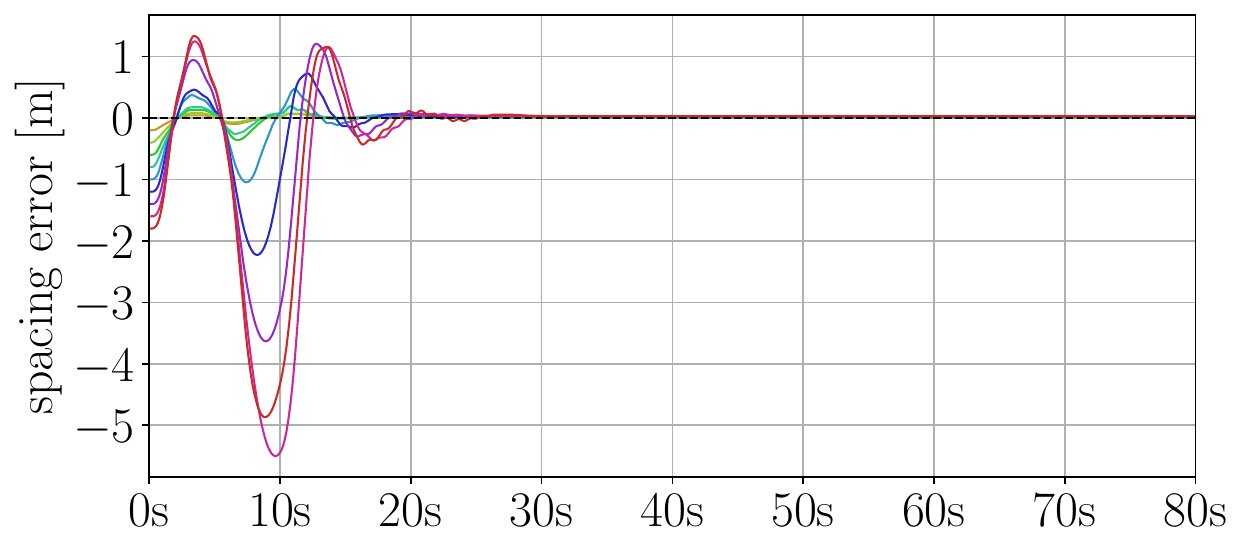}
        \label{subfig:TPF_uncertainties_Antiwindup}
    }
    \caption{Impact of uncertainties in the parameters of the vehicles for the \protect\ac{TPF} topology: a) conventional consensus strategy; b)  \protect\ac{SRL} method; and c) our proposed \protect\ac{RRL} approach.}
    \label{fig:TPF_uncertainties}
\end{figure}

\begin{figure}[!ht]
    \centering
    \subfigure[\protect\ac{TPFL} topology, Zheng et al. \cite{Zheng2015Stability}.]{
        \includegraphics[width=.32\linewidth]{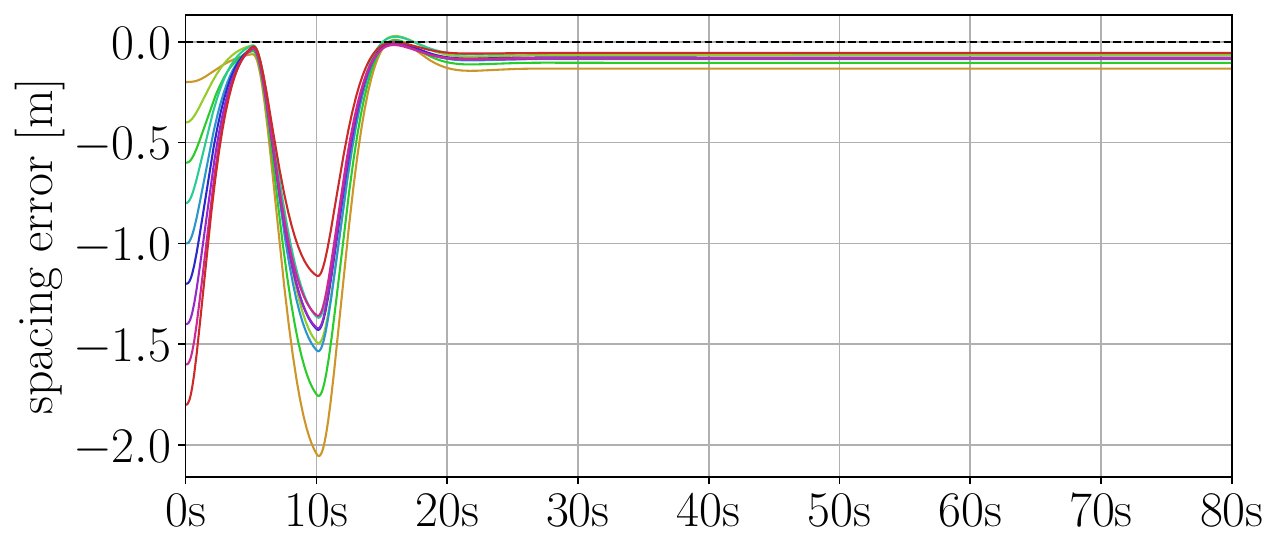}
        \label{subfig:TPFL_uncertainties_Zheng}
    }
    \subfigure[\protect\ac{TPFL} topology, \protect\ac{SRL}.]{
        \includegraphics[width=.32\linewidth]{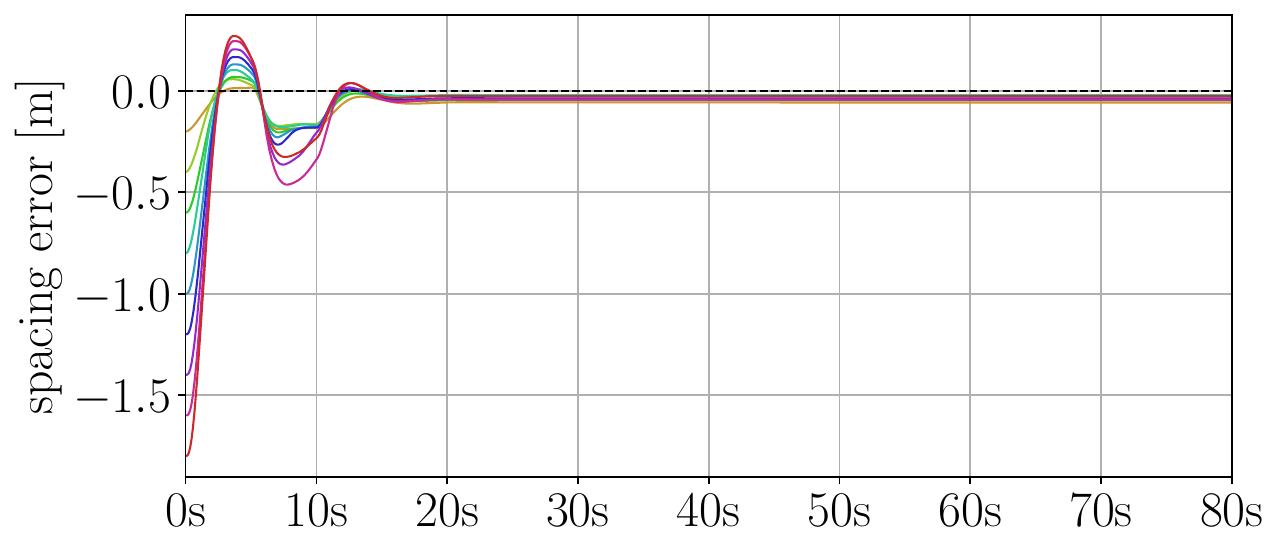}
        \label{subfig:TPFL_uncertainties_RL}
    }
    \subfigure[\protect\ac{TPFL} topology, proposed \protect\ac{RRL}.]{
        \includegraphics[width=.32\linewidth]{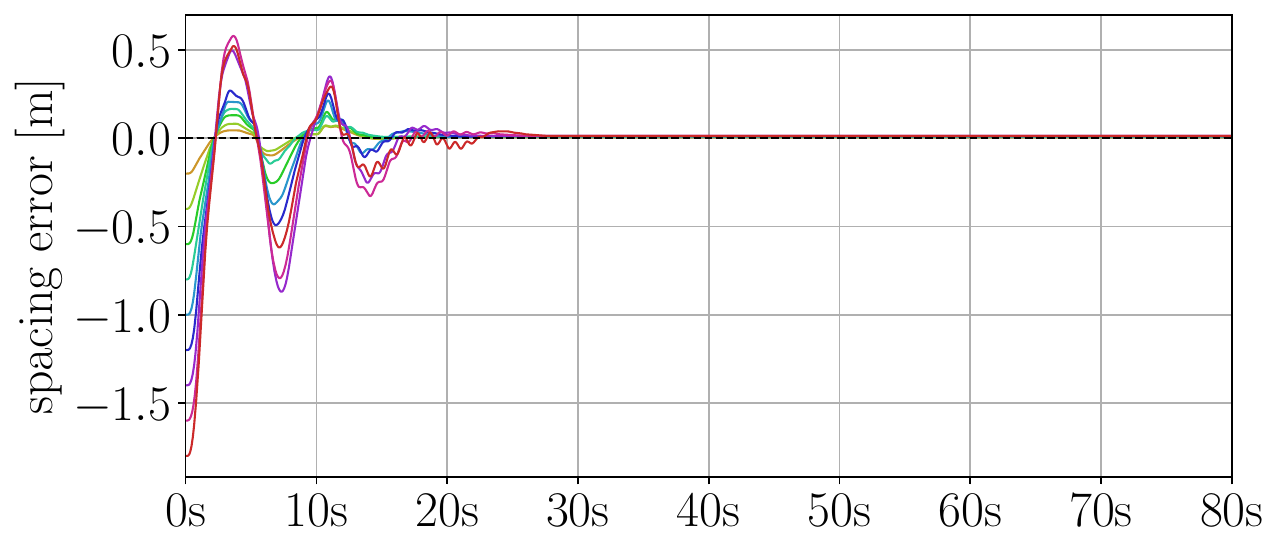}
        \label{subfig:TPFL_uncertainties_Antiwindup}
    }
    \caption{Impact of uncertainties in the parameters of the vehicles for the \protect\ac{TPFL} topology: a) conventional consensus strategy; b)  \protect\ac{SRL} method; and c) our proposed \protect\ac{RRL} approach.}
    \label{fig:TPFL_uncertainties}
\end{figure}

\rev{Without loss of generality, we have incorporated for all trials uncertainties into two parameters that present significant variations among ground vehicles, the mass and the power-train time constant. For the mass, we have randomly chosen values in the interval $\mass_{i}  =  \n{1500} + \n{100}i \pm \n[kg]{300}$, while for the time constant we used $\inertialdelay_{i} = \n{0.3} + \n{0.02}i \pm \n{0.1}$.
Figures \ref{fig:PF_uncertainties} to \ref{fig:TPFL_uncertainties} presents the controllers' responses for all evaluated techniques and all previously mention topologies.

At this point, it is interesting to perceive that both, the consensus protocol in Zheng et al. \cite{Zheng2015Stability} and the \ac{SRL} algorithm fail, while our proposed \ac{RRL} with integral action policy provides a null steady-state error.
In Zheng et al.\cite{Zheng2019Cooperative}, the authors demonstrate that their consensus protocol is also stable for heterogeneous teams of vehicles, as long as their communication topology can be modeled as \acd{DAGs}, which is the formal definition for all topologies in Fig.~\ref{fig:topologies}, but they can't handle uncertain parameters. Also, such stability depends mainly on the value of $\inertialdelay_{i}$ and the used controller gains.
On the other hand, the \ac{SRL} method was incapable of reducing the steady-state error since its network was trained using the feedback linearization with the nominal values in Table \ref{tab:model_parameters_nominal}. Without a strategy to compensate for all uncanceled nonlinear terms in Eq.~\ref{eq:nonlinear_system}, the policy can't be generalized to heterogeneous uncertain platoons. Our method, however, rejects the disturbance caused by the uncertainty in the parameters, ensuring null error.}

%%%%%%%%%%%%%%%%%%%%%%%%%%%%%%%%%%%%%%%%%%%%%%%%%%%%%%%%%%%%%%
\subsection{Analysis under external disturbances}

Next, we have considered the existence of external disturbances in the platoon. Here, the main goal is to demonstrate the effectiveness of the integral action term of our proposed topology, once again considering all topologies of Fig.~\ref{fig:topologies}.
To do so, let us increase the road slope angle $\slope$ in Eq.~\eqref{eq:nonlinear_system} to emulate the existence of an ascent ramp to be overcome by the vehicle team. 
\rev{For all trials, we modeled this inclination as:
\begin{align*}
    \slope =
    \begin{cases}
        ~~\n[degrees]{0}, & \text{for~~} ~~ \ipos{i} \leq ~\n[m]{135},\\
        \n[degrees]{10}, & \text{otherwise}.\\
    \end{cases}
\end{align*}
\noindent Therefore, with the given acceleration profile, the leader reaches the disturbances in about after approximately \n[s]{35} of simulation.
}

\begin{figure}[!ht]
    \centering
    \subfigure[\protect\ac{PF} topology, Zheng et al. \cite{Zheng2015Stability}.]{
        \includegraphics[width=.32\linewidth]{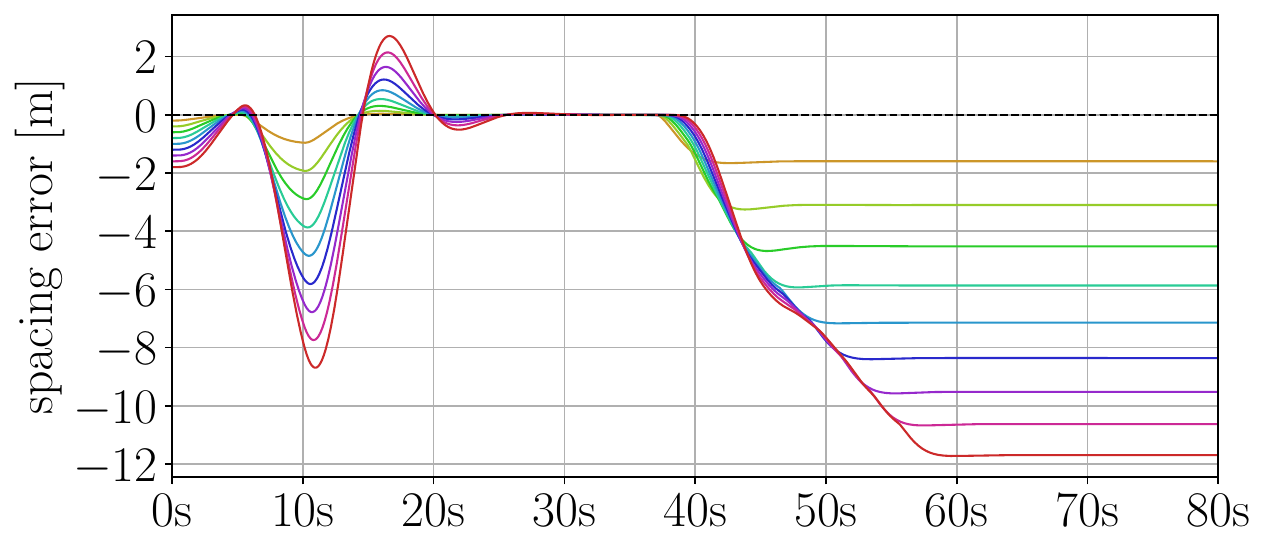}
        \label{subfig:PF_slope_Zheng}
    }
    \subfigure[\protect\ac{PF} topology, \protect\ac{SRL}.]{
        \includegraphics[width=.32\linewidth]{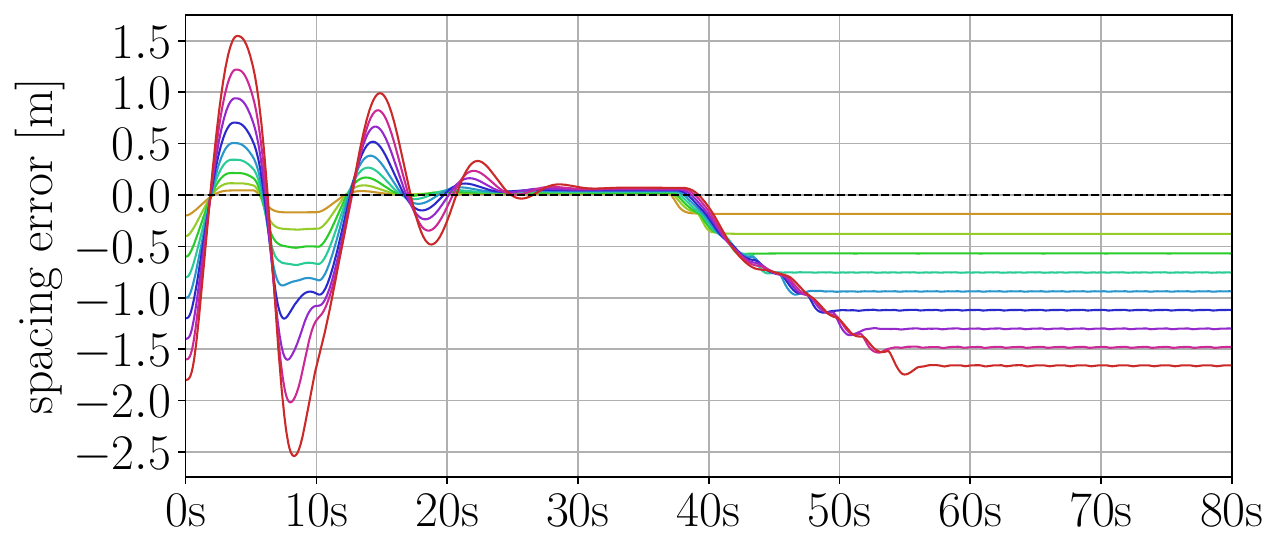}
        \label{subfig:PF_slope_RL}
    }
    \subfigure[\protect\ac{PF} topology, proposed \protect\ac{RRL}.]{
        \includegraphics[width=.32\linewidth]{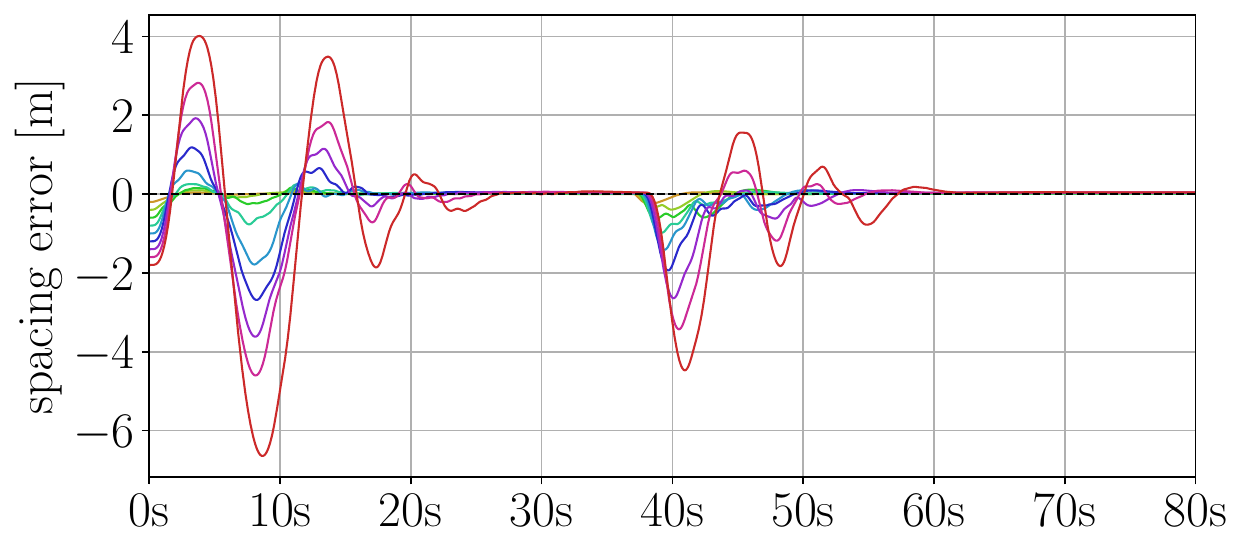}
        \label{subfig:PF_slope_Antiwindup}
    }
    \caption{Impact of a road slope of $\slope = \n[degrees]{10}$ for the \protect\ac{PF}: a) conventional consensus strategy; b)  \protect\ac{SRL} method; and c) our proposed \protect\ac{RRL} approach.}
    \label{fig:PF_slope}
\end{figure}

\begin{figure}[!ht]
    \centering
    \subfigure[\protect\ac{PFL} topology, Zheng et al. \cite{Zheng2015Stability}.]{
        \includegraphics[width=.32\linewidth]{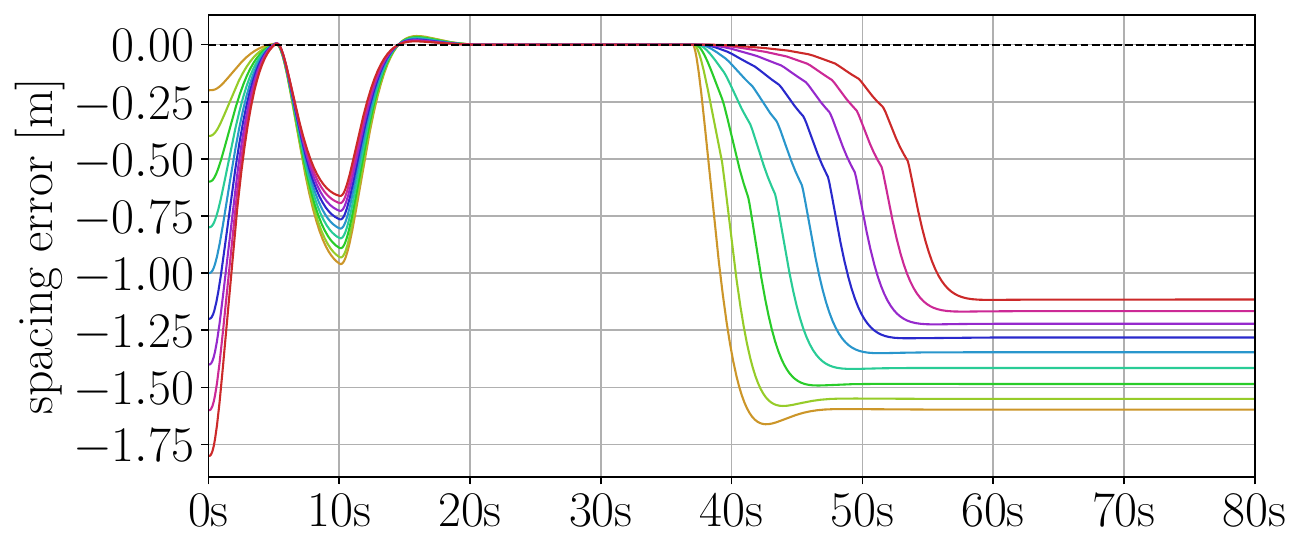}
        \label{subfig:PFL_slope_Zheng}
    }
    \subfigure[\protect\ac{PFL} topology, \protect\ac{SRL}.]{
        \includegraphics[width=.32\linewidth]{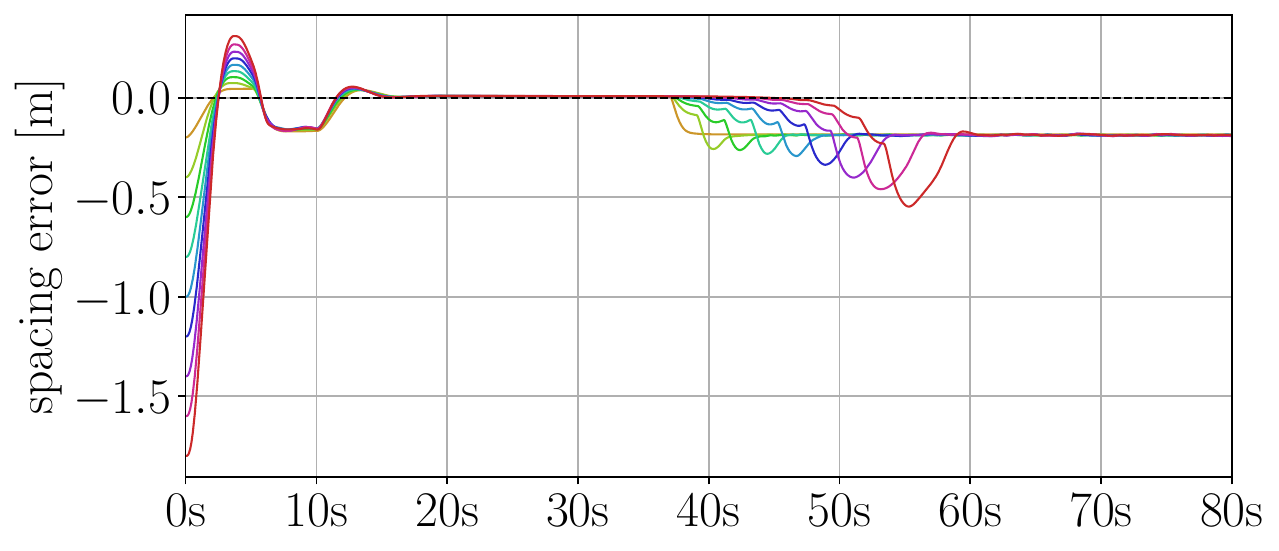}
        \label{subfig:PFL_slope_RL}
    }
    \subfigure[\protect\ac{PFL} topology, proposed \protect\ac{RRL}.]{
        \includegraphics[width=.32\linewidth]{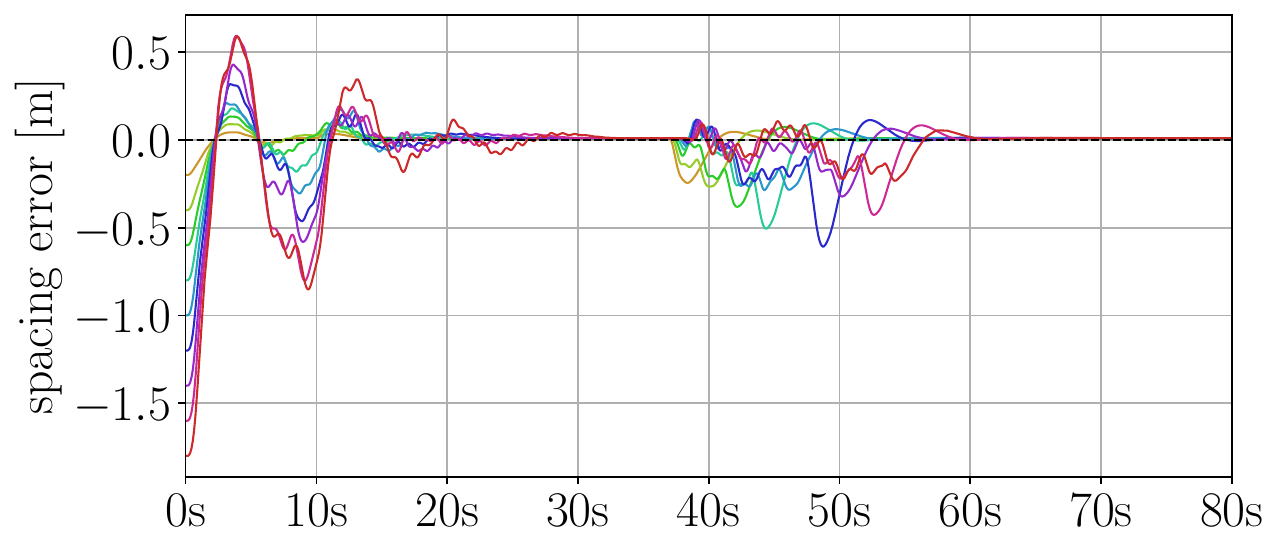}
        \label{subfig:PFL_slope_Antiwindup}
    }
    \caption{Impact of a road slope of $\slope = \n[degrees]{10}$ for the \protect\ac{PFL}: a) conventional consensus strategy; b)  \protect\ac{SRL} method; and c) our proposed \protect\ac{RRL} approach.}
    \label{fig:PFL_slope}
\end{figure}

\begin{figure}[!ht]
    \centering
    \subfigure[\protect\ac{TPF} topology, Zheng et al. \cite{Zheng2015Stability}.]{
        \includegraphics[width=.32\linewidth]{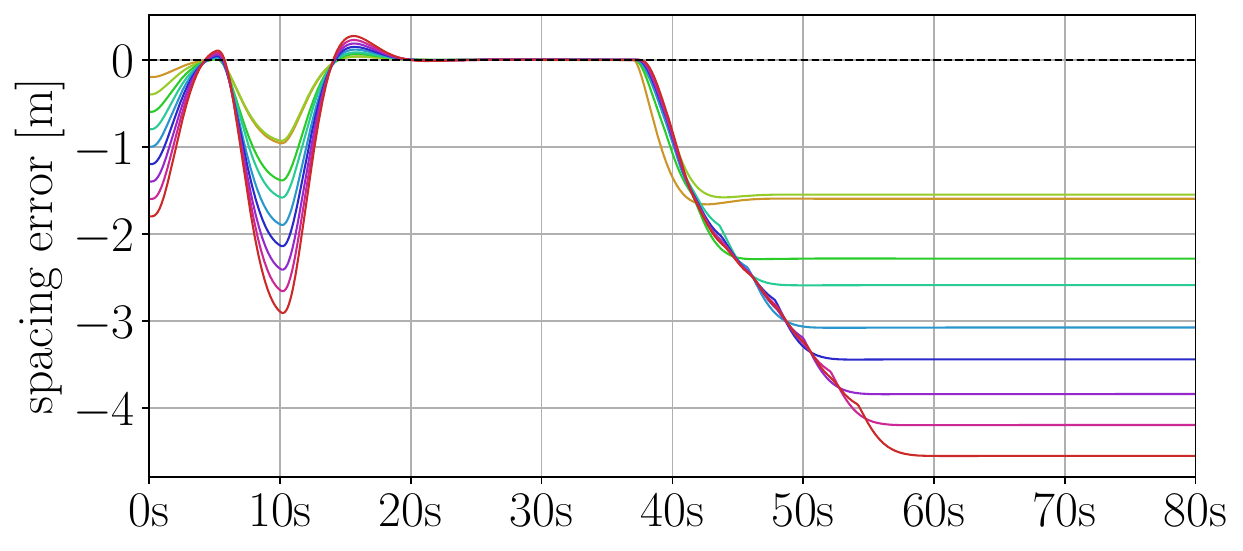}
        \label{subfig:TPF_slope_Zheng}
    }
    \subfigure[\protect\ac{TPF} topology, \protect\ac{SRL}.]{
        \includegraphics[width=.32\linewidth]{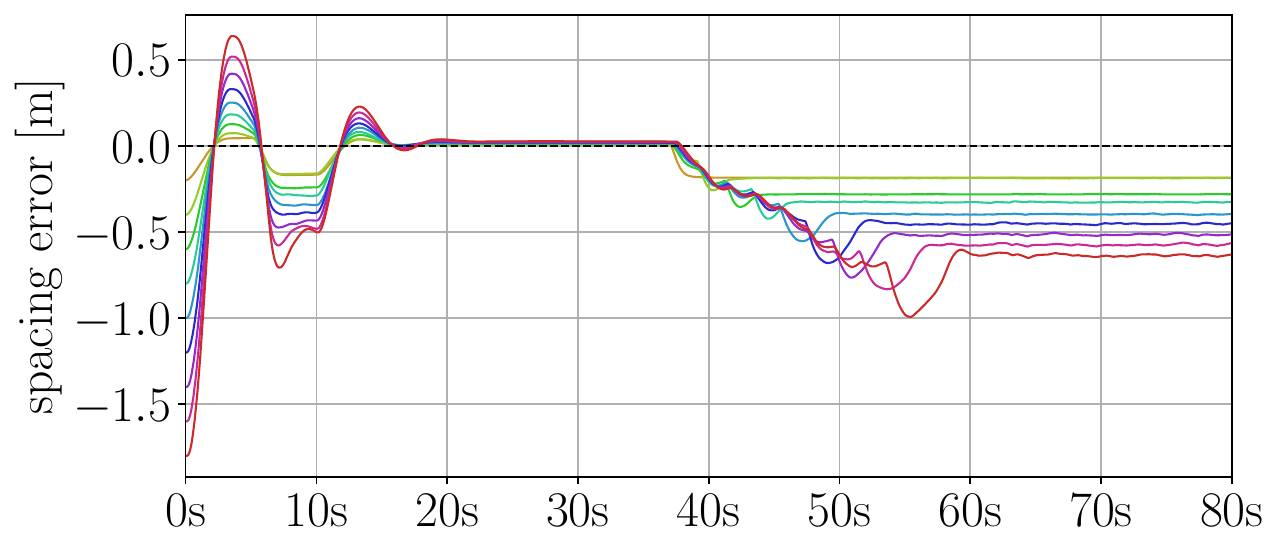}
        \label{subfig:TPF_slope_RL}
    }
    \subfigure[\protect\ac{TPF} topology, proposed \protect\ac{RRL}.]{
        \includegraphics[width=.32\linewidth]{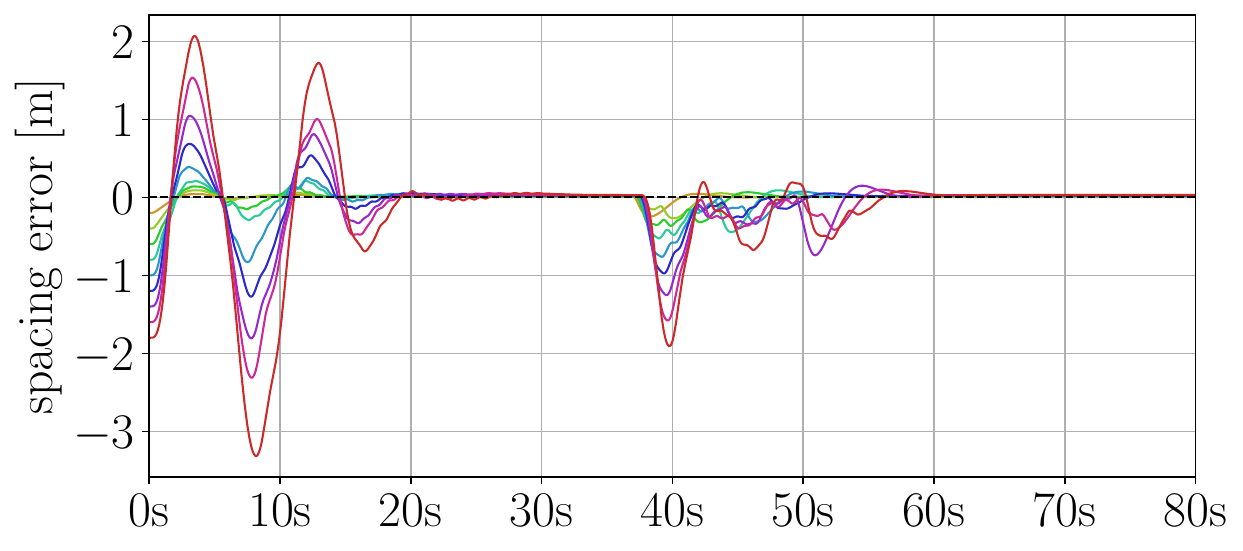}
        \label{subfig:TPF_slope_Antiwindup}
    }
    \caption{Impact of a road slope of $\slope = \n[degrees]{10}$ for the \protect\ac{TPF}: a) conventional consensus strategy; b)  \protect\ac{SRL} method; and c) our proposed \protect\ac{RRL} approach.}
    \label{fig:TPF_slope}
\end{figure}

\begin{figure}[!ht]
    \centering
    \subfigure[\protect\ac{TPFL} topology, Zheng et al. \cite{Zheng2015Stability}.]{
        \includegraphics[width=.32\linewidth]{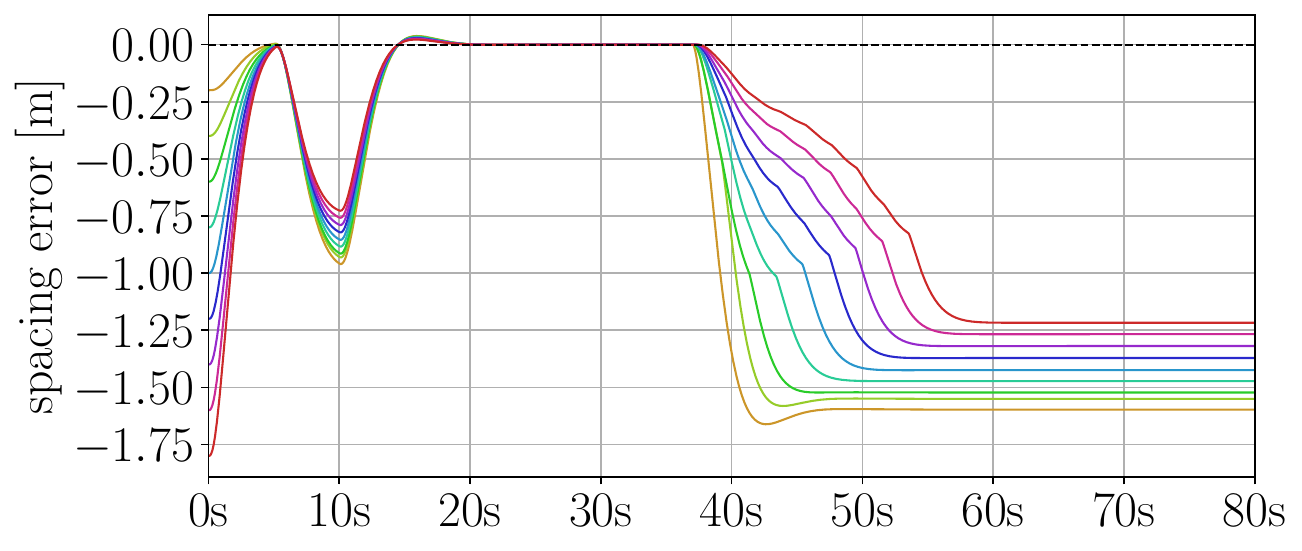}
        \label{subfig:TPFL_slope_Zheng}
    }
    \subfigure[\protect\ac{TPFL} topology, \protect\ac{SRL}.]{
        \includegraphics[width=.32\linewidth]{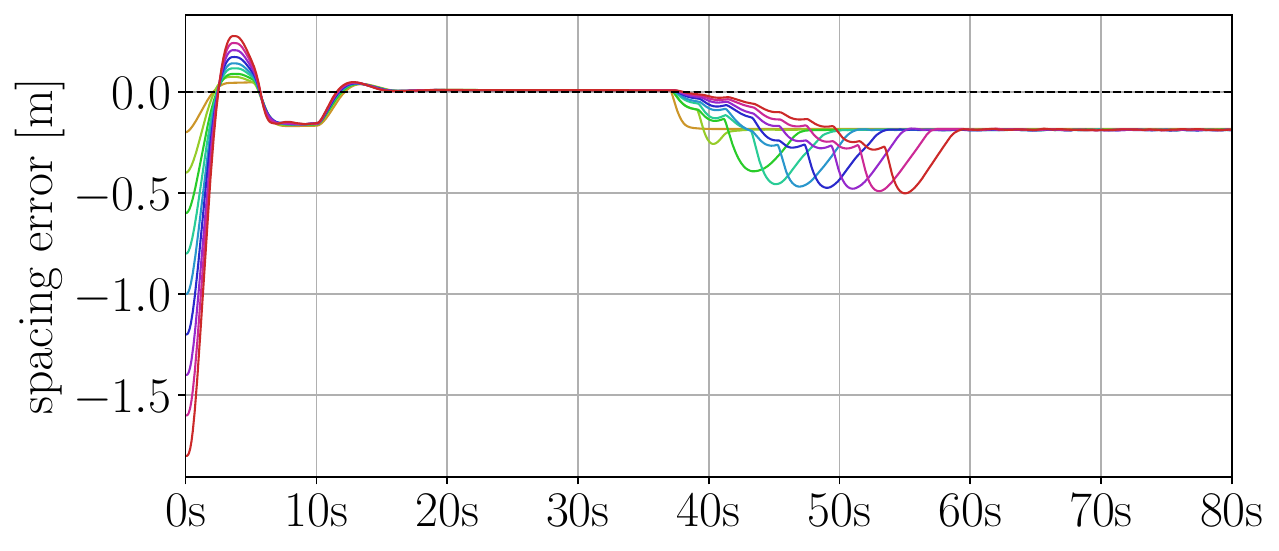}
        \label{subfig:TPFL_slope_RL}
    }
    \subfigure[\protect\ac{TPFL} topology, proposed \protect\ac{RRL}.]{
        \includegraphics[width=.32\linewidth]{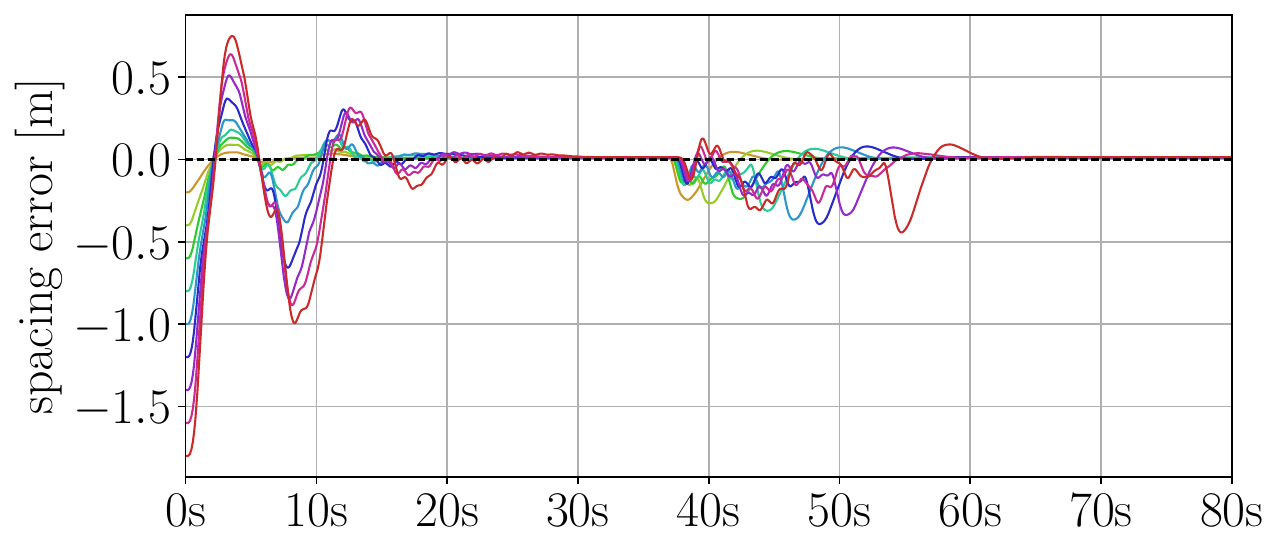}
        \label{subfig:TPFL_slope_Antiwindup}
    }
    \caption{Impact of a road slope of $\slope = \n[degrees]{10}$ for the \protect\ac{TPFL}: a) conventional consensus strategy; b)  \protect\ac{SRL} method; and c) our proposed \protect\ac{RRL} approach.}
    \label{fig:TPFL_slope}
\end{figure}

Figures \ref{fig:PF_slope}, \ref{fig:PFL_slope}, \ref{fig:TPF_slope} and \ref{fig:TPFL_slope} presents the step response of the platoon to the \ac{PF}, \ac{PFL}, \ac{TPF} and \ac{TPFL}, respectively. All of them show the susceptibility of the current state-of-the-art approaches to external forces acting on the system. This negative effect is more evident in the network topologies in which the agent doesn't communicate directly with the leader (\ac{PF} and \ac{TPF}). In leader-based hierarchies, the agent tends to perceive and respond more quickly to disturbances in the steady state, although, in practice, they are hard to implement for very long platoons due to real-world communication constraints.
On the other hand, our proposed method performed  more robustly to adverse road conditions. The same behavior may be expected with the introduction of wind, although ground vehicles are less susceptible to this type of disturbance.

\rev{Next, we also evaluate the cumulative impact of a topology change and external disturbances in the evaluated methods. The idea is to demonstrate how our proposed approach behaves in a more realistic traffic scenario. Figure \ref{fig:TPFL_slope_topology} presents a case in which the platoon composed of ten vehicles, starting with the \ac{TPFL} topology, is subjected to a road slope of $\slope = \n[degrees]{5}$ after \n[s]{40} of simulation. Similar to the results in Fig.~\ref{fig:TPFL_slope}, our approach is the only capable of completely rejecting the disturbance. But in the sequence, at \n[s]{80}, we change the current communication topology from \ac{TPFL} to \ac{PF}, and then, it is possible to observe that the controllers' response degrades significantly, with the exception of our method, which continues to reject disturbances.

\begin{figure}[!ht]
    \revcaption
    \centering
    \subfigure[\rev{\protect\ac{TPFL} topology, Zheng et al. \cite{Zheng2015Stability}.}]{
        \includegraphics[width=.32\linewidth]{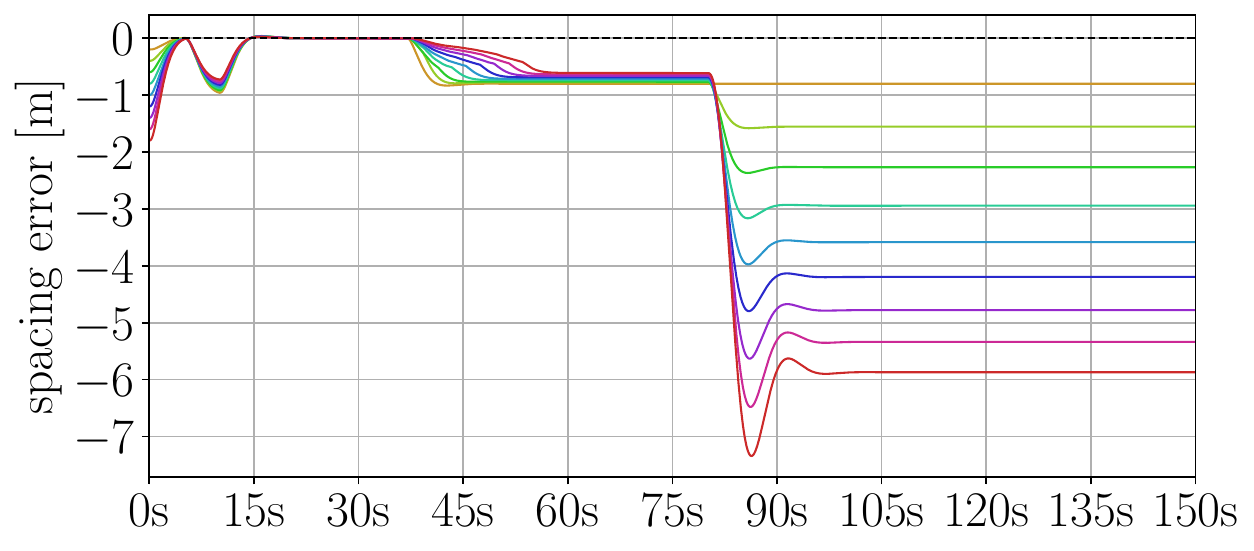}
        \label{subfig:TPFL_slope_topology_Zheng}
    }
    \subfigure[\rev{\protect\ac{TPFL} topology, \protect\ac{SRL}.}]{
        \includegraphics[width=.32\linewidth]{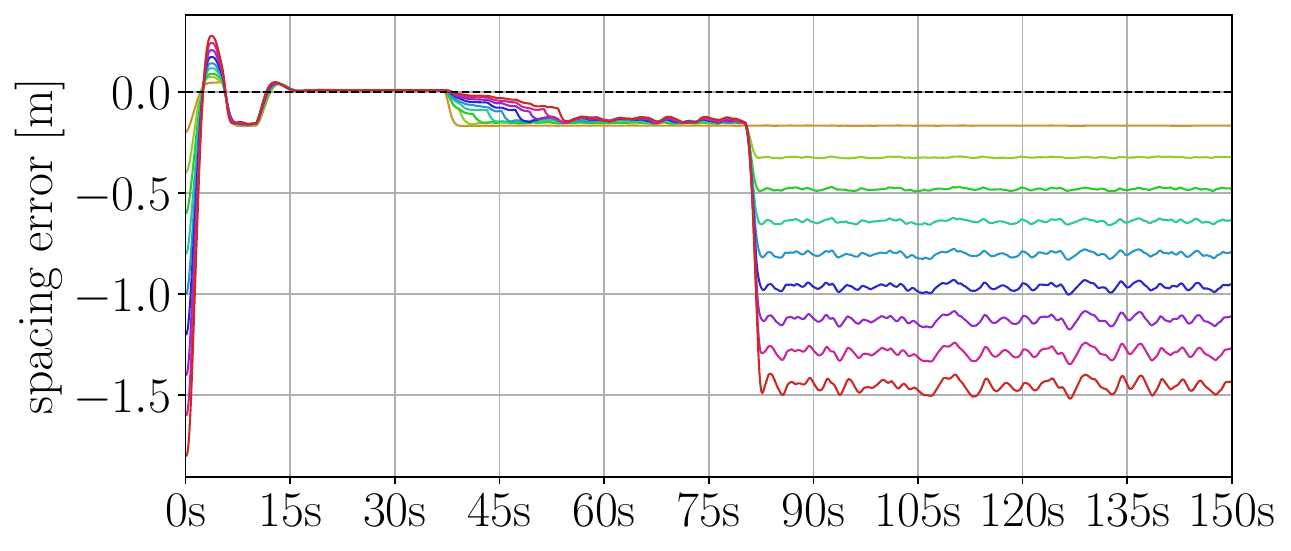}
        \label{subfig:TPFL_slope_topology_RL}
    }
    \subfigure[\rev{\protect\ac{TPFL} topology, proposed \protect\ac{RRL}.}]{
        \includegraphics[width=.32\linewidth]{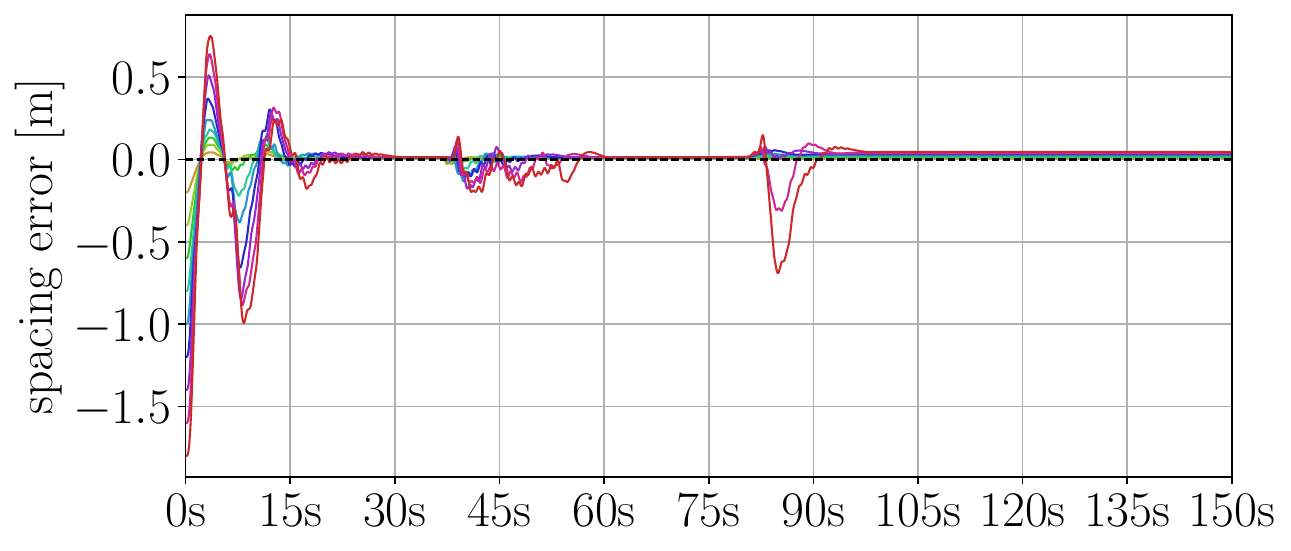}
        \label{subfig:TPFL_slope_topology_Antiwindup}
    }
    \caption{Impact of a road slope of $\slope = \n[degrees]{5}$ and switching topology from \protect\ac{TPFL} to \protect\ac{PF}: a) conventional consensus strategy;b)  \protect\ac{SRL} method; and c) our proposed \protect\ac{RRL} approach.}
    \label{fig:TPFL_slope_topology}
\end{figure}

This experiment is repeated for every pair of topologies. Because of the lack of space, instead of showing the simulation graphics, Table~\ref{tab:ise} compiles the integral square error (ISE) for the spacing error summed over all nine following vehicles in the platoon, as a figure of merit. The lines indicate the starting topology whereas the columns indicate the final topology. Due to the large oscillations in the transitory and the constant errors in the steady-state, Zheng and \protect\ac{SRL} provide very large values for the ISE. In other words, in the proposed metric the inability to provide convergence of the error was severely penalized. In every case tested, the proposed methodology performed better providing lower values of the ISE. This was possible because the proposed approach ensured zero error in the formation after the disturbances occurred. The analysis of Table~\ref{tab:ise} reveals that when the  \ac{SRL} starts with the \ac{PF} topology and then is forced to operate with other topologies the result is with the same order of magnitude than ours. However, when the \ac{SRL} starts with any topology other than \ac{PF} the results are far inferior, two or three orders of magnitude above the proposed. This denotes a clear deficiency of the conventional methods in the literature, failing to absorb the combined adverse effects of external disturbances and untrained topologies. 

\begin{table}[!ht]
    \npproductsign{\!\cdot\!}
    \centering
    \revcaption
    \caption{ISE of the spacing error for a platoon with ten vehicles subject to a road slope of $\slope = \n[degrees]{5}$ after \n[s]{40} and a switching communication after \n[s]{80} for different combinations of topologies. Each line shows the starting topology for the platoon and the columns show the swapped topology.}
    \rev{
    \nprounddigits{2}
    \resizebox{\linewidth}{!}{
    \begin{tabular}{l|ccc|ccc|ccc|ccc}
        \addlinespace
        \hline
        & \multicolumn{3}{c|}{\bf to \ac{PF}} & \multicolumn{3}{c|}{\bf to \ac{PFL}} & \multicolumn{3}{c|}{\bf to \ac{TPF}} & \multicolumn{3}{c}{\bf to \ac{TPFL}} \\
        %\hline
        & Zheng \cite{Zheng2015Stability} & \ac{SRL} & Ours & Zheng \cite{Zheng2015Stability} & \ac{SRL} & Ours & Zheng \cite{Zheng2015Stability} & \ac{SRL} & Ours & Zheng \cite{Zheng2015Stability} & \ac{SRL} & Ours \\
        \hline
        \bf from \ac{PF} & \multicolumn{3}{c|}{--} & \n{9.8304e3} & \n{5.3299e2} & \bf\n{1.1205e2} & \n{1.2456e4} & \n{6.4847e2} & \bf\n{1.1248e2} & \n{9.8746e3} & \n{5.3163e2} & \bf\n{1.1210e2}\\
        \bf from \ac{PFL} & \n{1.8461e4} & \n{1.0264e3} & \bf\n{4.5885} & \multicolumn{3}{c|}{--} & \n{3.4260e3} & \n{1.6324e2} & \bf\n{1.5769} & \n{9.3142e2} & \n{4.1485e1} & \bf\n{1.2056}\\
        \bf from \ac{TPF} & \n{1.9791e4} & \n{1.0928e3} & \bf\n{5.6193} & \n{2.2168e3} & \n{1.0424e2} & \bf\n{2.4262} & \multicolumn{3}{c|}{--} & \n{2.2668e3} & \n{1.0377e2} & \bf\n{2.4770}\\
        \bf from \ac{TPFL} & \n{1.8494e4} & \n{1.0261e3} & \bf\n{4.5955} & \n{9.1589e2} & \n{4.2312e1} & \bf\n{9.3764} & \n{3.4611e3} & \n{1.6326e2} & \bf\n{1.3586} & \multicolumn{3}{c}{--}\\
        \hline
    \end{tabular}
    }
    }
    \label{tab:ise}
\end{table}
}

%%%%%%%%%%%%%%%%%%%%%%%%%%%%%%%%%%%%%%%%%%%%%%%%%%%%%%%%%%%%%%
\rev{
\subsection{Experiments using the CARLA simulator}

Finally, in this last experiment, we have implemented our strategy with a heterogeneous team composed of the leader and nine followers in the \emph{CARLA} simulator\footnote{\rev{\url{https://carla.org/}}} \cite{Dosovitskiy17}, an open-source tool for autonomous driving research that provides a client/server interface allowing for external communication with many languages, including Python. Figure~\ref{fig:carla_simulator} illustrates the simulation environment.
To stress our policy, we have incorporated all the disturbing effects previously discussed in the same trial. In addition to the imperfect dynamic cancellation (a consequence of not knowing the exact parameters of the vehicles inside the simulator), we added a \n[degrees]{5} slope to the road and used the \ac{TPFL} communication topology instead of the \ac{PF} (used in the training stage). The world file employed in the tests was the \emph{Town06}, a scenario with a long straight road. To avoid modifying it, the road slope was emulated by applying an external force applied to the vehicles' center of mass proportional to their masses.

\begin{figure}[!ht]
    \revcaption
    \centering
    \includegraphics[width=.6\linewidth]{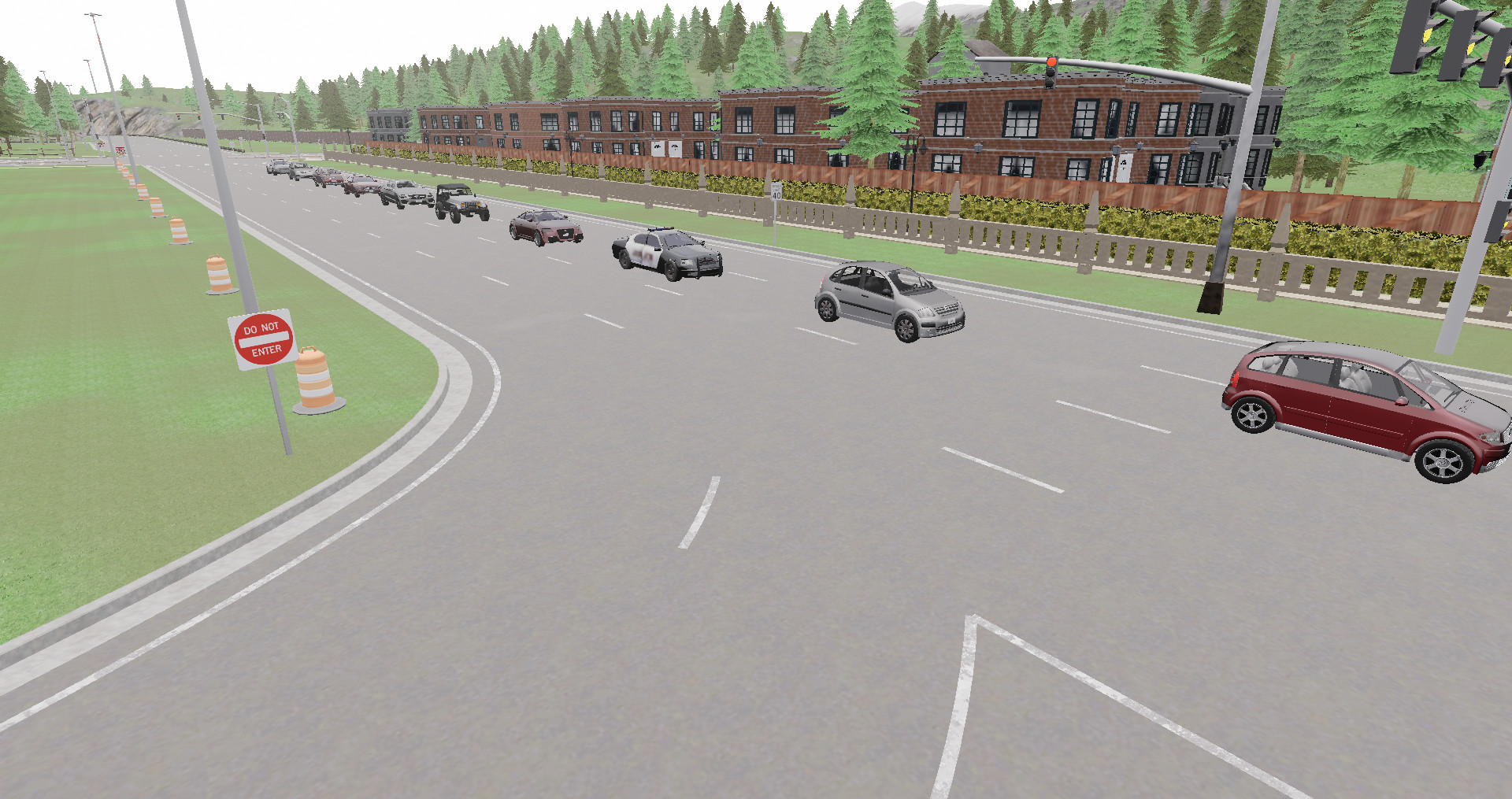}
    \caption{Heterogeneous vehicular platoon with one leader and nine followers running in the \emph{CARLA} simulator, subjected to the \protect\ac{TPFL} topology, uncertainties in the parameters and a road slope of $\slope = \n[degrees]{5}$.}
    \label{fig:carla_simulator}
\end{figure}

To simulate a more realistic traffic scenario, the leader's acceleration profile was now given by:
\begin{align*}
    \iuin{0}(t) =
    \begin{cases}
        \n[m/s^2]{2.5}, & \text{if~~} ~~\n[s]{0} < t \leq ~\n[s]{3},\\
        \n[m/s^2]{3.}, & \text{if~~} \n[s]{32} < t \leq \n[s]{34},\\ 
        \n[m/s^2]{0.}, & \text{otherwise}.\\
    \end{cases}
\end{align*}

As it can be seen in Fig.~\ref{fig:TPFL_slopeCARLA}, all three methods were able to stabilize the platoon, but similarly to our previous simulations, our proposed method was shown to be the most robust to the disturbances and uncertainties.

\begin{figure}[!ht]
    \revcaption
    \centering
    \subfigure[\rev{\protect\ac{TPFL} topology, Zheng et al. \cite{Zheng2015Stability}.}]{
        \includegraphics[width=.32\linewidth]{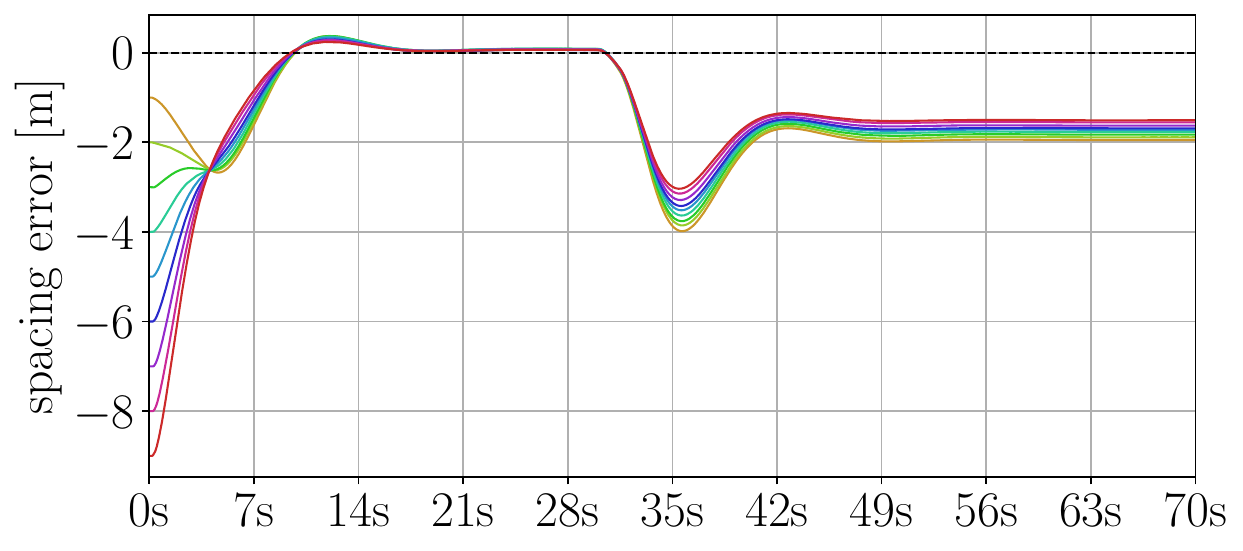}
        \label{subfig:TPFL_slopeCARLA_Zheng}
    }
    \subfigure[\rev{\protect\ac{TPFL} topology, \protect\ac{SRL}.}]{
        \includegraphics[width=.32\linewidth]{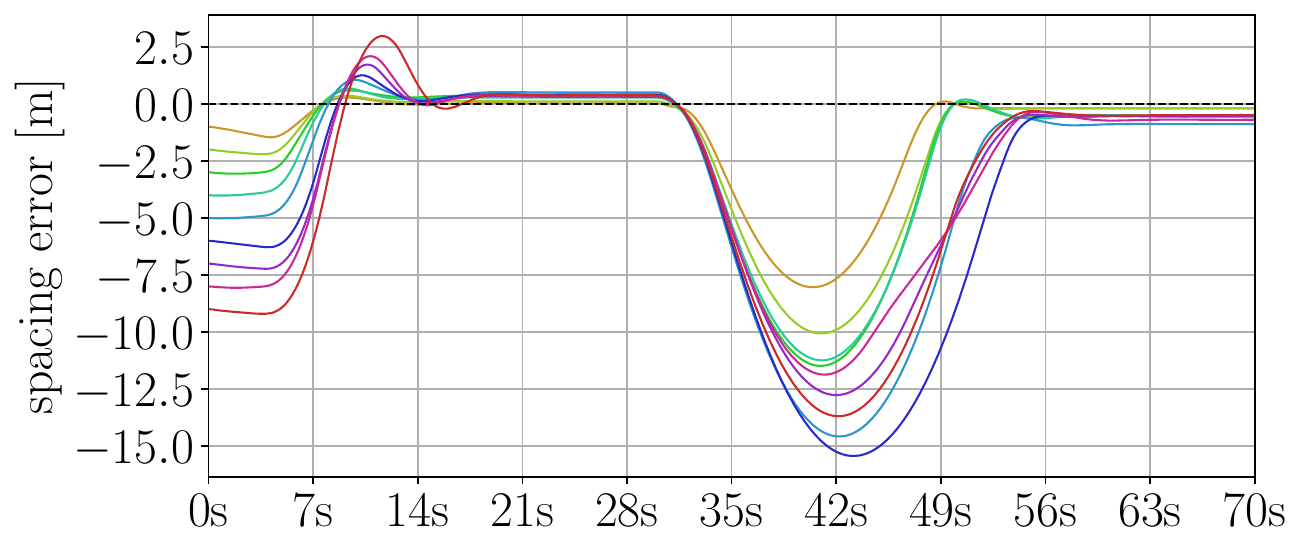}
        \label{subfig:TPFL_slopeCARLA_RL}
    }
    \subfigure[\rev{\protect\ac{TPFL} topology, proposed \protect\ac{RRL}.}]{
        \includegraphics[width=.32\linewidth]{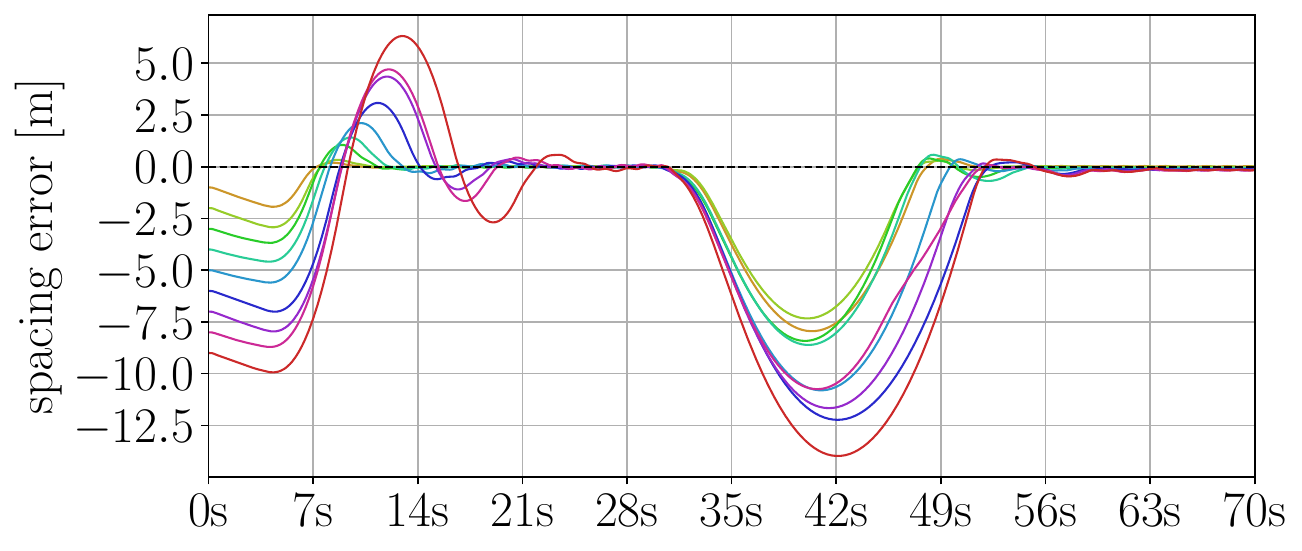}
        \label{subfig:TPFL_slopeCARLA_Antiwindup}
    }
    \caption{Test in the \emph{CARLA} simulator with a road slope of $\slope = \n[degrees]{5}$ and \protect\ac{TPFL} topology: a) conventional consensus strategy; b) \protect\ac{SRL} method; and c) our proposed \protect\ac{RRL} approach.}
    \label{fig:TPFL_slopeCARLA}
\end{figure}

}

%%%%%%%%%%%%%%%%%%%%%%%%%%%%%%%%%%%%%%%%%%%%%%%%%%%%%%%%%%%%%%
%%%%%%%%%%%%%%%%%%%%%%%%%%%%%%%%%%%%%%%%%%%%%%%%%%%%%%%%%%%%%%
\section{Conclusion and future work}
\label{sec:conclusion}

In this paper, we have proposed a \acd{DRL} approach for the longitudinal spacing control of vehicular platoons subject to external disturbances, parametric uncertainties, and different communication topologies. The main idea was to generalize a \ace{DRL} trained police by incorporating on it an integral action term, capable of rejecting these model misspecifications. 
As a first contribution, we have modeled the \ac{CNN} input as a scalar whose value $\statev$ was given by the summation of the errors in position, speed, and acceleration between vehicle $i$ and the average virtual state given by the states in $\neighSet_{i}$. With this, we were able to train the \ac{DRL} algorithm in an environment with only two agents communicating via \ac{PF} topology, making training more simple and generalizable. However, when applying this trained policy to simulations with different topologies, the incorporation of new neighbors brings unmodeled uncertainties to the system.
Therefore, as a second contribution, to reject disturbances caused by this simplification, and by other terms such as road slope, wind, and uncertainties in the vehicle parameters, we have added an integral term to reach null error in a steady-state condition. We have also discussed some mild conditions and assumptions, under which the nonlinear policy can guarantee zero error and time-domain string stability, once the platoon is modeled as a linear SISO system. 
Simulated experiments have shown that this strategy increased the robustness of the spacing control to external disturbances and the resiliency of the platoon to different communication topologies when compared to the current literature.

In future works, we intend to deepen the study of the stability conditions for the neural network, providing less conservative conditions, trying to incorporate these criteria in the training itself, and stipulating tighter bounds for the nonlinearities. The bi-directional topologies are another interesting scenarios to expand our research and we expect that our proposed approach can also be extended to platoons with latero-directional behavior, where we can incorporate lateral disturbances such as sideslip, and curvature constraints. \rev{Another important extension of the work is in the direction of improving security against time-delay communication, message falsification, and external attacks.} Finally, we can investigate how other reward functions aimed to optimize energy consumption or string stability, affect our method.

\pagebreak
\newpage

%%%%%%%%%%%%%%%%%%%%%%%%%%%%%%%%%%%%%%%%%%%%%%%%%%%%%%%%%%%%%%
\subsection*{Author contributions}

A. Alves Neto has implemented the algorithm, performed the simulations, and wrote the paper. L. A. Mozelli proposed the modeling, worked on the formalization, and also wrote the paper.

%%%%%%%%%%%%%%%%%%%%%%%%%%%%%%%%%%%%%%%%%%%%%%%%%%%%%%%%%%%%%%
\subsection*{Financial disclosure}

This study was financed by the \ac{CAPES} (Finance Codes 001 and 88887.136349/2017-00), \ac{CNPq} (Process 310446/2021-0), and \ac{FAPEMIG}.

%%%%%%%%%%%%%%%%%%%%%%%%%%%%%%%%%%%%%%%%%%%%%%%%%%%%%%%%%%%%%%
\subsection*{Conflict of interest}

The authors declare no potential conflict of interests.

%%%%%%%%%%%%%%%%%%%%%%%%%%%%%%%%%%%%%%%%%%%%%%%%%%%%%%%%%%%%%%
%\nocite{*}% Show all bib entries - both cited and uncited; comment this line to view only cited bib entries;
\bibliography{rl_platoon}%

%%%%%%%%%%%%%%%%%%%%%%%%%%%%%%%%%%%%%%%%%%%%%%%%%%%%%%%%%%%%%%
% \clearpage

% \section*{Author Biography}

% \begin{biography}{\includegraphics[width=66pt,height=86pt]{bios/armando.pdf}}{\textbf{Armando Alves Neto} received the B.S.E. degree in Automation and Control Engineering from the Universidade Federal de Minas Gerais in 2006, and S.M. and Ph.D. degrees in Computer Science from UFMG in 2008 and 2012, respectively. He is an Assistant Professor in the Department of Electronic Engineering at UFMG. Research interests include real-time motion planning, multi-agent control, robust control, and collision avoidance strategies.}
% \end{biography}

% \bigskip

% \begin{biography}{\includegraphics[width=66pt,height=86pt]{bios/mozelli.pdf}}{\textbf{Leonardo Amaral Mozelli} is currently an Associate Professor in the Department of Electronics Engineering at Federal University of Minas Gerais (UFMG), Belo Horizonte, Brazil. He teaches in the interlinked areas of signal processing, dynamic systems, and automatic control.  In a broad sense, his research interests include: adaptive signal processing, non-linear and robust control theory, multi-agent systems, and technologies for sustainable development.}
% \end{biography}

\end{document}